\documentclass[twocolumn,prd,nofootinbib,superscriptaddress]{revtex4-2}
\usepackage{color}
\usepackage{amsmath}
\usepackage{mathrsfs}
\usepackage{graphicx}
\usepackage{booktabs}
\usepackage{bm}
\usepackage{tikz}

\newcommand\apjl{ApJL}
\newcommand\apjs{ApJS}

\newcommand\cqg{CQG}

\newcommand\SkipNeuralNets[1]{}
\newcommand\skipme[1]{}
\newcommand\mysub[1]{\subsubsection{#1}}

\newcommand\qmstateproduct[2]{\left\langle#1|#2\right\rangle}

\newcommand\Y[1]{{{}_{#1}Y}}

\newcommand\optional[1]{}
\newcommand\unit[1]{{\rm #1}}

\newcommand{\llmfig}{ Generated with LLM assistance per Appendix~\ref{app:llm}.}

\newcommand{\cmNcalDemo}{100}
\newcommand{\cmNcalDefault}{100}
\newcommand{\cmNcalRecommended}{300}
\newcommand{\cmNcalMaxFactor}{8}

\newcommand{\cmSplineNodes}{10}

\newcommand{\cmScalingSnr}{20}
\newcommand{\cmScalingNdet}{3}
\newcommand{\cmSigmaTarget}{0.1}
\newcommand{\cmSigmaPerPercent}{1.0}

\newcommand{\cmPhaseIndepMax}{5}

\newcommand{\cmNcalRefAmp}{1}
\newcommand{\cmNcalRefPhase}{1}
\newcommand{\cmNcalRef}{171}

\newcommand{\cmNcalGwtcFour}{5824}
\newcommand{\cmNcalBandLow}{10^{2}}
\newcommand{\cmNcalBandHigh}{10^{4}}
\newcommand{\cmCeilingLo}{2.6}
\newcommand{\cmCeilingHi}{3.0}

\newcommand{\cmMemPerDraw}{0.28}
\newcommand{\cmMemAtRecommended}{84}

\newcommand{\cmSelfTermWidthLo}{1}
\newcommand{\cmSelfTermWidthHi}{18}
\newcommand{\cmSelfTermSnrLo}{10}
\newcommand{\cmSelfTermSnrHi}{35}
\newcommand{\cmSelfTermBiasLo}{0.1}
\newcommand{\cmSelfTermBiasHi}{190}
\newcommand{\cmSelfTermBiasMidSnr}{20}
\newcommand{\cmSelfTermBiasMidLo}{1.1}
\newcommand{\cmSelfTermBiasMidHi}{61}
\newcommand{\cmSelfTermCompleteTol}{0.05}
\newcommand{\cmSelfTermCompleteWorst}{0.6}

\newcommand{\cmSelfTermGWSnr}{8}
\newcommand{\cmSelfTermGWShift}{-0.07}
\newcommand{\cmSelfTermGWShiftErr}{0.04}
\newcommand{\cmSelfTermGWPred}{-0.04}
\newcommand{\cmCalBoundAmp}{2}
\newcommand{\cmCalBoundPhase}{2}
\newcommand{\cmCalBoundFmax}{2}
\newcommand{\cmGWfourInJS}{4\times10^{-3}}

\newcommand{\cmGWfourNeffCal}{69}

\newcommand{\cmGWfourCostRatio}{6.2}

\newcommand{\tieNsky}{100}
\newcommand{\tieCornerFmax}{895.875}
\newcommand{\tieCornerRate}{2048}
\newcommand{\tieNsnr}{10}
\newcommand{\tieRate}{4096}
\newcommand{\tieFmax}{1792}
\newcommand{\tieFmin}{13.33}

\newcommand{\mtFineFactor}{32}

\newcommand{\toaAccuracyBudget}{0.02} %
\newcommand{\toaNativeQRateHz}{4096} %
\newcommand{\toaPregridFactor}{8} %
\newcommand{\toaPregridRateHz}{32768} %
\newcommand{\toaValidationSnrMax}{652.31} %

\newcommand{\irQScanRho}{20}

\newcommand{\maLnZSEOB}{482.1}
\newcommand{\maLnZPhenomD}{479.2}
\newcommand{\maDeltaLnZ}{2.9}
\newcommand{\maBayesFactor}{19}
\newcommand{\maWeightSEOB}{0.95}
\newcommand{\maWeightPhenomD}{0.05}
\newcommand{\skyCompleteAnchors}{30,000}
\newcommand{\skyCompleteDraws}{300,000}

\newcommand{\skyCompletePerAnchor}{10}
\newcommand{\skyExportEss}{10}
\newcommand{\skySavedAnchors}{5,000}

\newcommand{\gwgNreps}{4}
\newcommand{\gwgNdraws}{20{,}000}

\newcommand{\eccPolicyAnchorX}{0.05}
\newcommand{\eccPolicyBandLow}{20}
\newcommand{\eccPolicyBandHigh}{40}

\newcommand{\eccUniformMOne}{20}
\newcommand{\eccUniformMTwo}{16}
\newcommand{\eccUniformETruth}{0.1}
\newcommand{\eccUniformEllTruth}{1}
\newcommand{\eccUniformSNR}{35}
\newcommand{\eccUniformFNative}{10}
\newcommand{\eccUniformFShared}{20}
\newcommand{\eccUniformEMin}{0}
\newcommand{\eccUniformEMax}{0.25}
\newcommand{\eccUniformPosteriorN}{4000}
\newcommand{\eccUniformCI}{90}
\newcommand{\eccUniformIteration}{7}

\newcommand{\eccSpinDirectOffsetA}{+1.25}
\newcommand{\eccSpinDirectOffsetB}{-1.34}

\newcommand{\etcDurationS}{8192}
\newcommand{\etcFmin}{5}
\newcommand{\etcSrate}{4096}
\newcommand{\etcLmax}{4}
\newcommand{\etcMassOne}{1.40}
\newcommand{\etcMassTwo}{1.35}
\newcommand{\etcEarlyRate}{256}
\newcommand{\etcTailS}{400}
\newcommand{\etcAttenDb}{160}
\newcommand{\etcUVMarginS}{32}
\newcommand{\etcGpuModel}{RTX PRO 4000 Blackwell}
\newcommand{\etcGpuMemGB}{24}
\newcommand{\etcAccFmin}{10}
\newcommand{\etcAccDurationS}{2048}
\newcommand{\etcAccTailS}{100}
\newcommand{\etcAccRho}{1000}
\newcommand{\etcDlnLMax}{\ensuremath{2.3\times10^{-4}}}
\newcommand{\etcDlnLGate}{0.1}
\newcommand{\etcDThreeGate}{0.01}
\newcommand{\etcQErr}{\ensuremath{1.1\times10^{-5}}}
\newcommand{\etcUVErr}{\ensuremath{8.4\times10^{-8}}}
\newcommand{\etcPeakShift}{\ensuremath{2\times10^{-9}}}
\newcommand{\etcDeviceErr}{\ensuremath{6\times10^{-15}}}
\newcommand{\etcProdSOne}{498}
\newcommand{\etcMultiSOne}{85}
\newcommand{\etcRatioOne}{4.5}
\newcommand{\etcProdRssOne}{18.7}
\newcommand{\etcMultiRssOne}{5.0}
\newcommand{\etcProdSTwo}{522}
\newcommand{\etcMultiSTwo}{79}
\newcommand{\etcRatioTwo}{6.1}
\newcommand{\etcProdRssTwo}{19.1}
\newcommand{\etcMultiRssTwo}{7.1}
\newcommand{\etcProdSThree}{663}
\newcommand{\etcMultiSThree}{86}
\newcommand{\etcRatioThree}{7.7}
\newcommand{\etcProdRssThree}{20.0}
\newcommand{\etcMultiRssThree}{8.5}
\newcommand{\etcRepeats}{5}
\newcommand{\etcJobRatioMin}{1.8}
\newcommand{\etcJobRatioMax}{1.9}
\newcommand{\etcMultiSMin}{79}
\newcommand{\etcMultiSMax}{86}
\newcommand{\etcRatioMin}{4.5}
\newcommand{\etcRatioMax}{7.7}
\newcommand{\etcRssRatioMin}{2.4}
\newcommand{\etcRssRatioMax}{3.7}
\newcommand{\etcWaveformMinS}{37}
\newcommand{\etcWaveformMaxS}{42}
\newcommand{\etcXpProdSOne}{463}
\newcommand{\etcXpMultiSOne}{25.7}
\newcommand{\etcXpRatioOne}{18.6}
\newcommand{\etcXpProdSTwo}{737}
\newcommand{\etcXpMultiSTwo}{31.6}
\newcommand{\etcXpRatioTwo}{23.3}
\newcommand{\etcXpProdSThree}{908}
\newcommand{\etcXpMultiSThree}{35.1}
\newcommand{\etcXpRatioThree}{28.5}
\newcommand{\etcXpRepeats}{5}
\newcommand{\etcXpMultiSMin}{26}
\newcommand{\etcXpMultiSMax}{35}
\newcommand{\etcXpRatioMin}{19}
\newcommand{\etcXpRatioMax}{29}
\newcommand{\etcXpJobRatioMin}{2.0}
\newcommand{\etcXpJobRatioMax}{2.4}
\newcommand{\etcConvFracMin}{\ensuremath{1.7\times10^{-4}}}
\newcommand{\etcConvFracMax}{\ensuremath{1.2\times10^{-3}}}
\newcommand{\etcConvRhoLo}{300}
\newcommand{\etcConvNatsLoMin}{7}
\newcommand{\etcConvNatsLoMax}{55}
\newcommand{\etcConvRhoHi}{1000}
\newcommand{\etcConvNatsHiMin}{83}
\newcommand{\etcConvNatsHiMax}{610}
\newcommand{\etcConvPerMFrac}{\ensuremath{9.4\times10^{-4}}}
\newcommand{\etcConvMbandFrac}{\ensuremath{1.6\times10^{-4}}}
\newcommand{\etcConvStandoffFrac}{\ensuremath{2.9\times10^{-4}}}
\newcommand{\etcConvSameFrac}{\ensuremath{2\times10^{-16}}}
\newcommand{\etcConvMassesA}{36+29}
\newcommand{\etcConvMassesB}{30+6}
\newcommand{\etcConvMassesC}{10+4}
\newcommand{\etcGFiveRatio}{5.1}
\newcommand{\etcGFivePairs}{6}
\newcommand{\gxpLalVersion}{6.2.0}
\newcommand{\gxpPrecVersion}{102}
\newcommand{\gxpFdStep}{\ensuremath{10^{-7}}}
\newcommand{\gxpValPoints}{50}
\newcommand{\gxpValBbh}{20}
\newcommand{\gxpValNsbh}{10}
\newcommand{\gxpValBns}{20}
\newcommand{\gxpValFminLo}{10}
\newcommand{\gxpValFminHi}{20}
\newcommand{\gxpValDurLo}{8}
\newcommand{\gxpValDurHi}{256}
\newcommand{\gxpValSrate}{4096}
\newcommand{\gxpMisTwoTwo}{\ensuremath{6.7\times10^{-16}}}
\newcommand{\gxpErrTwoTwo}{\ensuremath{1.1\times10^{-10}}}
\newcommand{\gxpMisTwoOne}{\ensuremath{1.0\times10^{-15}}}
\newcommand{\gxpErrTwoOne}{\ensuremath{1.6\times10^{-11}}}
\newcommand{\gxpMisThreeThree}{\ensuremath{8.9\times10^{-16}}}
\newcommand{\gxpErrThreeThree}{\ensuremath{3.3\times10^{-10}}}
\newcommand{\gxpMisFourFour}{\ensuremath{1.0\times10^{-15}}}
\newcommand{\gxpErrFourFour}{\ensuremath{7.0\times10^{-10}}}
\newcommand{\gxpMisThreeTwo}{\ensuremath{8.8\times10^{-7}}}
\newcommand{\gxpErrThreeTwo}{\ensuremath{5.9\times10^{-3}}}
\newcommand{\gxpErrMedThreeTwo}{\ensuremath{1.3\times10^{-7}}}
\newcommand{\gxpMisExactMax}{\ensuremath{1.0\times10^{-15}}}
\newcommand{\gxpErrExactMax}{\ensuremath{7.0\times10^{-10}}}
\newcommand{\gxpCupyNumpy}{\ensuremath{1.6\times10^{-13}}}
\newcommand{\gxpSOnePoints}{50}
\newcommand{\gxpSOneMis}{\ensuremath{1.2\times10^{-15}}}
\newcommand{\gxpSOneErr}{\ensuremath{6.4\times10^{-12}}}
\newcommand{\gxpSOneCupyNumpy}{\ensuremath{6.6\times10^{-12}}}
\newcommand{\gxpGpuModel}{RTX PRO 4000 Blackwell}
\newcommand{\gxpCpuModel}{Xeon E5-2650 v4}
\newcommand{\gxpBenchFmin}{20}
\newcommand{\gxpBatch}{100}
\newcommand{\gxpShortDur}{8}
\newcommand{\gxpShortSrate}{2048}
\newcommand{\gxpLongDur}{256}
\newcommand{\gxpLongSrate}{4096}
\newcommand{\gxpXLongDur}{2048}
\newcommand{\gxpXLongSrate}{4096}
\newcommand{\gxpCupyShortOne}{12}
\newcommand{\gxpNumpyShortOne}{22}
\newcommand{\gxpLalShortOne}{192}
\newcommand{\gxpCupyShortBatch}{945}
\newcommand{\gxpNumpyShortBatch}{54}
\newcommand{\gxpLalShortBatch}{126}
\newcommand{\gxpPoolShortBatch}{0.53}
\newcommand{\gxpSpeedShort}{7.5}
\newcommand{\gxpCupyLongOne}{11}
\newcommand{\gxpNumpyLongOne}{0.9}
\newcommand{\gxpLalLongOne}{5.2}
\newcommand{\gxpCupyLongBatch}{46}
\newcommand{\gxpNumpyLongBatch}{0.7}
\newcommand{\gxpLalLongBatch}{2.8}
\newcommand{\gxpPoolLongBatch}{6.6}
\newcommand{\gxpSpeedLong}{17}
\newcommand{\gxpCupyLongTen}{42}
\newcommand{\gxpFloorMs}{87}
\newcommand{\gxpCupyXLongTen}{5.0}
\newcommand{\gxpPoolXLongTen}{6.2}
\newcommand{\gxpLalXLongMin}{0.3}
\newcommand{\gxpLalXLongMax}{0.6}
\newcommand{\gxpOutXLongGB}{34}
\newcommand{\gxpRepeatMax}{1.9}

\newcommand{\pqCalEss}{9.13}
\newcommand\response[1]{{#1}}
\newcommand\mc{{{\cal M}_c}}

\def\RIT{Center for Computational Relativity and Gravitation, Rochester Institute of Technology, Rochester, New York
  14623, USA}

\def\UTAustin{University of Texas, Austin, TX 78712, USA}

\def\Caltech{California Institute of Technology}

\begin{document}

\title{RIFT for the O4 era: operating points, calibration marginalization, and cross-code validation}
\author{R. O'Shaughnessy}
\affiliation{\RIT}
\author{A. Jan}
\affiliation{\Caltech}
\affiliation{\UTAustin}
\author{J. Lange}
\affiliation{Cardiff University}
\author{R. Mechum}
\affiliation{\RIT}
\author{N. Malagon}
\affiliation{\RIT}
\begin{abstract}
The Rapid Iterative FiTting (RIFT) parameter  inference algorithm  provides a simulation-based-inference approach to  efficient, highly-parallelized
parameter inference for GW sources.
In this paper,  we describe improvements deployed during preparations for and analysis of O4 observations
(versions 0.0.17.3--0.0.18.0) that broaden RIFT's reach for single-event
inference with contemporary instruments.
We provide additional options to increase compatibility and comparability with other codes, notably including
(a) enhancing RIFT's capabilities for calibration marginalization;
(b) managing our likelihood's internal, output, and time-interpolated representation;
(c) introducing new coordinates for (transverse) spin and eccentric mean anomaly, enabling more robust inference of
parameters with subtle and complex correlations to others,
and (d) revised operating point choices better suited to the loudest, clearest signals.
We identify and reconcile the most substantial sources of head-to-head likelihood differences between RIFT and
other production inference codes, and describe automated waveform symmetry checks that detect upstream model
regressions before they bias inference.
Using the previously-reported RIFT/\texttt{asimov} interface for reproducible settings consistent with past work, we demonstrate
these capabilities end to end.
\end{abstract}
\maketitle

\section{Introduction}
Ground-based gravitational-wave (GW) detectors---Advanced LIGO
\cite{2015CQGra..32g4001L,Capote:O4performance,Soni:O4aDetChar}, Virgo
\cite{gw-detectors-Virgo-original-preferred,2015CQGra..32b4001A}, and KAGRA
\cite{2021PTEP.2021eA101A}---continue to identify and characterize coalescing compact binaries
\cite{LIGO-O3-O3bcatalog,LIGO-O3-O3a_final-catalog,LIGO-O4a-cbc-catalog_results,LIGO-O4b-CBC-catalog},
using Bayesian inference to reconstruct the properties of each source in the census by comparing their observations with
the predictions  of waveform models \cite{gw-astro-PE-lalinference-v1,gwastro-pe-bilby-2018,gwastro-PENR-RIFT,gwastro-RIFT-Update,gwastro-RIFT_FinerNet,RevModPhys.94.025001}.
Many different software packages have tackled this challenge,  implementing a variety of algorithms, simplifications,
approximations, and acceleration techniques
\cite{gw-astro-PE-lalinference-v1,gwastro-pe-bilby-2018,gwastro-PENR-RIFT,gw-astro-mergers-dingo-part2,2019PASP..131b4503B,2022PhRvD.106l3015R,2023ApJ...958..129W};
see \cite{RevModPhys.94.025001,LIGO-O4b-CBC-methods} and references therein.
Fewer source parameter inference codes have been applied to production-scale interpretation of GW observations including
whole-catalog reanalyses
\cite{LIGO-O3-O3bcatalog,2023ApJ...946...59N,2022arXiv220102252O,gw-astro-mergers-dingo-part2,gwastro-mergers-TousifGWTC3,gwastro-mergers-ManningReanalyzeO3}.
Too, careful analysis of GW sources requires large-scale investigation of multiple sources of systematic uncertainty,
including the impact of omitted physics \cite{2024PhRvD.109d3037D,2016ApJ...820....7L};
differences between different approximations to general relativity \cite{gwastro-RIFT-systematics-AnjaliAasim-2020,gwastro-mergers-bns-systematics-Yelikar2024};
and propagation of uncertainties about the detectors themselves including calibration error \cite{Farr2014Calibration,2020PhRvD.102l2004P}.
As the census grows in scale, as previously-exceptional configurations become more frequent, and particularly as
cross-code validation finally becomes possible at scale, the developers of inference packages have the opportunity to
reflect, reassess, and improve upon their configuration.
In this work, we describe incremental but impactful updates to one such parameter inference engine: RIFT.

The Rapid Iterative FiTting (RIFT) \cite{gwastro-PENR-RIFT,gwastro-RIFT_FinerNet,gwastro-RIFT-Update}
performs Bayesian inference through a mix of Monte Carlo quadrature; iterative simulation-based inference  to
successively approximate pertinent likelihoods \cite{pmlr-v89-papamakarios19a,2020PNAS..11730055C}; and
target RIFT's iterative refinement.
Unlike the most popular approaches for gravitational wave parameter inference, which rely on Markov chains
within either Markov Chain Monte Carlo or nested sampling codes (see \cite{RevModPhys.94.025001} for a recent review),
this simulation-based-inference method natively reduces the number of low-level costly simulations needed to carry out
inference (in the GW context, the costly simulations involve evaluating a GW signal model).
RIFT's inference decomposes into two key factors: evaluating a marginal likelihood over intrinsic parameters; then
fitting and sampling from that likelihood.
For the first task, RIFT uses a clever factorization of the likelihood
\cite{gwastro-PE-AlternativeArchitectures,gwastro-RIFT-Update}, allowing extremely efficient evaluation over all
extrinsic parameters (e.g., position and orientation) after a single waveform-dependent precomputation.
This precomputation allows RIFT to  incorportate additional physics or systematics while retaining extremely low cost; for example,  a recent
study \cite{gwastro-RIFT-roboto-scaling} demonstrated how to efficiently perform inference for long signals which account for the
earth's rotation and finite-size detector response.
For the second task, RIFT's iterative and intrinsic inference leverages domain experience, adopting well-chosen coordinates in which
the marginal or extrinsic likelihood is simple and well-behaved.
In this work, we enhance RIFT's capability with new intrinsic coordinates; incorporate new physics into its
precompute/marginalize step;
use systematic numerical comparisons against other codes' low-level likelihood
implementations to identify residual difference; and
provide operating point changes and corner-case bugfixes to improve its performance
over the whole census.
Specifically, we first introduce new fitting coordinates for transverse spins and eccentric orbital phase
that account for correlations with other intrinsic parameters.
Second, we incorporate calibration marginalization into the analysis workflow  \cite{2020PhRvD.102l2004P,gwastro-pe-bilby-2018},
with an option to marginalize jointly over calibration and extrinsic parameters.
Third, applying the tools developed in a previous study \cite{gwastro-RIFT-roboto-scaling}, we identify and address subtle time-interpolation
and data conditioning differences between different codes as a potential source of future discrepancy.
Finally, based on this and other experience, we  revise our preferred operating points for future RIFT inference.

This investigation is motivated by ongoing reanalyses of the GW census, which have unveiled significant differences between some analyses \cite{gwastro-mergers-ManningReanalyzeO3,gwastro-mergers-rift_asimov_O3-Fernando2024,gwastro-pe-eccentric-Malagon2026,LIGO-O4-GW231123}.  Some individual
cases  have been tracked back to user error (inconsistent or inappropriate settings) in some cases \cite{gwastro-mergers-ManningReanalyzeO3,gwastro-RIFT_FinerNet}; and waveform systematics
in others \cite{gwastro-mergers-rift_asimov_O3-Fernando2024,gwastro-mergers-TousifGWTC3,LIGO-O4-GW231123}.  However, because very few analyses have systematically re-examined large parts of the catalog with a single unified code
framework (but cf. \cite{gwastro-pe-eccentric-Malagon2026,gwastro-mergers-ManningReanalyzeO3,gwastro-mergers-TousifGWTC3,gw-astro-mergers-dingo-part2,2023ApJ...946...59N}), inevitably some of these discrepancies in the literature involve results produced with
different inference engines.   In this work, we investigate specific sources of difference between inference codes
through controlled likelihood comparisons.  We also provide richer, more informative outputs when performing calibration marginalization, enabling head-to-head
comparison with other inference engines on a sample-by-sample basis.
Finally, because the increasing GW census has unveiled more exceptional systems, including increasingly clear signals,
we adopt improved operating point choices.
Previous editions of RIFT have conservatively optimized for robust inference about poorly constrained observations,
adopting conservative choices ill-suited to these exceptional events. In this work, we provide simple tools which better
configure RIFT for the loudest, clearest signals.

This paper is organized as follows.
Section~\ref{sec:review} reviews the RIFT likelihood and the existing interfaces needed
for this discussion. Section~\ref{sec:core} traces calibration marginalization
from earlier posterior reweighting through the current pipeline and in-loop
methods to end-to-end demonstrations. Section~\ref{sec:core_operating}
examines operating points and their validation limits.
Section~\ref{sec:conclude} summarizes the supported conclusions.
The appendices document %
cross-code likelihood comparisons (Appendix~\ref{ap:likelihood_code_comparison});
the calibration error budget
(Appendix~\ref{ap:calmarg}), redundancy checks (Appendix~\ref{ap:checks});
waveform-symmetry diagnostics (Appendix~\ref{ap:waveform_checks});
and new capability to perform multibanding (Appendix~\ref{app:early-multirate}) particularly via
native GPU-accelerated waveform models (Appendix~\ref{app:gpu-xphm}).

\section{RIFT essentials}
\label{sec:review}

A coalescing compact binary can be completely characterized by its intrinsic
and extrinsic parameters.   The intrinsic parameters are necessary to characterize a binary's orbital inspiral and
merger trajectory, up to
spacetime symmetries.  For a quasicircular binary, this trajectory is specified by the binary's  (detector-frame) masses $m_{i,z}$, spins, and any quantities
characterizing matter in the system.  For an eccentric binary, these parameters must be supplemented by the orbital
eccentricity and mean anomaly.
Conversely, the extrinsic parameters are seven quantities reflecting flat spacetime symmetries: seven numbers needed to characterize its spacetime location and orientation.
We express masses in solar mass units and
 dimensionless spins in terms of Cartesian components $\chi_{i,x},\chi_{i,y}, \chi_{i,z}$, expressed
relative to a frame with $\hat{\mathbf{z}}=\hat{\mathbf{L}}$ and (for simplicity) at the orbital frequency corresponding to the earliest
time of computational interest (e.g., an orbital frequency of $\simeq 10 \unit{Hz}$).  We will use $\lambda,\theta$ to
refer to intrinsic and extrinsic parameters, respectively.

At a high level of abstraction, RIFT can be understood as a two-stage iterative process to interpret gravitational wave observations $d$ via comparison to
predicted gravitational wave signals $h(\bm{\lambda}, \bm\theta)$.   In one stage,
RIFT computes a marginal likelihood
\begin{equation}
 {\cal L}\response{({\bm \lambda})}\equiv\int  {\cal L}_{\rm full}(\bm{\lambda} ,\bm\theta )p(\bm\theta )d\bm\theta
\end{equation}
from the likelihood ${\cal L}_{\rm full}(\bm{\lambda} ,\theta ) $ of the gravitational wave signal in the multi-detector network,
accounting for detector response, and a prior $p(\bm\theta)$ over extrinsic parameters; see  \cite{gwastro-PE-AlternativeArchitectures,gwastro-PENR-RIFT} for a more detailed
specification.  This calculation is performed in parallel on a large number of candidate intrinsic parameters ${\bm
  \lambda}_\alpha$ by a program denoted as ILE (because it integrates the GW likelihood over extrinsic parameters).
In the second stage, RIFT interpolates and integrate.   Specifically, first it interpolates the likelihood information
accumulated from all previous iterations, making an approximation to ${\cal L}(\lambda)$ based on its
accumulated archived knowledge of marginal likelihood evaluations
$(\lambda_\alpha,{\cal L}_\alpha)$.  Second, using this approximation, it deduces the (detector-frame) posterior distribution
\begin{equation}
\label{eq:post}
p_{\rm post}=\frac{{\cal L}(\bm{\lambda} )p(\bm{\lambda})}{\int d\bm{\lambda} {\cal L}(\bm{\lambda} ) p(\bm{\lambda} )}.
\end{equation}
where prior $p(\bm{\lambda})$ is the prior on intrinsic parameters like mass and spin.

At the end of the iterative calculation, RIFT performs one final pass of ILE over posterior intrinsic draws $\lambda_k$,
fairly drawing some (fixed) number of extrinsic parameters $\theta_{k,p}$ from the  underlying ILE Monte
Carlo integral weighted samples.   These combined samples ($\lambda_k,\theta_{k,p}$) provide a full posterior for all
intrinsic and extrinsic parameters.   Postprocessing each sample provides derived parameters, including the source-frame masses $m_{i}$.

This section is organized as follows.  In Section \ref{sec:sub:L} we describe how  RIFT evaluates the gravitational wave
likelihood ${\cal  L}_{\rm full}$.  In Section \ref{sec:sub:Lmarg}, we describe how ILE performs Monte Carlo integration
to compute the marginal likelihood ${\cal L}$.   In Section \ref{sec:sub:CIP}, we outline how CIP interpolates the
likelihood, then performs Monte Carlo integration to produce fair draws from the (estimated) posterior distribution.
Sections~\ref{sec:rf:transverse:methods} and~\ref{sec:eccentric:coordinates}
describe fitting coordinates for transverse spins and eccentricity.
Then in Section \ref{sec:sub:explore} we explain how RIFT actually uses ILE and CIP iteratively, augmented with
other tools and using binary-inspiral-specific coordinate systems that are well-adapted to GW source parameter
inference, and describe  infrastructure used to efficiently validate RIFT code and operating-point
choices with standard PP tests.
In Section \ref{sec:sub:external:Waveforms} we summarize RIFT's updated interface to external waveforms, including key
features needed to ensure compatibility with previous and other work.
In Section \ref{sec:sub:waveform-model-marginalization} we describe the native RIFT implementation of a
previously-described RIFT-based technique to marginalize over different waveform models \cite{gwastro-RIFT-systematics-AnjaliAasim-2020}.
The earlier calibration workflow and its replacement are developed together in
Section~\ref{sec:core}.

\subsection{Review of the RIFT likelihood}
\label{sec:sub:L}
As described in previous work \cite{gwastro-PE-AlternativeArchitectures,gwastro-RIFT-Update}, RIFT uses physical
insight to decompose the overall inference calculation  into two stages: inference and marginal likelihoods for a fixed
physical binary, and inference about different physical binaries.
At a high level, RIFT expresses gravitational wave signals $h(t)$ in terms of physical basis signals $h_{lm}(t)$, associated
with a (spin-weighted) spherical harmonic decomposition of radiation in all possible emission directions.  This
decomposition allows RIFT to compute  cross-correlations between this basis and each detector's data; the likelihood for
arbitrary source orientations,  sky positions, and distances follows by a weighted average of these cross-correlation timeseries.

Using notation established in previous studies
\cite{gwastro-PENR-RIFT,gwastro-PENR-RIFT-GPU,gwastro-PE-AlternativeArchitectures}, the RIFT likelihood is expressed in
terms of a (spin-weighted) spherical harmonic decomposition of the complex gravitational wave strain
\begin{align} \label{eq:strain}
h(t,\vartheta,\phi;\bm{\lambda}) =  h_+(t,\vartheta,\phi;\bm{\lambda}) -
                                i h_\times (t,\vartheta,\phi;\bm{\lambda}) \, ,
\end{align}
Customarily its real and imaginary parts are denoted its two fundamental polarizations $h_+$ and $h_\times$.
Here, $t$ denotes time, $\vartheta$ and $\phi$ are the polar and azimuthal angles
for the direction of gravitational wave propagation away from the source.
The complex gravitational-wave strain can be written in terms of
spin-weighted spherical harmonics $\Y{-2}_{\ell m} \left(\vartheta, \phi \right)$ as
\begin{align} \label{eq:strain_mode}
h(t,\vartheta,\phi;\bm{\lambda}) =
\sum_{\ell=2}^{\infty} \sum_{m=-\ell}^{\ell} \frac{D_{\rm ref}}{D} h^{\ell m}(t;\bm{\lambda}) \Y{-2}_{\ell m} \left(\vartheta, \phi \right) \, ,
\end{align}
where the sum includes all available harmonic modes $h^{\ell m}(t;\pmb{\bm{\lambda}})$ made available by the model;  where
$D_{\rm ref}$ is a fiducial reference distance; and where $D$, the luminosity distance to the  source, is one of the
extrinsic parameters.
Each detector has (assumed constant) response functions $F_{+}$ and $F_\times$, such that the time-dependent strain
response has the form  $h_k(t) =F_{+,k} h_+(t_k) +
  F_{\times,k}h_\times(t_k)$ for the detector response $h_k$,
where $t_k=t_c - \vec{x}_k \cdot \hat{n}$ is the arrival time at the $k$th detector (at position $\vec{x}_k$)
for a plane wave propagating along $\hat{n}$ \cite{gwastro-PE-AlternativeArchitectures}.
We then substitute these expressions for $h_k$ into the likelihood function
thereby generating~\cite{gwastro-PE-AlternativeArchitectures}
\begin{widetext}
\begin{align}
\ln {\cal L}(\bm{\lambda}, \theta)
&= (D_{\rm ref}/D) \text{Re} \sum_k \sum_{\ell m}(F_k \Y{-2}_{\ell m})^* Q_{k,lm}(\bm{\lambda},t_k)\nonumber \\
&   -\frac{(D_{\rm ref}/D)^2}{4}\sum_k \sum_{\ell m \ell' m'}
\left[
{
|F_k|^2 [\Y{-2}_{\ell m}]^*\Y{-2}_{\ell'm'} U_{k,\ell m,\ell' m'}(\bm{\lambda})
}
 {
+  \text{Re} \left( F_k^2 \Y{-2}_{\ell m} \Y{-2}_{\ell'm'} V_{k,\ell m,\ell'm'} \right)
}
\right]
\label{eq:def:lnL:Decomposed}
\end{align}
\end{widetext}
where
 where $F_k = F_{+,k} - i F_{\times,k}$ are the
complex-valued detector
response functions of the $k$th detector \cite{gwastro-PE-AlternativeArchitectures} and
the quantities $Q,U,V$ depend on $h$ and the data as
\begin{subequations}
\label{eq:QUV}
\begin{align}
Q_{k,\ell m}(\bm{\lambda},t_k) &\equiv \qmstateproduct{h_{\ell m}(\bm{\lambda},t_k)}{d}_k \nonumber\\
&= 2 \int_{|f|>f_{\rm low}} \frac{df}{S_{n,k}(|f|)} e^{2\pi i f t_k} \tilde{h}_{\ell m}^*(\bm{\lambda};f) \tilde{d}(f)\ , \\
{ U_{k,\ell m,\ell' m'}(\bm{\lambda})}& = \qmstateproduct{h_{\ell m}}{h_{\ell'm'}}_k\ , \\
V_{k,\ell m,\ell' m'}(\bm{\lambda})& = \qmstateproduct{h_{\ell m}^*}{h_{\ell'm'}}_k  \ .
\end{align}
\end{subequations}
The likelihood can be equivalently expressed as
\begin{align}
\ln {\cal L} &= \frac{D_{\rm ref}}{D} \text{Re}[ (F Y)^\dag Q]  \nonumber \\
 & - \frac{D_{\rm ref}^2}{4 D^2} [ (FY)^\dag U FY + (FY)^TV FY]
\label{eq:lnL:MatrixForm}
\end{align}
where this symbolic expression employs an implicit index-summation convention such that all naturally paired
  indices are contracted.  The result is an array of shape (time)$\times$(extrinsic).

For each candidate set of waveform parameters $\bm \lambda$, RIFT computes the inner product arrays $U,V$ and the inner
product timeseries $Q$.  Particularly for low-mass binaries, these precomputed quantities can be costly to evaluate
(once, for each $\bm\lambda$, as
their calculation  involves manipulating long timeseries.  By contrast, when constructing an extrinsic likelhood versus
$\bm\theta$ given these quantities, the reduced quantities $Q$ (interpolated as needed to account for detector response
from the original $Q$ inner product) and $\ln {\cal L}$
only need to be evaluated over a very short range of times, allowing for massive reduction in computational cost; see
\cite{gwastro-PENR-RIFT-GPU} for a detailed discussion.
Time marginalization is performed by direct quadrature; the correlation sampling rate and the
integration nodes need not be the same.  Paper~1 \cite{gwastro-RIFT-roboto-scaling}, especially
Appendix~D (``Quadrature over the extrinsic angles, time and distance''), derives the time-sampling
requirements and validates the available representations against a direct Fourier reference.

\noindent \emph{Time resolution}: RIFT stores the detector--mode correlations
$Q_{k,\ell m}(t)$ on a discrete time grid.  The original likelihood used a continuous-time
representation \cite{gwastro-PE-AlternativeArchitectures}; historical production configurations
instead evaluated detector arrival times by nearest-sample lookup.  Starting with version
0.0.17.8, users can choose the internal and output time resolutions separately.  Data import can
\emph{upsample} a common, downsampled input segment, retaining compatibility with the data
conditioning used by another inference engine.  Conversely, after selecting the other extrinsic parameters,
the conditional time likelihood can also be interpolated to generate final samples on a denser
grid (or even continuously), otherwise carrying out all stages in the marginal likelihood calculation using a coarser
method.   Starting with 0.0.17.14, however, RIFT can also improve the underlying quality of the integration, using
superior quadrature and interpolation to increase the accuracy of the estimate of each $\int {\cal L}(t) dt$.

The marginal time likelihood is built from $Q_{k,\ell m}$ evaluated at the detector arrival times -- generally
incommensurate with our fiducial time sampling.
RIFT now provides mutiple methods to interpolation $Q_{k,\ell m}$ onto a uniformly sampled grid, including
nearest-neighbor interpolation (the previous default); cubic interpolation; sinc interpolation; and band-limited
spectral reconstruction (for high-quality global convergence) followed by cubic interpolation (for fast, local accuracy).
All interpolation choices leave $U,V$ unchanged and are completely compatible with changing the input or output sampling
rate.

The marginal time likelihood $\int {\cal L}(t)dt$ likewise requires some interpolation or quadrature scheme to perform
the integral over uniformly-sampled likelihoods.
The likelihood's peak narrows as SNR increases, with width $\sigma_t\propto 1/\rho$, so for any integration spacing $\Delta t$
eventually a sufficiently loud peak will be narrower than the grid spacing: %
$\Delta t/\sigma_t\propto\rho$ at fixed sampling rate.
In the previous paper (Paper 1, Appendix D), we described methods to address the high-amplitude limit reliably using a
locally gaussian approximation, as well as different methods to perform this quadrature reliably.
Appendix~\ref{ap:likelihood_code_comparison} below relates these
choices to controlled cross-code likelihood comparisons.

To summarize, the internal sampling rate determines the input information available for  use for likelihood
calculations;
fractional-time interpolation evaluates the detector
correlations $Q_{k,\ell m}(t)$ between grid points; and
resampling the
final time likelihood onto a denser output grid enables even more accurate marginal likelihoods (as well as
continuously-sampled time posteriors).
RIFT previously recommended
at least $4096\,\mathrm{Hz}$ to resolve narrow likelihood peaks, to address all these concerns.
For long signals the full-rate data also set the cost and memory of the per-point precompute.
Appendix~\ref{app:early-multirate} describes an opt-in prototype that keeps the full rate only near
coalescence.

\subsection{Evaluating the marginalized likelihood}
\label{sec:sub:Lmarg}
Given the likelihood ${\cal L}_{\rm full}(\bm\lambda,\bm\theta)$, RIFT evaluates the marginal likelihood via an adaptive Monte Carlo
integrator:
\begin{align}
\label{eq:lnL:MonteCarlo}
{\cal L}(\bm\lambda) \simeq \frac{1}{N} \sum_k {\cal L}_{\rm full}(\bm\lambda,\theta_k) p(\bm\theta_k)/p_s(\bm\theta_k)
\end{align}
RIFT provides multiple  adaptive Monte Carlo techniques \cite{gwastro-RIFT-Update} to perform this and any other integral.    Two techniques -- the original
sampler \cite{gwastro-PE-AlternativeArchitectures} and a GPU-accelerated variant \cite{gwastro-RIFT-Update} -- involve an (adaptive) sampling prior $p_s$ which has
 product form, consistent with standard Cartesian adaptive integrators
 \cite{Lepage:1980dq,2021JCoPh.43910386L,book-mm-NumericalRecipies}.
After a large block of evaluations, each one-dimensional marginal sampling prior can be updated to more
closely conform to the support of the integrand.
A third implementation uses Gaussian mixture models (GMMs) to generate sampling priors for any user-specified subset of
dimensions.  This implementation was well-suited to very high-amplitude sources with well-localized
intrinsic parameters.
RIFT also provides other integration techniques, including normalizing flow integrators, ``portfolio'' integrators which
combine multiple different configurations, and ``oracles'' which help target sampling (e.g., by hill-climbing) in a
non-Markovian way \cite{gwastro-RIFT_FinerNet}.
However, due to its ubiquitously robust performance, almost all RIFT calculations currently use the adaptive volume (AV)
integrator \cite{gwastro-RIFT_FinerNet,gw-astro-mergers-VTiwariPE}.
\subsection{Likelihood interpolation and posterior distributions}
\label{sec:sub:CIP}
To estimate ${\cal L}$ from discrete samples $\lambda_\alpha,{\cal L}_\alpha$, RIFT previously  deploys two unstructured
interpolation techniques:  Gaussian process regression \cite{book-Rasmussen-GP} and random forest regression
\cite{mm-stat-RandomForests}.
Given the extremely high training and evaluation cost for Gaussian processes (as well as the finicky training necessary
to optimize their model hyperparameters), in practice we continue to almost exclusively
recommend using random forest regression for production work.
In this edition, however, we also enable additional interpolation techniques including random-forest PCA, RBF
\cite{mm-stat-Buhmann-RBF,mm-stat-scipy}, and a gradient-boosted (xgboost) fit \cite{mm-stat-XGBoost}.

Given the likelihood, fair samples from the posterior distribution are generated by the following two-step process,
described in the RIFT paper.
First, using the likelihood estimate $\hat{\cal L}_{\rm marg}$ and the same adaptive Monte Carlo integrator described above,
we perform the Monte Carlo integral $\int d{\bm \lambda} \hat{\cal L}_{\rm marg} p({\bm \lambda})$, producing sample
locations ${\bm \lambda}_k$ and associated weights $w_k= \hat{\cal L}_{\rm marg} p(\lambda)/p_s(\lambda)$.
Second, we make a fair draw from these weighted samples.

In typical practice, the Monte Carlo integration(s) used to generate the posterior are  performed in
parallel,  with each CIP worker instance terminating once a threshold $n_{\rm eff}$ is reached, where as described at length in
Wofford et al \cite{gwastro-RIFT-Update} RIFT uses an effective sample estimate
\begin{align}
\label{eq:neff}
n_{\rm eff} = \frac{\sum p_k}{\max \{p_k \} } %
\end{align}
expressed for convenience in terms of normalized sample probabilities $p_k = w_k/\sum_q w_q$.
The value $1/n_{\rm eff}$ is the largest discontinuous jump in the estimator $\hat{P}(<x) = \sum_k p_k \theta(x_k-x)$
for any one-dimensional cumulative probability distribution $P(<x)$ derived from the full samples.
The posterior is the union of fairly drawn samples from each CIP worker, where each worker has been configured to
fairly draw (a constant multiple of)  $n_{\rm eff}$ samples under the assumption that the $n_{\rm eff}$ threshold has
been reached  (i.e., rather than a user-supplied maximum evaluation count, to regulate runtime).
In this work, unless otherwise noted we use $n_{\rm eff}/N$ to characterize our sampling efficiency.
\subsection{Coordinates for transverse-spin inference}
\label{sec:rf:transverse:methods}
RIFT's likelihood interpolation can be improved by expressing the underlying training data $(\lambda_\alpha, {\cal
  L}_\alpha)$ using a coordinate system in which key correlations are well captured; the likelihood varies smoothly;
and the coordinates and sampler do not share common degrees of freedom.  The first two requirements are generic to any
interpolation engine: they reduce the complexity that the interpolator needs to encode.  The last requirement, however,
is a particular feature of RIFT's reliance on fast, robust RF interpolation: at root, these interpolators are per-component
stepwise along input-coordinate planes, leading to highly undesirable behavior when the sampler and the fit have common
coordinate representations.  (The latter feature motivated the introduction of the inspiral-motivated $\mu_1,\mu_2$
coordinate system for mass and aligned spin adopted  in previous work \cite{gwastro-RIFT-Update}.)

In previous work, we largely performed interpolation of transverse spin with independent cartesian coordinates
($\chi_{i,x},\chi_{i,y}$).  For much the same reason that individual component spins only encode the dominant
aligned-spin parameter in correlation ($\chi_{\rm eff}$), these coordinates do not leverage the strong dependence of the
waveform on the physical total transverse spin ${\mathbf S}_\perp$.   For example, the total angular momentum direction
depends on $\boldsymbol J=\boldsymbol L_N+\boldsymbol S$, and is encoded naturally in the gravitational wave signal and
our posterior \cite{gwastro-pe-systemframe}.
Because RIFT explicitly splits intrinsic and extrinsic coordinates and uses ${\mathbf L}$-frame waveforms, we cannot
conveniently adopt natural J-frame variables used in other work, without adding considerable complexity to our data
products and inter-code communication. For this
reason, we adopt a simpler alternative: we interpolate the likelihood using a set of coordinates that includes ${\mathbf S}_\perp$.

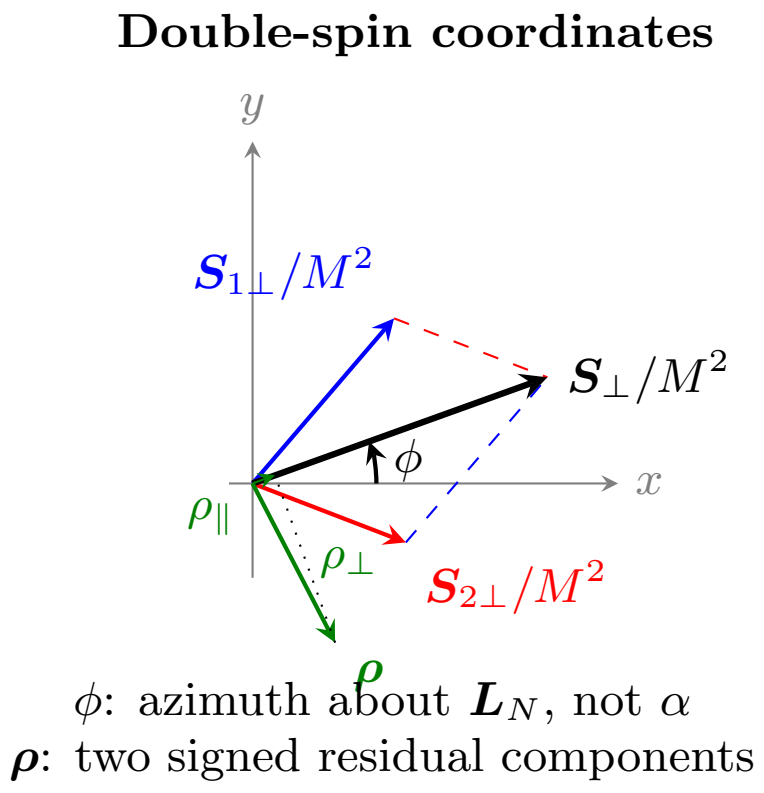
\begin{figure}[t]
\centering
\begin{tikzpicture}[>=stealth,scale=0.78]
\begin{scope}
\node[anchor=west] at (-1.3,3.8) {\textbf{Double-spin coordinates}};
\draw[->,gray] (-0.2,0)--(3.1,0) node[right] {$x$};
\draw[->,gray] (0,-0.8)--(0,2.9) node[above] {$y$};
\draw[->,thick,blue] (0,0)--(1.2,1.4) node[above left] {$\boldsymbol S_{1\perp}/M^2$};
\draw[->,thick,red] (0,0)--(1.3,-0.5) node[below right] {$\boldsymbol S_{2\perp}/M^2$};
\draw[dashed,red] (1.2,1.4)--(2.5,0.9);
\draw[dashed,blue] (1.3,-0.5)--(2.5,0.9);
\draw[->,very thick] (0,0)--(2.5,0.9) node[right] {$\boldsymbol S_\perp/M^2$};
\draw[->,thick] (1.05,0) arc[start angle=0,end angle=19.8,radius=1.05];
\node at (1.32,0.20) {$\phi$};
\draw[->,thick,green!50!black] (0,0)--(0.7,-1.35) node[below right] {$\boldsymbol\rho$};
\draw[dotted] (0.7,-1.35)--(0.19,0.07);
\draw[->,green!50!black] (0,0)--(0.19,0.07);
\node[green!50!black] at (-0.35,-0.28) {$\rho_\parallel$};
\node[green!50!black] at (0.82,-0.67) {$\rho_\perp$};
\node[align=center] at (1.1,-2.1) {$\phi$: azimuth about $\boldsymbol L_N$, not $\alpha$\\$\boldsymbol\rho$: two signed residual components};
\end{scope}
\end{tikzpicture}
\caption{\label{fig:rf:precession:geometry}The double-spin coordinate
construction in the $L$-frame transverse plane. Polar coordinates
$(R,\phi)$ describe the mass-weighted spin sum. The curved arrow marks
$\phi$ from the positive $x$ axis to $\boldsymbol S_\perp/M^2$.
The signed components
$(\rho_\parallel,\rho_\perp)$ of the residual vector complete the four
coordinates and recover both component spins. Arrow lengths are schematic;
this is a coordinate diagram, not a posterior plot or a change of frame.\llmfig}
\end{figure}

Specifically, let $M=m_1+m_2$ and
$\boldsymbol S_i=m_i^2\boldsymbol\chi_i$. In terms of the transverse spins (relative to  the instantaneous orbital direction
$\hat{\boldsymbol L}_N$ at the reference frequency), we define
\begin{align}
 a&=(m_1/M)^2,\qquad b=(m_2/M)^2,\qquad h=\sqrt{a^2+b^2},\\
 \boldsymbol T&=\frac{\boldsymbol S_\perp}{M^2}
     =a\boldsymbol\chi_{1\perp}+b\boldsymbol\chi_{2\perp},\\
 \boldsymbol\rho&=
 \frac{-b\boldsymbol\chi_{1\perp}+a\boldsymbol\chi_{2\perp}}{h}.
\end{align}
The magnitude and cylindrical-polar angle of $\mathbf{T}$ define two of our coordinates  ( $\boldsymbol T=R\boldsymbol e_R$, with
$\boldsymbol e_R=(\cos\phi,\sin\phi)$) and the projection of the remaining antisymmetric spin combination provides the
remaining two coordinates, expressed as cartesian projections relative to the direction of ${\mathbf T}$: if
$\boldsymbol e_\phi=(-\sin\phi,\cos\phi)$, then the coordinates are
\begin{equation}
 \left(R,\phi,\rho_\parallel,\rho_\perp\right)
 =\left(\frac{|\boldsymbol S_\perp|}{M^2},\phi,
 \boldsymbol\rho\cdot\boldsymbol e_R,
 \boldsymbol\rho\cdot\boldsymbol e_\phi\right).
 \label{eq:rf:geometric4}
\end{equation}
Figure~\ref{fig:rf:precession:geometry} shows the construction.
The angle $\phi$ is
about $\boldsymbol L_N$, not the accumulated precession phase $\alpha$
about $\boldsymbol J$.

Appendix~\ref{ap:rf:additional:coordinates} describes another family of precession-derived coordinates which have
demonstrated value in our experiments.

\subsection{New coordinates for eccentric inference}
\label{sec:eccentric:coordinates}
Eccentricity produces a periodic modulation of the gravitational-wave signal
as the binary moves between pericenter and apocenter. The data constrain the
phase of this modulation where the signal is best measured.
Waveform models specify an orbital angle at an earlier reference frequency.
The phase accumulated between these frequencies depends on eccentricity and
masses. As these parameters vary, the initial angle must change to keep
the modulation in phase in the sensitive band. This correlation can wrap
around the initial-angle interval several times, appearing as repeated
bands of high likelihood.
We account for this accumulated phase when choosing coordinates for the
likelihood fit.

Let $e_0$ and $\ell_0$ be the waveform's native eccentricity and orbital
angle at its reference frequency $f_0$.
This angle locates the binary along its eccentric orbit; its precise
definition depends on the waveform model.
Choose a later frequency $f_a$ and estimate the phase
advance $A$ from $f_0$ to $f_a$.
This estimate depends on eccentricity and masses, but is independent of
$\ell_0$.
We use the shifted angle $\theta$ for fitting and recover $\ell_0$ before
evaluating the waveform:
\begin{align}
 \theta&=\operatorname{wrap}_{2\pi}[\ell_0+A],
 &\ell_0&=\operatorname{wrap}_{2\pi}[\theta-A].
 \label{eq:ecc:angular:shear}
\end{align}
Here $\operatorname{wrap}_{2\pi}$ reduces an angle to $[0,2\pi)$.
At fixed eccentricity and masses, this transformation shifts every angle
by the same amount. Its Jacobian is unity, so a uniform angular prior
remains uniform. For a nonuniform prior, we evaluate the original density
at the recovered $\ell_0$. The waveform still uses $e_0$ and $\ell_0$ at
$f_0$; only the coordinates used for inference change.

Our default model for eccentricity evolution uses the Peters equations
\cite{Peters:1964}. Define the detector-frame total mass $M=m_1+m_2$,
the symmetric mass ratio $\eta=m_1m_2/M^2$, and
$x=(\pi GMf/c^3)^{2/3}$, where $f$ is twice the orbit-averaged orbital
frequency. The orbit-averaged evolution under quadrupole gravitational
radiation is
\begin{align}
 \dot x&=\frac{64\eta c^3}{5GM}x^5
 \frac{1+73e^2/24+37e^4/96}{(1-e^2)^{7/2}},
 \label{eq:ecc:peters:x}\\
 \dot e&=-\frac{304\eta c^3}{15GM}x^4 e
 \frac{1+121e^2/304}{(1-e^2)^{5/2}}.
 \label{eq:ecc:peters:e}
\end{align}
In this Newtonian description, the mean anomaly advances at
$\dot\ell=c^3x^{3/2}/GM$.
We obtain $A$ by integrating this rate from $f_0$ to $f_a$ along the
Peters evolution. Appendix~\ref{ap:eccentric:high_e} describes a more
relativistic approximation based on radial orbital cycles and the energy
radiated per orbit.

To illustrate more concretely how this calculation works in a tractable
form, consider the small-eccentricity solution
$e(x)=e_0(x/x_0)^{-19/12}+O(e_0^3)$.
Integrating the mean-anomaly rate from $x_0=x(f_0)$ to $x_a=x(f_a)$ gives,
to quadratic order in $e_0$ and with $R=x_a/x_0$,
\begin{align}
 A&=\Phi_{\rm circ}-\kappa e_0^2,\label{eq:ecc:advance}\\
 \Phi_{\rm circ}&=\frac{x_0^{-5/2}}{32\eta}(1-R^{-5/2}),\\
 \kappa&=\frac{785x_0^{-5/2}}{8704\eta}(1-R^{-17/3}).
\end{align}
The first term is the phase accumulated by a circular inspiral.
The second accounts for the faster inspiral, and hence fewer accumulated
cycles, at nonzero eccentricity \cite{Moore:2016eccphase}.
The phase estimate need only capture enough of the evolution to reduce
the winding of the high-likelihood bands. For this angular shift,
subtracting the same $A$ recovers $\ell_0$ regardless of the estimate's
physical accuracy. The relativistic construction also converts the native
orbital angle to a radial angle, with a generally nonunit Jacobian as
described in Appendix~\ref{ap:eccentric:high_e}.
The reader is free to choose the full Peters evolution,
its small-eccentricity limit, or this relativistic approximation.

We choose $f_a$ once at the start of an analysis, using the initial
detector-frame total-mass estimate $M_{\rm build}$.
The default choice starts from
$f_{\rm mass}=x_\star^{3/2}c^3/(\pi GM_{\rm build})$, with
$x_\star=\eccPolicyAnchorX{}$.
This selects a common dimensionless orbital scale for systems of different
masses. We use $f_{\rm mass}$ within the nominal
$[\eccPolicyBandLow{},\eccPolicyBandHigh{}]\,\mathrm{Hz}$ interval,
cap it at the upper edge, and respect the detector's analysis frequency
bounds. If $f_{\rm mass}$ falls below this interval, or the resulting
$f_a$ precedes $f_0$, we retain the native angle by setting $f_a=f_0$.
For sufficiently high masses, the chosen weak-field orbital scale lies
below the prescribed frequency interval. Keeping the native angle avoids extrapolating
the phase approximation to a later, more relativistic part of the inspiral.
The same chosen frequency is used throughout fitting and the placement
of new likelihood points. This prescription provides a consistent default;
it has not been optimized for detector sensitivity or for the accuracy
of the phase approximation.

\begin{figure*}[t]
\centering
\includegraphics[width=0.495\textwidth]{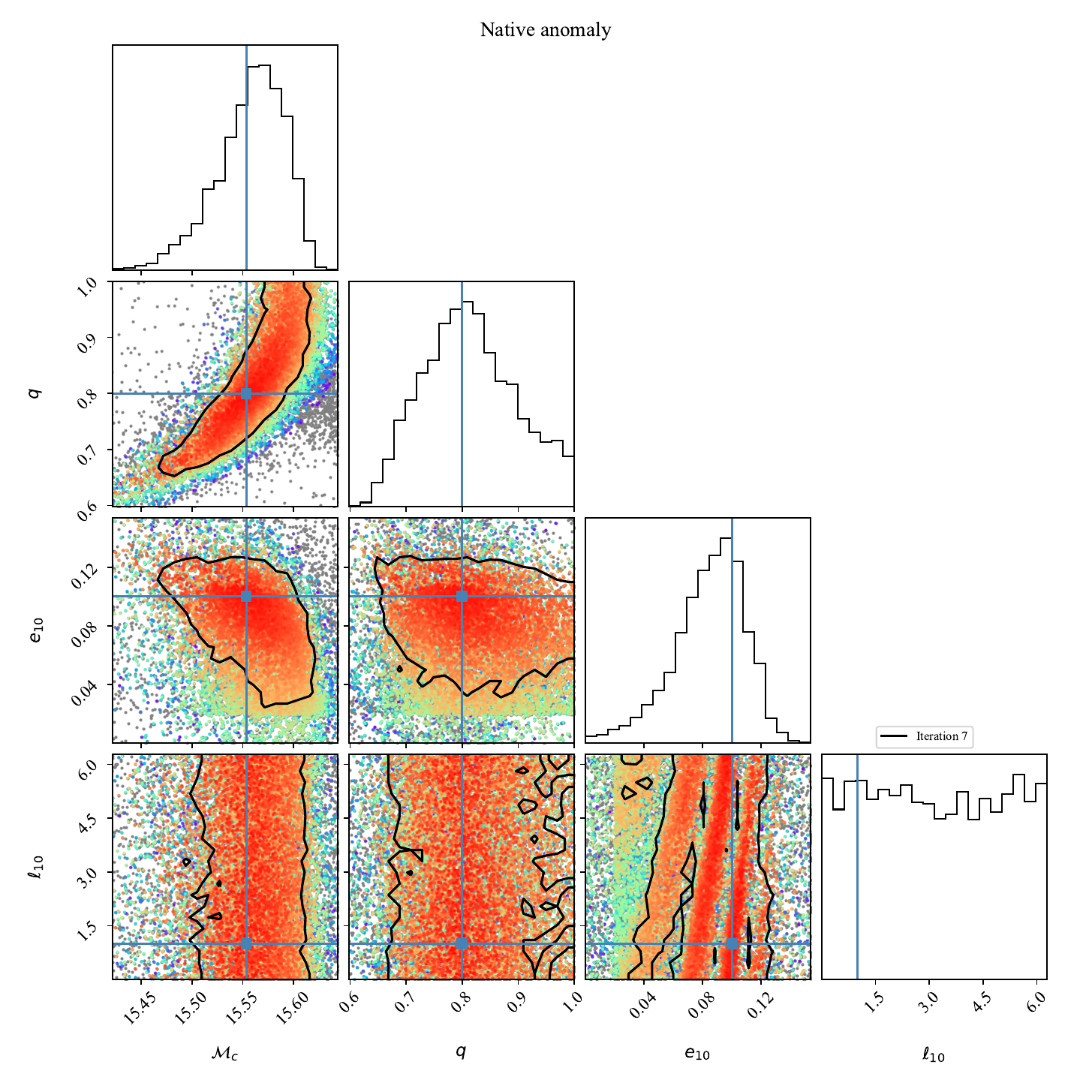}\hfill
\includegraphics[width=0.495\textwidth]{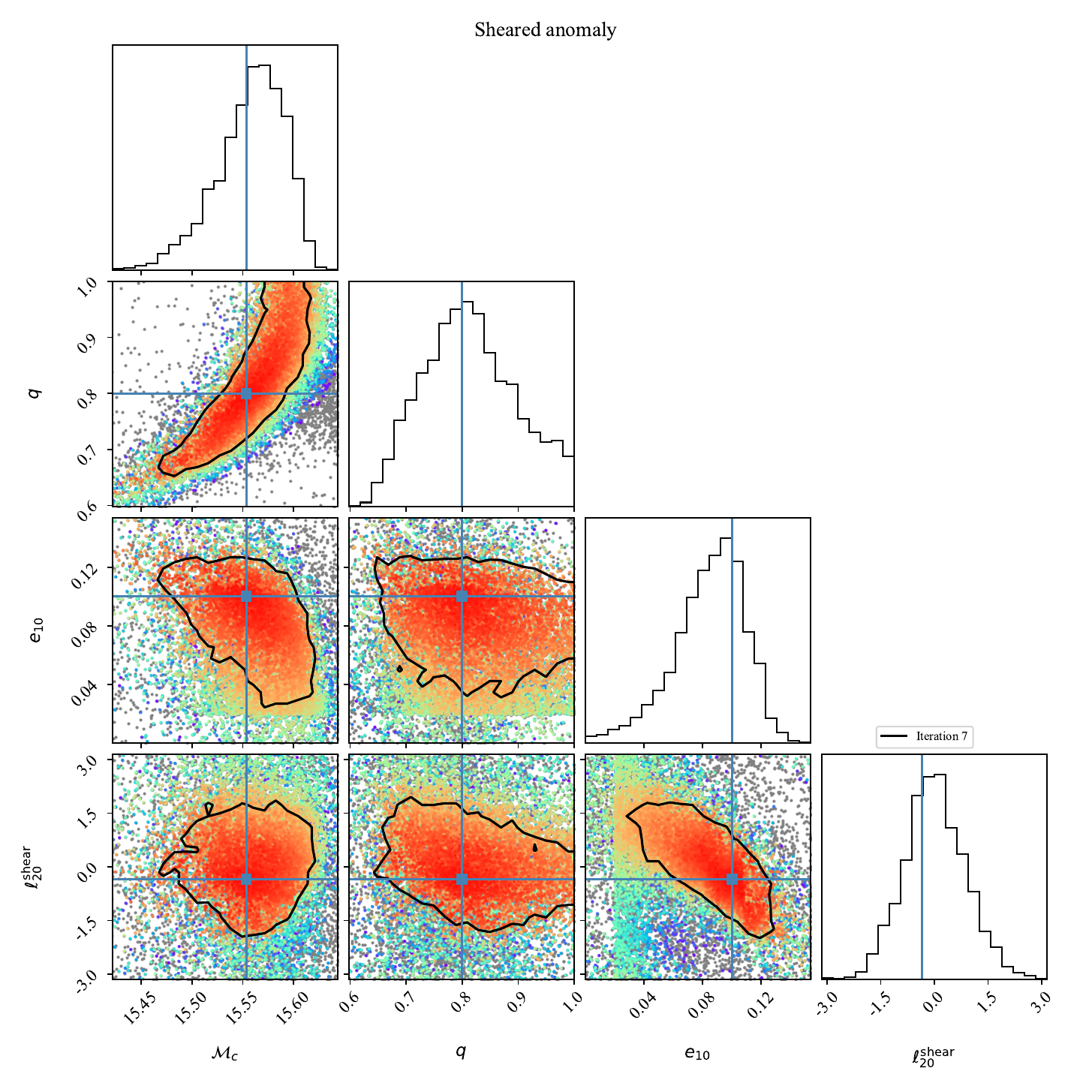}
\caption{\label{fig:ecc:seob:bands}
\textbf{Eccentric coordinates in a zero-spin injection with a uniform eccentricity prior.}
Native anomaly $\ell_{10}$ (left) and shifted fitting angle
$\ell_{20}^{\mathrm{shear}}$ (right), alongside chirp mass $\mc$, mass ratio
$q=m_2/m_1\leq1$, and native eccentricity $e_{10}$.
The same \eccUniformPosteriorN{} intrinsic posterior samples from iteration
\eccUniformIteration{} give the black histograms and \eccUniformCI{}\%
contours in both panels.
Colored points show measured marginalized likelihoods, with higher-likelihood
points drawn on top of lower-likelihood points and a gray background
of likelihood evaluations.
Blue lines mark the injection.
Only the angular coordinate changes between panels.
\llmfig}
\end{figure*}

Figure~\ref{fig:ecc:seob:bands} compares the two angle choices in a
zero-spin SEOBNRv5EHM injection with detector-frame component masses
$\eccUniformMOne{}+\eccUniformMTwo{}\,M_\odot$, $e_{10}=\eccUniformETruth{}$,
$\ell_{10}=\eccUniformEllTruth{}$, and network signal-to-noise ratio
$\eccUniformSNR{}$ in zero-noise H1--L1--V1 data.
Both panels show the same evaluated likelihood points and posterior samples.
The left panel uses the native anomaly at $\eccUniformFNative{}\,\mathrm{Hz}$;
the right uses the shifted angle of Eq.~\eqref{eq:ecc:angular:shear} at the chosen frequency
$f_a=\eccUniformFShared{}\,\mathrm{Hz}$.
Here we integrate the full eccentricity-dependent Peters evolution to evaluate
$A$, retaining Eq.~\eqref{eq:ecc:advance} as its small-eccentricity limit.
We denote this fitting angle by $\ell_{20}^{\mathrm{shear}}$ to distinguish
it from an anomaly evolved with the waveform model.
The left panel shows repeated likelihood bands. Shifting the angle changes
their arrangement without changing the mass or eccentricity of any point.
The run uses a uniform eccentricity prior on
$[\eccUniformEMin{},\eccUniformEMax{}]$ and the shifted angle for sampling,
fitting, and placing new likelihood points.
Proposed negative eccentricities are reflected to positive values.

\subsection{Other core RIFT features}

\label{sec:sub:explore}

\noindent \emph{Supplementary exploration methods}:
For expedient convergence, RIFT has two additional methods to explore the parameter space: dithering and incremental
dimensionality, both described in detail in previous work \cite{gwastro-RIFT-Update,gwastro-RIFT_FinerNet}.  In one
approach, RIFT's ``puffball'' code selectively adds a
 second candidate grid
derived from and supplementing the first.  In this second grid, points are generated by performing dithering on (or
randomization of) arbitrary
combinations of parameters, then rejecting unphysical combinations.
This supplementary grid helps RIFT cover the posterior and continue exploring away from the posterior
peak at every iteration.
Appendix~\ref{sec:transverse_seed_climb} describes an optional likelihood-guided
search for additional transverse-spin grid points.
In the second approach, RIFT adopts different parameterizations and/or priors at each stage. Adopting different spin
priors early on helps RIFT ensure that transverse spins are well-explored in early iterations, providing sufficient
training data for the transverse degrees of freedom.  At the same time, using a likelihood parameterization that omits
selected degrees of freedom (i.e.,the same transverse spins; subdominant tidal parameters; et cetera) ensures these
subdominant degrees of freedom are not spuriously overfit while RIFT accumulates training data in early bootstrap iterations.
The appropriate dimensional hierarchy depends on the physics involved (e.g., configurations with high
mass; BHNS with strong precession; NS-NS binaries with tides; et cetera) but is well-motivated from simple Fisher matrix
arguments.

\noindent \emph{Source-dependent pipeline tuning}
In the workflow described here, operating-point choices such as the number of likelihood
grid points evaluated by ILE and the number of CIP workers used per iteration are set when the pipeline is
built. We do not adaptively reconfigure them during an analysis.

\noindent \emph{Standard validation infrastructure}
Though RIFT's source tree includes multiple internal validation tests, typical users will be most interested in the
classic probability-probability (PP) test \cite{gwastro-skyloc-Sidery2013}.
For RIFT, we formulate a PP test as a direct constraint on the cumulative distribution function.  Specifically, using
RIFT on each source $k$ in a fair sample of sources drawn from a specific (intrinsic plus extrinsic) prior with true parameters $\mathbf{\lambda}_k$, we estimate
the fraction of the posterior distributions which is below the true source value $\lambda_{k,\alpha}$   [$\hat{P}_{k,\alpha}(<\lambda_{k,\alpha})$] for each intrinsic parameter $\alpha$.  After reindexing the sources so $\hat{P}_{k,\alpha}(\lambda_{k,\alpha})$ increases with $k$ for some fixed $\alpha$,  a
plot of $k/N$ versus $\hat{P}_k(\lambda_{k,\alpha})$ for both mass parameters can be compared with the expected result
($P(<p)=p$) and binomial uncertainty interval.

\noindent \emph{External user-provided likelihoods and priors in ILE, CIP}
\label{sec:sub:external_lnL}
RIFT provides mechanisms for users to  supplement the GW
likelihood ${\cal L}$ with an external likelihood ${\cal L}_{\rm ext}$ (e.g., to perform inference using ${\cal L}\times {\cal L}_{\rm ext}$), without needing to access or modify RIFT itself.
Additionally (albeit somewhat redundantly in principle), users can override any of the priors employed in inference
within CIP, using a similar framework.

\noindent \emph{Multiple Monte Carlo samplers}: Previous work \cite{gwastro-RIFT_FinerNet} thoroughly describes the multiple
integration schemes available within RIFT.   With the recent release, we only perform maintenance on pre-existing and
previously-reported samplers.   Notably, the GMM sampler is now GPU-accelerated throughout, and has a substantially
improved strategy for adaptive tempering.  However, the GMM sampler fundamentally remains limited by its choice of GMM
model: rather than adaptively learn the number and nature of the best-fitting model, the user must determine this choice
themselves. While useful for low- and high-amplitude sources, the GMM sampler remains too
human-guidance-intensive for large-scale automated production inference.  We defer the implementation of flexible
mixture models whose model complexity adapts to the data to future work.

\subsection{External interfaces: Waveforms}
\label{sec:sub:external:Waveforms}

Previous work describes RIFT's interfaces to \texttt{lalsimulation} and \texttt{gwsignal}
\cite{gwastro-RIFT_FinerNet}; for the underlying software see \cite{lalsuite,ligo-gwsignal}.
Other RIFT investigations used the noncircular effective-one-body model TEOBResumSDALI
\cite{2024PhRvD.110h4001N} for unbound encounters
\cite{gwastro-hyperbolic-Henshaw2025}. Here we update RIFT's waveform interface to
access supported models including IMRPhenomXPNR.

\noindent \emph{Numerical relativity library}:
RIFT has long been able to use numerical-relativity (NR) waveforms directly as templates, via an in-house
package, \texttt{NRWaveformCatalogManager}, that presents simulations from several groups --- including RIT,
SXS, and Georgia Tech (MAYA) --- through a single uniform interface, and that can evaluate the strain
$h(t,\vartheta,\phi)$ along an arbitrary line of sight.  After more than a decade of internal use across many
RIFT analyses, we are releasing \texttt{NRWaveformCatalogManager} publicly, together with interface and
catalog updates (a modular rewrite with a plugin registry, per-simulation metadata indexed for fast query,
and OSDF/pelican-based data access for use off the home filesystem).
The source repository \cite{gwastro-NRWaveformCatalogManager-repo} documents RIT and MAYA data layouts
and retains an archived SXS interface; availability of catalog simulation files is separate from the code.

\skipme{
\begin{figure*}
  \includegraphics[width=\textwidth]{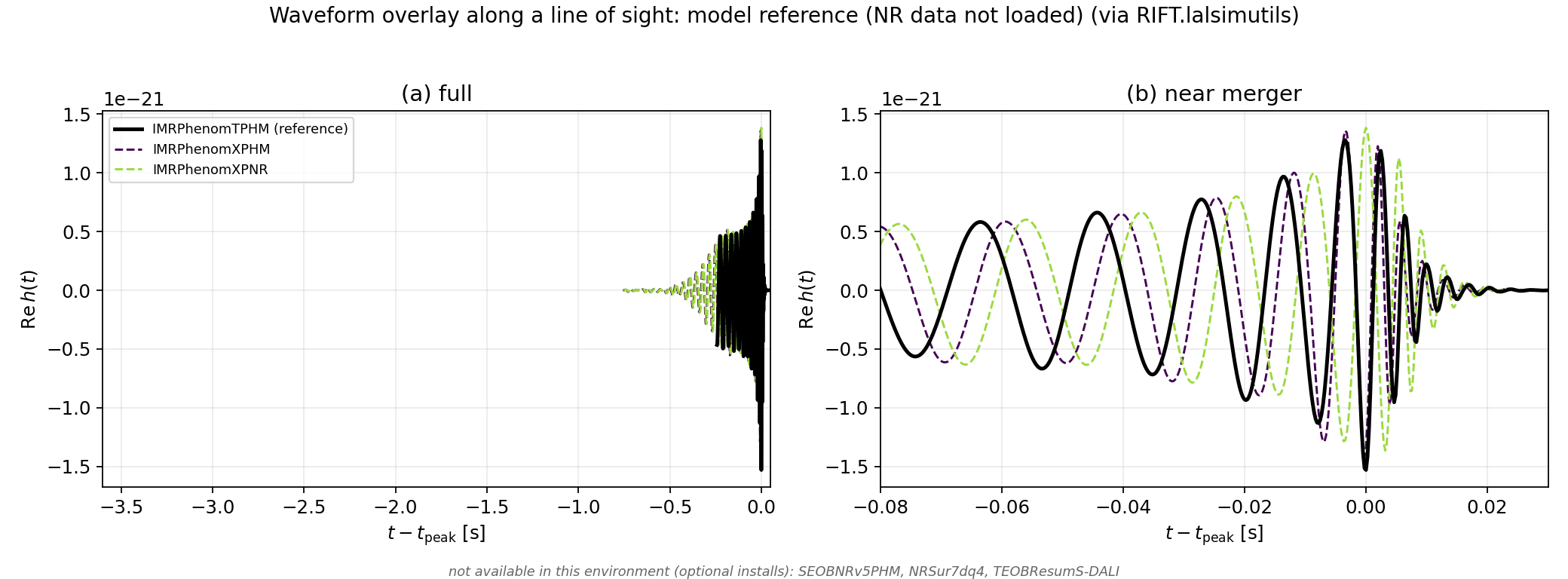}
  \caption{\label{fig:waveforms}\textbf{Overlaying NR and model waveforms along a line of sight.}
    The complex strain $\mathrm{Re}\,h(t)$ for a fiducial precessing binary, peak-aligned, generated through
    RIFT's uniform waveform interface (\texttt{RIFT.lalsimutils}); \emph{(a)} the full inspiral--merger and
    \emph{(b)} a zoom near merger.  The reference track (black) is a numerical-relativity simulation loaded
    with an NR catalog interface when NR data is available; the figure shown here,
    produced without NR archives, instead uses the time-domain model IMRPhenomTPHM as the reference, overlaid
    with IMRPhenomXPHM and the NR-tuned IMRPhenomXPNR.  Models requiring optional installs not present in the
    build environment (SEOBNRv5PHM, NRSur7dq4, TEOBResumS-DALI) are attempted and skipped with a note,
    documenting the supported stack as a moving target.%
    \llmfig}
\end{figure*}
}

\subsection{Marginalization over waveform models}
\label{sec:sub:waveform-model-marginalization}

Waveform uncertainty can be represented by a discrete model index $m$ with prior probability
$\pi_m$.  At fixed intrinsic parameters, the model-marginalized likelihood is the linear mixture
\begin{equation}
\label{eq:waveform-model-marginalization}
{\cal L}_{\rm mix}(\lambda)=\sum_m \pi_m {\cal L}_m(\lambda).
\end{equation}
A previous study developed and demonstrated this  point-by-point RIFT construction for assessing and
marginalizing waveform systematics \cite{gwastro-RIFT-systematics-AnjaliAasim-2020}.  In this work, we expand this
proof-of-concept study into an easy-to-use option for any RIFT analysis.

During each iteration, every configured waveform model is evaluated at the same proposed intrinsic
points.  Both the extrinsic likelihood integral and the intrinsic-posterior integral use RIFT's
adaptive-volume sampler in log-likelihood space.  Repeated Monte Carlo evaluations of one model estimate the same quantity and are first
pooled in linear likelihood using their sample budgets.  The per-model estimates are then combined
according to Eq.~\eqref{eq:waveform-model-marginalization}.

\begin{figure}
\includegraphics[width=\columnwidth]{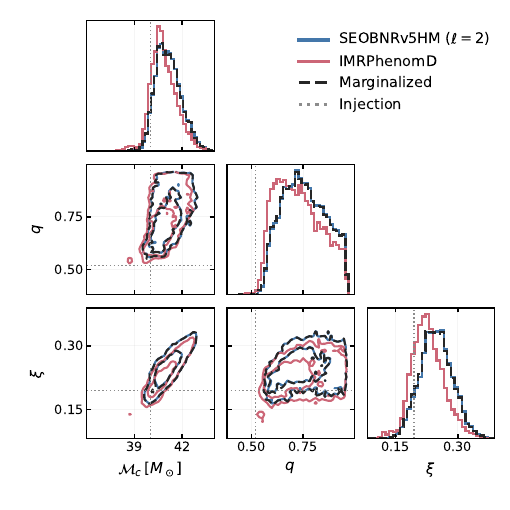}
\caption{\label{fig:multiapprox:seed1}
\textbf{A shared-grid waveform-model marginalization run.}
Preliminary single-seed intrinsic posteriors in $({\cal M}_c,q,\xi)$ for an injection generated with
SEOBNRv5HM modes through $\ell=4$.  Recovery uses SEOBNRv5HM ($\ell=2$; blue) and IMRPhenomD
(red), neither of which contains the injected higher-mode content; the evidence-weighted marginalized
posterior is dashed black. Diagonal panels show marginal densities; joint contours
enclose 50\% and 90\% of each component's histogram probability over its full
histogram domain. Dotted gray lines mark the injection.
The two models are not equally supported: $\ln Z=\maLnZSEOB$ for SEOBNRv5HM ($\ell=2$) against
$\maLnZPhenomD$ for IMRPhenomD, a difference $\Delta\ln Z=\maDeltaLnZ$ and hence a Bayes factor of
roughly $\maBayesFactor{}:1$.  The mixture weights of Eq.~\eqref{eq:waveform-model-posterior-weight}
are therefore $\maWeightSEOB$ and $\maWeightPhenomD$, which is why the marginalized curve nearly
coincides with the favoured component rather than sitting between the two.
\llmfig}
\end{figure}

Let $Z_m=\int d\lambda\,{\cal L}_m(\lambda)p(\lambda)$ be the evidence of model $m$ and
$p_m(\lambda)={\cal L}_m(\lambda)p(\lambda)/Z_m$ its posterior.  Inserting
Eq.~\eqref{eq:waveform-model-marginalization} into Eq.~\eqref{eq:post} gives
\begin{equation}
\label{eq:waveform-model-posterior-weight}
p_{\rm mix}(\lambda)=\frac{{\cal L}_{\rm mix}(\lambda)p(\lambda)}{\sum_n \pi_n Z_n}=\sum_m w_m p_m(\lambda),
\qquad w_m=\frac{\pi_m Z_m}{\sum_n \pi_n Z_n}.
\end{equation}
The pointwise likelihood average and the evidence-weighted posterior average are therefore analytically equivalent.
The samples in Figure \ref{fig:multiapprox:seed1} use the evidence-weighted form, which keeps each model's posterior
available for comparison.  This form is equivalent to running with ${\cal L}_{\rm mix}$, as in
\cite{gwastro-RIFT-systematics-AnjaliAasim-2020}.

Figure \ref{fig:multiapprox:seed1} shows the construction running end to end on a synthetic
aligned-spin binary.  The injection is generated with SEOBNRv5HM through $\ell=4$, while both recovery
models are restricted to $\ell=2$, so neither contains the injected higher-mode content: this is a
deliberate and well-understood instance of missing physics, and the realistic case, since in practice
we never analyze data whose waveform we possess.
Both recovered posteriors are correspondingly
displaced from the injected values in $q$ and $\xi$, by design.  The evidence favors one model
by $\Delta\ln Z=\maDeltaLnZ$, so the marginalized posterior resembles the result from that model exclusively.

\noindent\emph{Shared-grid waveform-model marginalization demonstration.}
\label{sec:results:waveform-model-marginalization}

We illustrate this native implementation of the Jan--Yelikar construction with a preregistered
synthetic aligned-spin binary.  The injection is
generated with SEOBNRv5HM modes through $\ell=4$, while the recovery compares SEOBNRv5HM
($\ell=2$) and IMRPhenomD on their common $\ell=2$ content.  Thus neither recovery model contains all the injected
physics.  Each model is evaluated on the same intrinsic points in-loop, and the result is compared
against separate single-model pipelines mixed only after inference.  The registered diagnostics
report the model evidences and weights, source-parameter intervals, pairwise Jensen--Shannon
divergences, complete-model coverage, compute cost, shared-grid occupancy and prior-boundary contact,
and the spread across independent Monte Carlo seeds.

Figure~\ref{fig:multiapprox:seed1} illustrates a single seed; it does not establish robustness across seeds, complete-model
coverage, or a grid-placement advantage over mixing separately completed model runs;
those comparisons require matched multi-seed and full-physics tests.

\section{Calibration marginalization}
\label{sec:core}

We trace the change from posterior reweighting to native, in-loop calibration
marginalization and then demonstrate the resulting workflow on GW240426\_031451.
This section separates the earlier approach, the current implementation,
and its demonstrations.

\subsection{Previous calibration marginalization}
\label{sec:sub:external:calmarg}

As described in previous work \cite{gwastro-RIFT_FinerNet}, RIFT marginalizes over calibration uncertainty by fixed-size Monte
Carlo.
The approach evolved from the external postprocessing used for O3 through a batched O4b variant before
the pipeline and in-loop implementations below.

\noindent\emph{Pre-O4b editions}: In versions 0.0.15 and earlier, RIFT  performs calibration reweighting as a
postprocessing step to its final extrinsic samples by rejection sampling.  Following previous work RIFT adopts a single fixed set of calibration
realizations for all extrinsic samples. Using tools obtained from \texttt{bilby} \cite{2020PhRvD.102l2004P}, it generates a weight $W_\alpha$ for each sample $x_\alpha$ derived from the
RIFT-estimated gravitational wave strain $h(x_\alpha)$ associated with that sample and a fixed number $N_{\rm cal}$
realizations of the detector calibration uncertainty, then performing rejection sampling.  This approach does not
provide calibration parameters for the final samples.

\noindent\emph{O4b edition}: RIFT performs calibration reweighting on the
final extrinsic samples in batches of size $N_{B}$, such that each batch has $N_{\rm cal}$ independent calibration
realizations.  The algorithm proceeds as described above.  At the end, the user can request that the final samples
include calibration-parameter realizations, using a random draw from the same code used to generate the calibration
weights $W_\alpha$.

\subsection{Current calibration marginalization: pipeline and in-loop}
\label{sec:calmarg}

The earlier post-hoc reweighting used the external \texttt{bilby} calibration likelihood
\cite{2020PhRvD.102l2004P,gwastro-pe-bilby-2018} and remained a separate,
operator-driven step through GWTC-3.
In versions 0.0.17.9 and above,  RIFT can perform calibration marginalization as part of its self-generated workflow.
In versions 0.0.18 and above, RIFT can alternatively performs the calibration marginalization
internally, self-consistently with its marginalization over extrinsic parameters and as part of computing
the marginal likelihood ${\cal L}$, as described below.
\mysub{In-loop marginalization}
Physically, calibration factors $\eta$ represent the correction that must be applied to the data to produce physical
strain ($d_{\rm {true}}=\eta d$).   This correction impacts both the data $d$ appearing in our likelihood and
(implicitly) our estimate of the noise power spectrum.   As a result, in practice the likelihood can most efficiently
incorporate calibration uncertainty by modifying the waveform by these factors as $\qmstateproduct{d-h/\eta}{d-h/\eta}$.
We therefore characterize residual calibration error in detector $k$ as a smooth
multiplicative frequency response applied to the template~\cite{Sun2020Calibration,Farr2014Calibration}
$h_k(f) \rightarrow \alpha_k(f;\bm{c})\, h_k(f)$,
\begin{align}
\label{eq:cal:model}
\alpha_k(f;\bm{c}) \equiv \eta_k(f;\bm{c})^{-1} = A_k(f;\bm{c})\, e^{i \phi_k(f;\bm{c})},
\end{align}
where the amplitude $A_k$ (represented as $1+\delta A_k$, with $\delta A_k$ a cubic spline) and phase $\phi_k$ are cubic splines in $\log f$ over $M$ nodes,
and the node vector $\bm{c}$ (all detectors' amplitude and phase nodes) has an independent-Gaussian prior on linear amplitude and phase
$p(\bm{c})$ whose per-node means and widths are read from the measured calibration-uncertainty envelopes.
The calibration-marginalized likelihood at fixed binary parameters is
\begin{align}
\label{eq:cal:Lmarg}
{\cal L}_{\rm marg}(\bm\lambda,\bm\theta)
 &= \int d\bm{c}\; p(\bm{c}) %
{\cal L}_{\rm full}(\bm\lambda,\bm\theta;\bm{c}) \nonumber\\
&\simeq \frac{1}{N_{\rm cal}} \sum_{c=1}^{N_{\rm cal}} w_c
     {\cal L}_{\rm full}(\bm\lambda,\bm\theta;\bm{c}_c),
\end{align}
a fixed-size Monte Carlo sum over $N_{\rm cal}$ realizations $\bm{c}_c$ drawn from a sampling
distribution $p_s(\bm{c})$ with importance weights $w_c = p(\bm{c}_c)/p_s(\bm{c}_c)$ ($w_c=1$ for prior
draws, the default).  Because the calibration factor enters multiplicatively, each realization
transforms the precomputed inner products of Eq.~(\ref{eq:lnL:MatrixForm}) rather than requiring a fresh
waveform.  In the cross-term the conjugate template factor is applied to the \emph{data},
$\tilde d \to \alpha_k^*(f;\bm c_c)\,\tilde d$, so a single template set serves every draw: the data--template timeseries is then linear
in the realization, $Q_{k,\ell m}(\bm c_c) = \qmstateproduct{h_{\ell m}}{\alpha_k^*(\bm c_c)\,d}_k$, and the
$N_{\rm cal}$ copies are packed contiguously and selected by a window offset, so the whole draw set
shares one precompute of $Q$.
By contrast, the template--template blocks are quadratic in the model and thus calibration:
\begin{align}
\label{eq:cal:selfterm}
U_{k,\ell m,\ell'm'}(\bm c_c)
 &= \qmstateproduct{\alpha_k(\bm c_c)\,h_{\ell m}}
   {\alpha_k(\bm c_c)\,h_{\ell'm'}}_k \nonumber\\
 &= 2\!\int\!\frac{df}{S_{n,k}(|f|)}\,|\alpha_k(f;\bm c_c)|^2 \nonumber\\
 &\quad\times\tilde h_{\ell m}^*\,\tilde h_{\ell'm'}\,,
\end{align}
and analogously $V_k(\bm c_c)$, so that the amplitude-profiled norm is $\qmstateproduct{\alpha_k h}{\alpha_k h}_k$
rather than the calibration-independent $\qmstateproduct{h}{h}_k$.
Substituting Eq.~(\ref{eq:cal:Lmarg}) into the extrinsic marginalization of
Eq.~(\ref{eq:lnL:MonteCarlo}) yields a joint Monte Carlo over both extrinsic parameters and calibration parameters
simultaneously.

To support scenarios with good amplitude calibration, we provide a fast approximation in which
we only incorporate the calibration correction into the data-template inner product
($Q$), retaining calibration-independent $U,V$.  This approximation slightly biases inference,
with an estimated evidence shift scaling as $\rho^4\sigma_\mu^2$ for small amplitude widths;
Appendix~\ref{ap:calmarg:selfterm} discusses this approximation in more detail.
To reduce the cost of the full calculation, we take advantage of the limited dimension of $\alpha$:
because $|\alpha_k(f;\bm c_c)|^2$ is a
squared amplitude spline, the set $\{|\alpha_k(\bm c_c)|^2\}_c$ spans a subspace of dimension $\le M(M+1)/2$,
\emph{independent of} $N_{\rm cal}$, allowing us to precompute a handful of basis-weighted template
blocks once and form each $U_k(\bm c_c),V_k(\bm c_c)$ as a cheap linear combination.

\mysub{Calibration Monte Carlo error and an adaptive calibration proposal}
The extrinsic sampler's reported variance is conditioned on the fixed draw set $\{\bm{c}_c\}$ and cannot
see the spread over calibration realizations, so the error budget for
$\ln{\cal L}_{\rm marg}$ acquires a separate calibration term, added in quadrature.  Writing the
fraction of the marginal likelihood carried by realization $c$ as
\begin{align}
\label{eq:cal:ac}
a_c = \frac{w_c\, {\cal L}_c}{\sum_{c'} w_{c'}\, {\cal L}_{c'}},
\qquad
{\cal L}_c \equiv \int d\bm\theta\, p(\bm\theta)\, {\cal L}_{\rm full}(\bm\lambda,\bm\theta;\bm{c}_c),
\end{align}
the calibration contribution is controlled by the effective number of contributing draws
$n^{\rm cal}_{\rm eff} = 1/\sum_c a_c^2$:
\begin{align}
\label{eq:cal:sigma}
\left(\sigma^{\rm cal}_{\ln{\cal L}}\right)^2
 \simeq N_{\rm cal}\, {\rm Var}_c(a_c)
 = \frac{1}{n^{\rm cal}_{\rm eff}} - \frac{1}{N_{\rm cal}} ,
\end{align}
i.e.\ $\sigma^{\rm cal}_{\ln{\cal L}} \simeq 1/\sqrt{n^{\rm cal}_{\rm eff}}$ when
$n^{\rm cal}_{\rm eff} \ll N_{\rm cal}$ (we verify this relation empirically to better than a percent
across full runs).
For modest-SNR sources a fixed calibration prior can be used; however, draws from
a fixed broad prior breaks down precisely when the data constrain calibration.

To enable inference for loud sources with calibration uncertainty, we provide a proof-of-concept demonstration for
adaptive sampling for calibration parameters.
Specifically, in parallel with CIP (our procedure to generate \emph{intrinsic} parameters for future investigation), we
operate a parallel set of tools to generate a sampling prior for subsequent \emph{calibration} inference: we harvest the
highest-likelihood points of the previous iteration, evaluate the per-realization responsibilities
$a_c$ there, fit a Gaussian proposal $p_s(\bm{c})$ in spline-node space, and seed the next iteration's
calibration draws from it.  The importance weights $w_c$ in Eq.~(\ref{eq:cal:Lmarg}) fold the proposal
back into the same estimator, so the marginalization remains unbiased while
$n^{\rm cal}_{\rm eff}$ is restored toward $N_{\rm cal}$.  %

\subsection{A real-event in-loop demonstration: GW240426\_031451}
\label{sec:results:calmarg}

For a realistic end-to-end demonstration we analyze GW240426\_031451, whose calibration envelope is narrow (amplitude $1\sigma\sim2\%$) and is
not informed by the data.  We perform two otherwise-identical full RIFT analyses --- adaptive grid, AV
extrinsic integrator, production operating point, \textsc{IMRPhenomXPHM} templates --- that differ only in
whether in-loop calibration marginalization is enabled.
 To keep the intrinsic space compact we restrict to nonspinning templates; the comparison is
internal (marginalized versus unmarginalized under identical settings), so this simplification does not
affect the conclusion. As an external comparison, we show the public GWTC-5.0
\textsc{Bilby} \textsc{IMRPhenomXPHM-SpinTaylor} posterior, which allows
nonzero spins~\cite{LVK:GWTC5PE1}.
We also ran zero-spin \textsc{Bilby} \textsc{IMRPhenomXPHM} through $\ell=3$
with public GWOSC strain and the released PSDs and calibration envelopes.
This run marginalizes over spline calibration responses.

\begin{figure}
\centering
\includegraphics[width=\columnwidth]{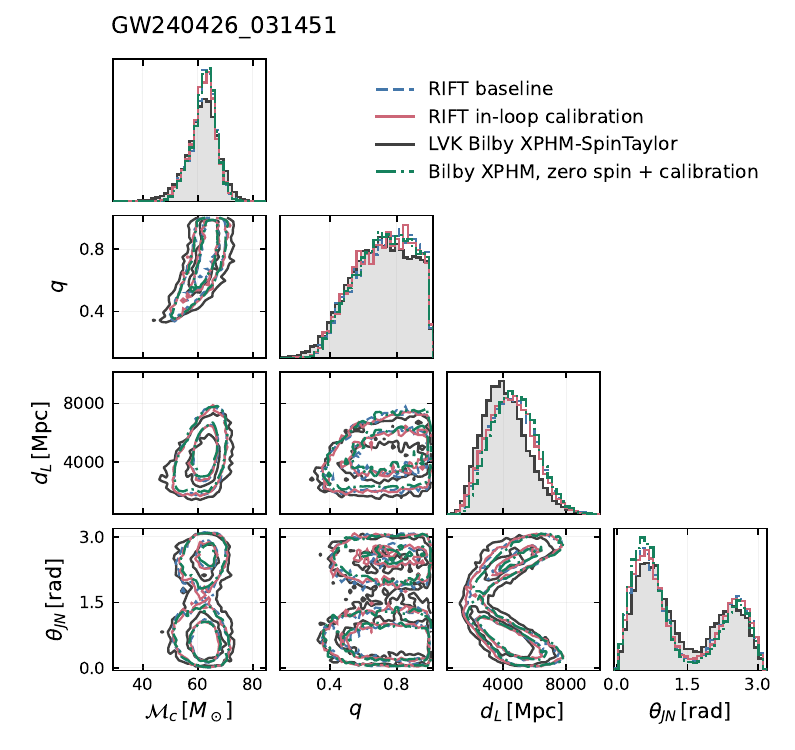}
\caption{\label{fig:calmarg}
\textbf{In-loop calibration marginalization is faithful on a well-calibrated event.}
Zero-spin \textsc{RIFT} \textsc{IMRPhenomXPHM} analyses of GW240426\_031451
without calibration marginalization (blue dashed) and with in-loop calibration marginalization (red solid).
The public GWTC-5.0 C00 \textsc{Bilby} \textsc{IMRPhenomXPHM-SpinTaylor}
posterior is shown in gray~\cite{LVK:GWTC5PE1}.
Our zero-spin \textsc{Bilby} \textsc{IMRPhenomXPHM} run with calibration
marginalization is shown in green dash-dot.
Diagonal panels show marginal densities; joint contours
enclose 50\% and 90\% of the histogram probability.
The parameters are chirp mass, mass ratio, luminosity distance and inclination.
\llmfig}
\end{figure}

The two \textsc{RIFT} posteriors in Fig.~\ref{fig:calmarg} are in excellent agreement.
Marginalizing over calibration leaves the source-parameter posterior statistically
unchanged: the chirp mass, mass ratio, luminosity distance, and inclination marginals of the two analyses
agree to a Jensen--Shannon divergence of $\le\cmGWfourInJS$~bits per parameter (Fig.~\ref{fig:calmarg}), at
a $\cmGWfourCostRatio\times$ cost in likelihood evaluations.  Throughout, the calibration Monte Carlo error
remains subdominant, with a healthy effective sample size (median
$n^{\rm cal}_{\rm eff}\simeq\cmGWfourNeffCal$ at $N_{\rm cal}=\cmNcalDemo$).

\begin{figure*}
\centering
\includegraphics[width=\textwidth]{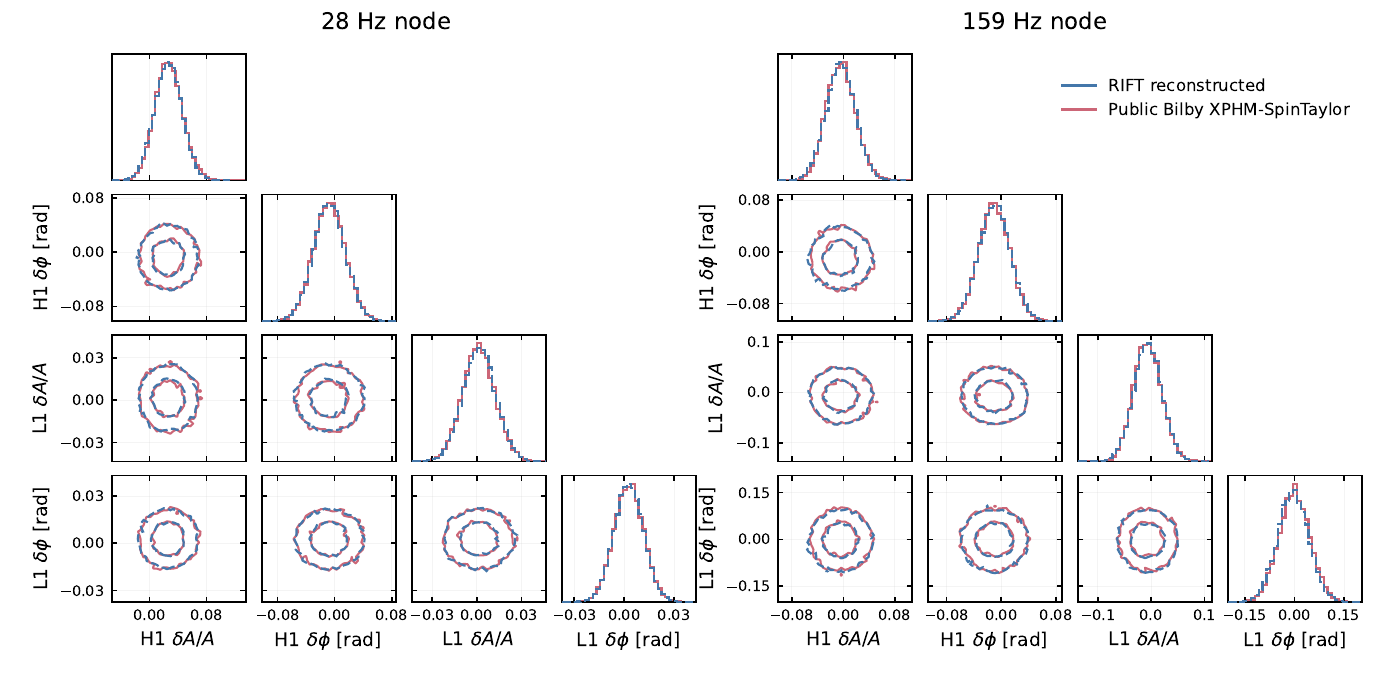}
\caption{\label{fig:calmarg_cal}
\textbf{Reconstructed calibration marginals compared with public PE.}
Calibration spline-node posteriors for GW240426\_031451 at a low-frequency node (left)
and a mid-frequency node (right), showing H1 and L1 amplitude and phase in
RIFT's data-side convention. RIFT postprocessing is blue. The public C00
\textsc{Bilby} \textsc{IMRPhenomXPHM-SpinTaylor} samples are red.
Diagonal panels show marginal densities. Joint contours enclose 50\% and
90\% of the unsmoothed histogram probability within each node.
Cross-node correlations are omitted.\llmfig}
\end{figure*}

Figure~\ref{fig:calmarg_cal} shows the recovered amplitude and phase spline-node
posteriors for GW240426\_031451 with the public GWTC-5.0 \textsc{Bilby}
\textsc{IMRPhenomXPHM-SpinTaylor} samples~\cite{LVK:GWTC5PE1}.
We convert the template corrections in the public samples to the data-side
convention used by the \textsc{RIFT} reconstruction.

\section{Diagnosing limitations of and improving upon operating point choices}
\label{sec:core_operating}
RIFT's  previous operating point choices, motivated by the first GW observations, were adopted empirically to balance efficiency and sufficient
accuracy for those sources.  As in our previous updates, ongoing operations with the growing GW census have identified
new areas where our operating point choices could be improved.  In this section, we describe the previous operating
point choice; illustrate the limitations of this choice with concrete examples; present a revised operating point; and
demonstrate the revised operating point mitigates the issue illustrated by these examples

\subsection{Extrinsic sampling and ILE termination threshold}
RIFT's marginal likelihoods and extrinsic samples are produced by a Monte Carlo integrator whose termination criteria
is based on reaching a threshold $n_{\rm eff}$.  Previously, we adopted a threshold $n_{\rm eff}=10$ for intrinsic
analysis and $n_{\rm eff}=4$ for the final extrinsic analysis.  The former choice was  sufficient to
achieve modest accuracy in the marginal likelihood (particularly when supplemented by many random nearby evaluations
with independent Monte Carlo errors); the latter choice was sufficient to produce multiple independent extrinsic
posterior samples.  Though
somewhat aggressive, this choice demonstrably worked well for most GW sources identified prior to O4b.
With the larger sample afforded by O4b, however, the limitations of this threshold
became apparent.

\begin{figure*}
\centering
\includegraphics[width=0.85\textwidth]{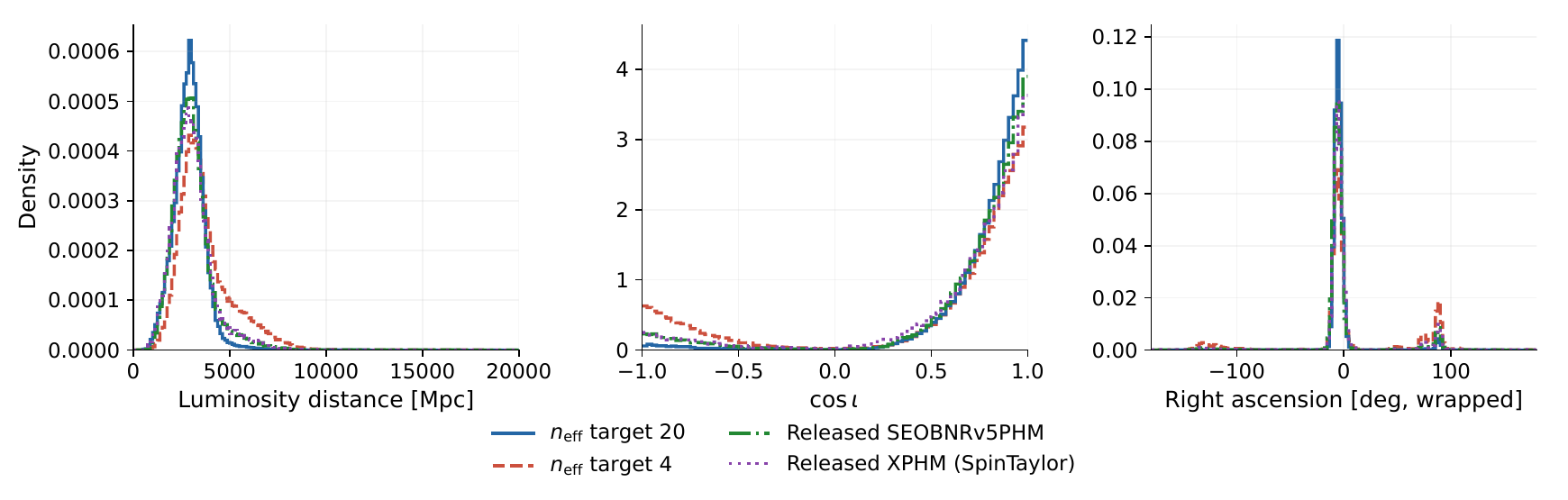}
\caption{\label{fig:review:gw241009em:neff}\textbf{The final-extrinsic
stopping target changes the recovered distance and orientation structure.}
Independent GW241009\_220455 \textsc{SEOBNRv5PHM} re-extractions with stopping
targets $n_{\rm eff}=20$ (blue) and 4 (red dashed).
Released GWTC-5.0 C00 \textsc{SEOBNRv5PHM} (green dash-dotted) and
C00 \textsc{IMRPhenomXPHM-SpinTaylor} (purple dotted) posteriors are overlaid
from the GW241009\_220455 combined release~\cite{LVK:GWTC5PE1}.
Panels show distance, inclination and wrapped right ascension.
\llmfig}
\end{figure*}

Figure~\ref{fig:review:gw241009em:neff} compares independent final-extrinsic
reconstructions with stopping targets $n_{\rm eff}=4$ and 20, generated from  a shared
intrinsic grid, for GW241009\_220455. Both analyses use the same strain, PSDs, waveform model (here, SEOBNRv5PHM), and
physical priors, without calibration reweighting or optional time resampling.  Up to the final iteration, both analyses
are identical: only the final stage differs.
The low-$n_{\rm eff}$ configuration includes a broad tail.
These results are reproducible: every analysis performed in this configuration will produce this result.
For comparison, we also show the released LVK IMRPhenomXPHM result (from bilby) and SEOBNRv5PHM result (from RIFT, with
$n_{\rm eff}=10$).
In the low-$n_{\rm eff}$ analysis, a small effective sample pool in each
conditional extrinsic integration inadequately represents sky, distance, and
inclination structure, even though the intrinsic posterior and marginal likelihood
appear stable.
By contrast, the high-$n_{\rm eff}$ analysis produces a skymap which matches the production configuration.

\subsection{Loud-event recovery example: GW250114 }
\label{sec:results:loud_event}

Previous RIFT choices for sampling rate, time interpolation and sky coordinates
produced suboptimal extrinsic posteriors for GW250114.
The LVK discovery and spectroscopy studies are Refs.~\cite{LIGO-O4-GW250114-anniversary_event,LIGO-O4-GW250114_spectroscopy}.
We illustrate the
limitations and their mitigation with two complementary tests: targeted
extrinsic integrations at fixed intrinsic parameters, and full extrinsic
re-extractions over an intrinsic posterior. Unless otherwise stated, we use
\textsc{SEOBNRv5PHM}, adaptive-volume (AV) integration, 4096 Hz detector data
and PSD estimates from the LVK C01 offline \textsc{Bilby} \textsc{SEOBNRv5PHM} analysis.
Public analysis-ready strain for GW250114 is available through GWOSC~\cite{GWOSC:O4DiscoveryData}.
The figure comparisons use \texttt{posterior\_samples\_SEOBNRv5PHM.h5}
from the GW250114 discovery release~\cite{LVK:GW250114DiscoveryData}.

\begin{figure*}
\centering
\includegraphics[width=0.75\textwidth]{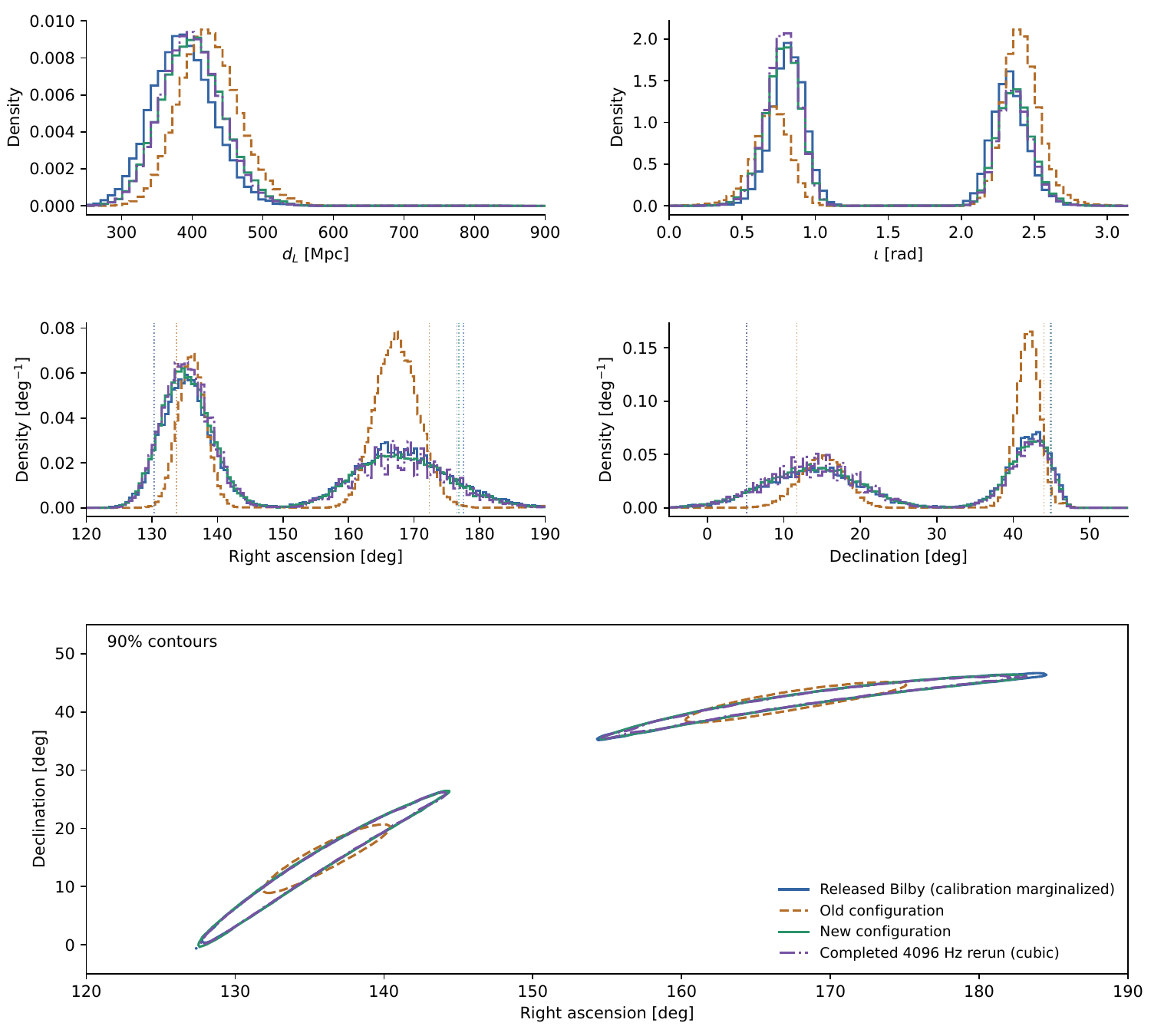}
\caption{\label{fig:review:s250114ax}
\textbf{Sky-coordinate recovery for GW250114.}
Released bilby \textsc{SEOBNRv5PHM} samples (blue), the old configuration
(brown dashed), the new configuration on the saved intrinsic grid (green),
and the completed cubic-interpolation rerun at 4096 Hz (purple dash-dot).
The upper panels show distance and inclination; the middle panels show
right ascension and declination, with dotted lines at the 5th and 95th
percentiles. The lower panel shows 90\% sky contours, all of which close
inside the displayed window.
The green curve uses corrected H1--L1-aligned coordinates, AV with full
adaptation and phase rotation, and 16384 Hz internal/output sampling on
$\skySavedAnchors$ intrinsic points.
The purple curve uses $\skyCompleteDraws$ conditional draws from all
$\skyCompleteAnchors$ points in its export grid, with
$\skyCompletePerAnchor$ draws per point and $n_{\rm eff}\ge\skyExportEss$.
These configurations change several settings and do not isolate the
sky-coordinate choice. RIFT precedes calibration reweighting; the released
bilby reference includes calibration uncertainty.\llmfig
}
\end{figure*}

Figure~\ref{fig:review:s250114ax} compares the sky, distance and inclination
posteriors in full re-extractions with a common intrinsic posterior and
the same final-stage likelihood, priors and sampling settings. The change is
the choice of sky coordinates in the final extrinsic integration. Aligning
the polar axis with the H1--L1 baseline makes the narrow timing ring
approximately a contour of constant polar angle, allowing the Cartesian
AV sampler to adapt more effectively to its geometry. The aligned-coordinate
extraction recovers the extended sky modes and their relative weights,
together with the associated distance and inclination distributions.
The standard, widely-adopted coordinate transformation and its inverse are described in
Appendix~\ref{ap:review:skyframe} for completeness.

\begin{figure*}
\centering
\includegraphics[width=\textwidth]{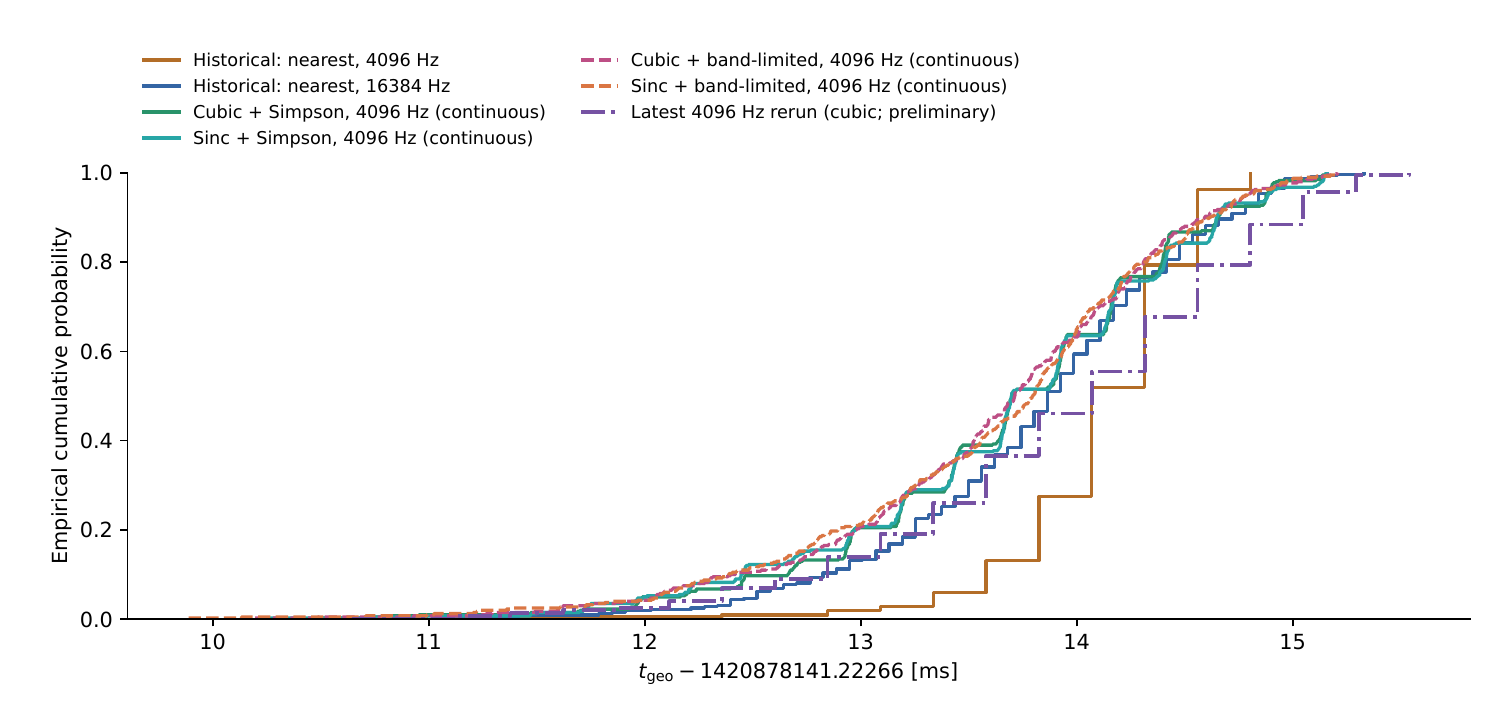}
\caption{\label{fig:review:s250114ax:time}
\textbf{Geocenter-time recovery: preliminary numerical comparison.} Empirical CDFs for historical nearest-sample
calculations at 4096 and 16384 Hz, and replacement cubic/sinc calculations
with Simpson/band-limited time quadrature and explicitly requested continuous
conditional-time export at 4096 Hz. The historical arms contain 320 fair
draws from 32 independent seeds each. Each replacement arm currently contains 400 fair draws from 4 integration seeds. The targeted integrations share one
intrinsic point; the replacement numerical arms share a pinned source,
data, priors, sampler settings and seed ensemble. Each displayed replacement
run passes the finite-output checks and reaches $n_{\rm eff}\ge100$ without
exhausting its evaluation budget. The purple dash-dot curve adds 1,200 conditional
draws from 120 of 30,000 intrinsic points in the latest full rerun
(preliminary), using cubic interpolation with default discrete time export.
It marginalizes over a scheduler-selected intrinsic subset and is not a
fixed-intrinsic numerical control. Historical curves retain their archived
time-label convention, so their differences from the replacements do not
isolate interpolation alone. Sinc denotes $Q_{\ell m}(t)$ interpolation,
separate from the choice of time quadrature.\llmfig}
\end{figure*}

Figure~\ref{fig:review:s250114ax:time} compares the recovered geocenter-time
distribution for different choices of $Q_{\ell m}(t)$ interpolation, time
quadrature and sampling rate. The historical calculation represents the
posterior on a coarse time lattice, under-resolving its structure even when
it identifies the arrival-time region. Discrete time export remains the
default. With continuous conditional-time export explicitly requested, cubic
interpolation and conditional time draws recover that structure at 4096 Hz,
with the historical 16384 Hz calculation providing a higher-rate comparison.
We compare Simpson
and band-limited time quadrature separately to check the integrated likelihood
weights: a smooth exported time distribution does not by itself establish
that the underlying time integral is resolved.

\subsection{Transverse spin recovery and interpolation}
\label{sec:rf:transverse:operating}
\begin{figure*}[t]
\centering
\includegraphics[width=0.8\textwidth]{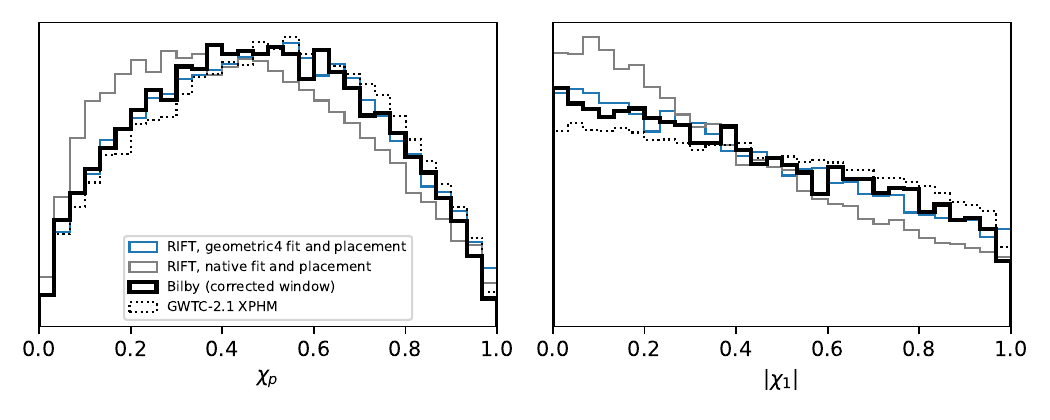}
\caption{\label{fig:results:gw150914_geometric4}
\textbf{With the geometric4 chart, a full GW150914 RIFT analysis recovers the
\textsc{bilby} $\chi_p$ marginal.}
Marginals of $\chi_p$ (left) and $|\chi_1|$ (right) from two complete RIFT
IMRPhenomXPHM analyses: geometric4 fit and point placement (blue) and the
native coordinates (gray). Black curves are the window-corrected
\textsc{bilby} analysis (solid) and the GWTC-2.1 IMRPhenomXPHM release
(dotted). Each RIFT curve pools $\gwgNreps$ final fits
($\gwgNdraws$ draws). The RIFT curves do not marginalize over calibration;
both references do.\llmfig}
\end{figure*}

Changing the RF interpolation coordinates can change the recovered posterior
even when the likelihood archive, physical prior and sampler are held fixed.
As an example, Figure~\ref{fig:results:gw150914_geometric4} shows the inferred posterior for $\chi_p$ for GW150914 using an analysis with the new coordinate system
end-to-end against (a) a RIFT analysis using the previous rift default settings, (b) a bilby reference (solid black) performed with
matched analysis settings; and (c) previously published results, which adopt slightly different settings.

\section{Conclusions}
\label{sec:conclude}
We extended RIFT's single-event inference capabilities for the O4 era with new intrinsic fitting coordinates,
calibration and waveform-model marginalization workflows, revised operating points, and controlled cross-code
likelihood comparisons. Building on our previous update \cite{gwastro-RIFT_FinerNet}, these developments
address both the range of analyses RIFT can perform and the accuracy of the source properties it recovers.

The new intrinsic coordinates change the fitting geometry without changing the physical source model.
This distinction matters because interpolation can otherwise obscure weakly constrained spin structure or
make a correlated eccentric likelihood unnecessarily difficult to explore.
We promoted calibration reweighting to a native pipeline stage and developed an alternative that marginalizes
calibration uncertainty inside the extrinsic likelihood integral.
The native waveform-model mixture extends the same principle to uncertainty
in the signal description. Together with the revised operating points, these capabilities help distinguish
features supported by the data from features sensitive to calibration, waveform choice or incomplete
extrinsic reconstruction. This distinction becomes more important as individual events constrain source
properties more tightly.

In Appendix~\ref{ap:likelihood_code_comparison} we provide extensive cross-code tests of our likelihood against another
implementation.  Different likelihood implementations and code paths invoke different assumptions, approximations, and
conditioning choices; in our appendix, we quantify the impact of these choices on the likelihood both numerically and
semianalytically, providing a framework to identify and adopt future operating-point choices to achieve a desired
fidelity to another code.

RIFT's computational efficiency has also improved through several optimizations beyond the methods emphasized
in the main text, including optional prototypes that give factor-of-several reductions in cost
(see Appendices~\ref{app:early-multirate} and~\ref{app:gpu-xphm} for the multirate and GPU-waveform benchmarks),
above and beyond the improvements
reported in our previous studies \cite{gwastro-RIFT_FinerNet,gwastro-RIFT-roboto-scaling}.
The gains depend on the waveform, signal duration, sampling rate and batching strategy; in the tested
configurations, combining multirate preparation with GPU waveform generation reduces the steady-state
cost per intrinsic point by more than an order of magnitude (Table~\ref{tab:early-multirate}).

Several opportunities identified in our previous study \cite{gwastro-RIFT_FinerNet} remain incompletely
exploited. Parsimonious placement could reduce redundant likelihood evaluations by concentrating new
points where they add information to the surrogate, while better use of the portfolio integrator could
combine complementary proposals and adapt their allocation to the likelihood being explored. Reusing
information between successive integrations, adapting worker allocation to measured demand, and reducing
repeated input and initialization costs offer further opportunities. The present improvements do not
exhaust these possibilities: turning them into reliable production gains requires coordinated choices of
placement, integration and scheduling, tested against posterior accuracy as well as time to solution.

While we propose several configurations which improve RIFT's accuracy, for simplicity we do not balance accuracy and
throughput.   Particularly if pushed to extreme choices, settings which on a per-ILE basis improve RIFT's accuracy (i.e., larger
values of $n_{\rm eff}$, for Monte Carlo integration, and $Q$, for time interpolation) can require substantially and
even prohibitively increased runtime and (GPU) memory footprint.   A larger footprint in particular can potentially
significantly restrict the pool of compatible resources: few machines will have the necessary capability.  A large-scale analysis campaign should carefully balance the
potentially increased accuracy these settings enable with the increased cost they require, and judiciously adopt
settings which allocate effort where the improved performance has demonstrable scientific benefit.  The cluster-specific
analysis and testing
needed to carefully develop and choose these operating points depends sensitively and dynamically on the available
computing resources, and is beyond the scope of the investigation performed here.

\begin{acknowledgments}
We thank the LVK parameter estimation group, particularly Nelson Christensen, Colm Talbot, Greg Ashton, and Nihar Gupte
for helpful discussions which  motivated some of the investigations described in this work, and Mattia Emma for helpful
feedback on the draft manuscript.
We thank Danilo Chiaramello for discussions of eccentric inference and its
coordinate dependence.
ROS acknowledges support from  NSF PHY 2012057, 2309172 and 2206321. ROS also acknowledges assistance from
  Claude/Codex during the implementation, operation, and drafting of the analyses in this manuscript,
  as described in Appendix~\ref{app:llm}. ROS acknowledges support from
  OpenAI's Academic Researchers program.  The authors are
  grateful for computational
  resources provided by the LIGO Laboratory and supported by National Science
  Foundation Grants PHY-0757058 and PHY-0823459. This research has made use of data or
  software obtained from the Gravitational Wave Open Science Center (gwosc.org),
  a service of the LIGO Scientific Collaboration, the Virgo Collaboration, and
  KAGRA. The O4b open-data release is described in Ref.~\cite{LIGO-O4b-opendata}.
  This material is based upon work supported by NSF's LIGO Laboratory
  which is a major facility fully funded by the National Science Foundation, as
  well as the Science and Technology Facilities Council (STFC) of the United
  Kingdom, the Max-Planck-Society (MPS), and the State of Niedersachsen/Germany
  for support of the construction of Advanced LIGO and construction and
  operation of the GEO600 detector. Additional support for Advanced LIGO was
  provided by the Australian Research Council. Virgo is funded, through the
  European Gravitational Observatory (EGO), by the French Centre National de
  Recherche Scientifique (CNRS), the Italian Istituto Nazionale di Fisica
  Nucleare (INFN) and the Dutch Nikhef, with contributions by institutions from
  Belgium, Germany, Greece, Hungary, Ireland, Japan, Monaco, Poland, Portugal,
  Spain. KAGRA is supported by Ministry of Education, Culture, Sports, Science
  and Technology (MEXT), Japan Society for the Promotion of Science (JSPS) in
  Japan; National Research Foundation (NRF) and Ministry of Science and ICT
  (MSIT) in Korea; Academia Sinica (AS) and National Science and Technology
  Council (NSTC) in Taiwan.
\end{acknowledgments}

\appendix

\clearpage

\section{Assessing likelihood differences between codes}
\label{ap:likelihood_code_comparison}

Independent parameter-estimation codes nominally evaluate the same Gaussian
gravitational-wave likelihood, but pointwise comparisons at fixed source
parameters are sensitive to several implementation conventions.  In this
appendix we describe the controlled comparison protocol used for RIFT--bilby
checks, emphasizing mechanisms rather than the chronological diagnostic history.
The differences we encountered fall into three broad classes: selecting a
different discrete strain segment, evaluating the same precomputed correlation
series at different detector arrival times, and applying different data
conditioning, frequency masks, or PSD conventions.  A separate prerequisite is
that both codes use the same waveform object, including its epoch, support,
conditioning, mode content, and tapering; otherwise waveform-generation
differences dominate before the likelihood implementation is being tested.

The relevant RIFT quantity is the likelihood-ratio exponent returned by the
maintained NoLoop path,
\texttt{DiscreteFactoredLogLikelihoodViaArrayVectorNoLoop},
\begin{equation}
  \ln \Lambda = \langle h|d\rangle - {1\over 2}\langle h|h\rangle .
\end{equation}
It does not include the Gaussian noise evidence
$-{1\over2}\langle d|d\rangle$ or determinant normalization terms.  The directly
comparable bilby quantity is therefore \texttt{log\_likelihood\_ratio()}, or
equivalently \texttt{log\_likelihood()}$-$\texttt{noise\_log\_likelihood()}.

\subsection{One-sample data offsets}

The first requirement is that both codes analyze the same detector samples.
Frame-reader backends can otherwise introduce a one-sample shift.  In common
\texttt{gwpy} installations, GWF data are read through \texttt{frameCPP} when
the \texttt{LDAStools.frameCPP} bindings are available, and otherwise through
\texttt{LALFrame}.  For a requested start time that is not on the sample grid,
these paths may select neighboring samples: if
\begin{equation}
  x=(t_{\rm start}-t_0)f_s ,
\end{equation}
then the \texttt{frameCPP} path floors the sample index, while the
\texttt{LALFrame} path takes the next sample for non-integer $x$.  This is a
backend convention, not an astrophysical effect.  Using a common frame backend,
or explicitly snapping the segment start to the sample grid before frame access,
removes this ambiguity.  RIFT now supports both \texttt{gwpy} and \texttt{frameCPP} import, allowing us to directly test
the impact of this (minor) change.

\subsection{Waveform-generation sampling rate}

A common data rate does not guarantee a common waveform-generation rate.
The pySEOBNR frequency-domain interface chooses its internal time step from the
requested maximum frequency and Fourier spacing.  If the cutoff is well below
Nyquist, it can generate a coarser time-domain waveform while returning the
same frequency grid.  For example, a $\tieRate\,\mathrm{Hz}$ data stream with
$f_{\max}=\tieCornerFmax\,\mathrm{Hz}$ selects
$\tieCornerRate\,\mathrm{Hz}$ generation in the frequency-domain call, while
RIFT's time-domain mode call retains $\tieRate\,\mathrm{Hz}$.
This introduces a waveform difference before detector-time interpolation is tested.
Figure~\ref{fig:likelihood_matched_taper}
uses the O4b/c setting $f_{\max}=\tieFmax\,\mathrm{Hz}$ with
$\tieRate\,\mathrm{Hz}$ data; both backends were checked to generate at that rate.

\subsection{Detector-time interpolation and accuracy}

Even after the waveform support is fixed, RIFT and bilby implement detector
arrival times differently in the low-level likelihood.  In the maintained
``main'' NoLoop path, RIFT samples the precomputed $\rho_{\ell m}$ time series
on integer detector-time bins, whereas bilby applies a continuous Fourier phase
for the arrival time.  Thus changing the sky position changes not only the
antenna factors, but also the fractional part of each detector's arrival time
relative to the sample grid.  This is not a normalization error; it is the
ordinary loss in matched-filter output incurred by evaluating a signal at a
nearby sampled time.

The expected scale is transparent.  If detector $i$ has SNR $\rho_i$ and its
arrival time is rounded by $\epsilon_i$, then
\begin{equation}
  \Delta \ln\Lambda \simeq
  \sum_i \rho_i^2\left[1-M_i(\epsilon_i)\right]
  \simeq
  2\pi^2\sum_i \rho_i^2 \epsilon_i^2 \langle f^2\rangle_i ,
\end{equation}
where $M_i$ is the normalized autocorrelation and $\langle f^2\rangle_i$ is
the PSD-weighted second frequency moment of the signal in that detector.  For
random sky positions the fractional sample phase is approximately uniform, so
$\langle\epsilon_i^2\rangle=1/(12 f_s^2)$ and
\begin{equation}
  \left\langle \Delta \ln\Lambda\right\rangle
  \simeq { \pi^2 \over 6}
  \sum_i \rho_i^2 {\langle f^2\rangle_i\over f_s^2}.
\end{equation}
Rounding to the nearest sample is one case of a general stencil error. RIFT
evaluates each detector correlation at a requested time from samples spaced by
$\delta t_Q$, through a stencil whose response to a Fourier component, relative to
exact translation by the fractional offset $u_i\delta t_Q$, is $H(\omega;u_i)$ with
$\omega=2\pi f\delta t_Q$. For a GW signal the fixed-time error is
\begin{equation}
  \Delta\ln\Lambda \simeq \sum_i \rho_i^2
  \left\langle 1-\mathrm{Re}\,H(2\pi f\delta t_Q;u_i)\right\rangle_i ,
  \label{eq:stencil-error}
\end{equation}
where $\langle\cdot\rangle_i$ is the PSD-weighted average over the signal spectrum
in detector $i$. A stencil exact for polynomials of degree $p-1$ has
$1-H=O(\omega^p)$, so
\begin{equation}
  \Delta\ln\Lambda \propto \rho^2\left(2\pi\sigma_f\,\delta t_Q\right)^p ,
  \qquad \delta t_Q=\frac{1}{N_Q f_s},
  \label{eq:stencil-scaling}
\end{equation}
with $\sigma_f$ the PSD-weighted frequency width. Nearest-sample lookup has $p=2$
and gives the expressions above; cubic interpolation has $p=4$
(App.~\ref{ap:likelihood_sampling_rate}). Spectral refinement of the correlation
table by $N_Q$ divides the error by $N_Q^p$. A finite sinc stencil follows no such
power law: its error is set by the stencil length and the correlation window.
The error depends on each detector's offset, so it varies with sky position.
For broad sky posteriors it partly averages out; for loud, well-localized sources
it can appear as a coherent pointwise offset. The JAX likelihood
\cite{gwastro-RIFT-roboto-scaling} has the same structure and the same remedies.

\begin{figure*}
\includegraphics[width=\textwidth]{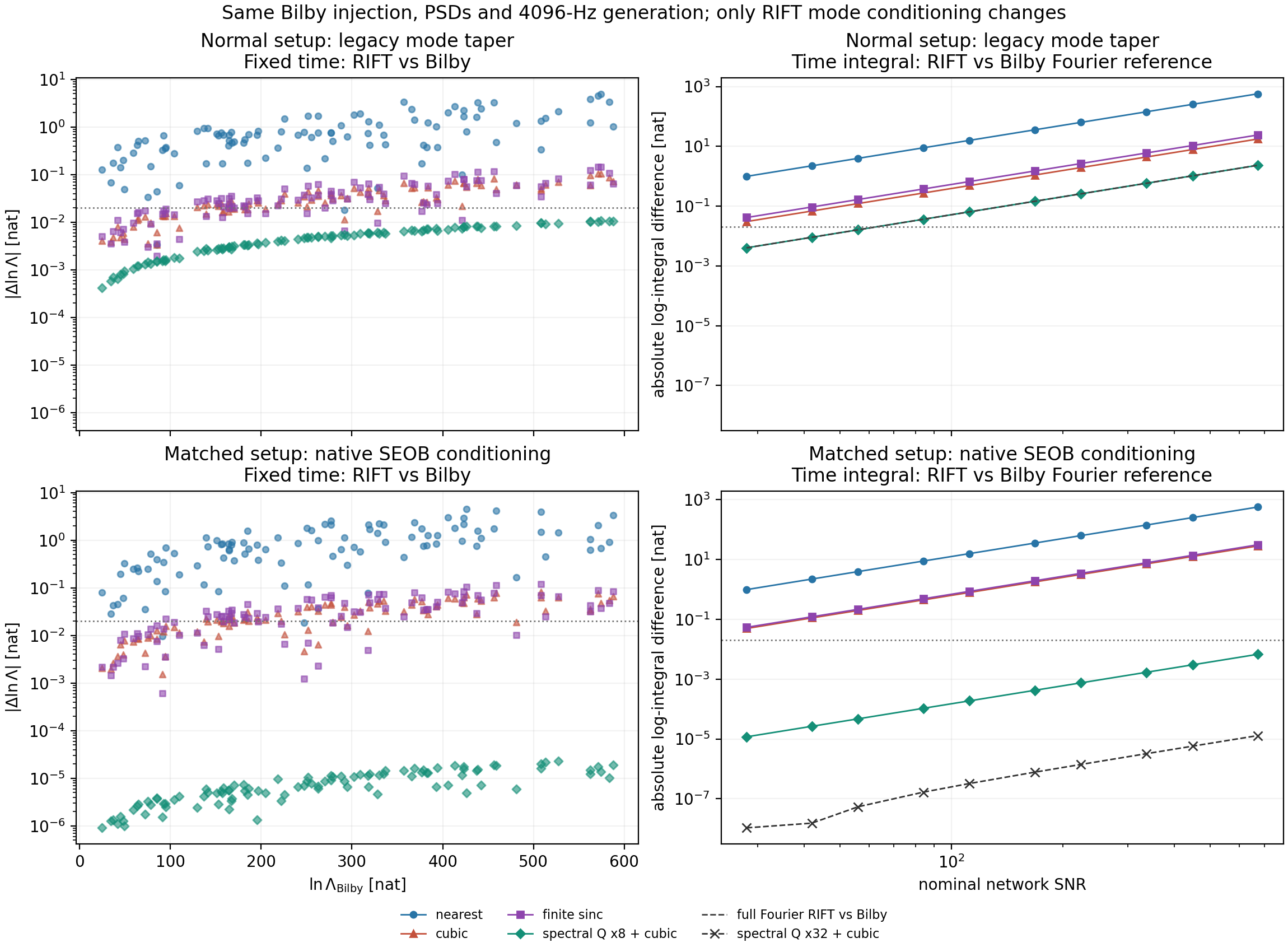}
\caption{
Time-interpolation accuracy and waveform conditioning with O4b/c sampling settings,
for one zero-noise H1/L1 SEOBNRv5PHM injection.
Both codes use $\tieRate\,\mathrm{Hz}$ data and waveform generation, a
$\tieFmax\,\mathrm{Hz}$ analysis cutoff and a $\tieFmin\,\mathrm{Hz}$ waveform start.
\emph{Top:} RIFT's normal mode taper. \emph{Bottom:} native SEOB conditioning.
\emph{Left:} absolute fixed-time RIFT--bilby likelihood-ratio differences at
$\tieNsky$ sky locations.
\emph{Right:} absolute time-integral residuals against a converged Fourier
reference using bilby's waveform, at $\tieNsnr$ amplitudes.
Nearest, cubic and finite sinc evaluate the native $\toaNativeQRateHz\,\mathrm{Hz}$
detector correlations; spectral $Q$ refinement resamples them to
$\toaPregridRateHz\,\mathrm{Hz}$ ($\toaPregridFactor$-fold) before cubic interpolation.
The dashed upper curve uses full Fourier reconstruction for RIFT's normal waveform;
the dashed lower curve uses $\mtFineFactor$-fold $Q$ refinement for the matched waveform.
Dotted lines mark the $\toaAccuracyBudget$-nat budget.\llmfig
}
\label{fig:likelihood_matched_taper}
\end{figure*}

Figure~\ref{fig:likelihood_matched_taper} compares these methods against bilby
at fixed time and against a converged Fourier reference for the time integral,
for one zero-noise H1/L1 \texttt{SEOBNRv5PHM} injection at the O4b/c operating
settings. Both codes receive the same in-memory data and PSD values, so the fixed-time
comparison isolates time representation and waveform conditioning. The ordering
follows Eq.~(\ref{eq:stencil-scaling}): nearest, then cubic, then refined cubic.
Finite sinc does not improve on cubic, so choosing sinc is not itself an accuracy
certificate (Paper~1, Appendix~D). Fixed-time accuracy also does not settle the
time integral: the terminal quadrature must itself be band-limited, or it
dominates the error. Spectral $Q$ refinement followed by cubic interpolation and
band-limited quadrature meets the $\toaAccuracyBudget$-nat budget through network
SNR $\toaValidationSnrMax$. These tests cover aligned spins and specified windows
and bandwidths; they do not rank interpolation methods in general.

\begin{figure}
\includegraphics[width=\columnwidth]{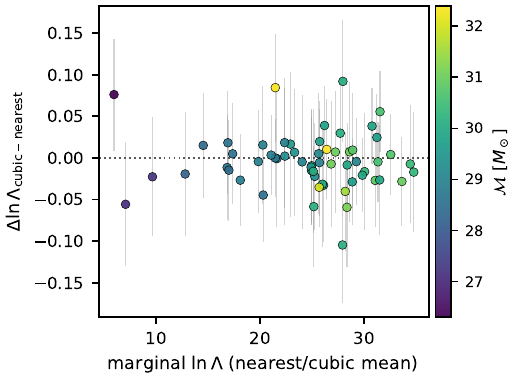}
\caption{
Propagation of the nearest-sample detector-time effect into ILE marginal
likelihoods.  Each point is a paired nearest/cubic NoLoop ILE run for the same
saved zero-noise H1/L1 data and the same template parameters, plotted against
the mean of the two marginal likelihood estimates and colored by chirp mass.
The injection for this
local reproducibility test is a nonspinning \texttt{TaylorT4} signal with
$(m_1,m_2)=(40.565894,30.012597)\,M_\odot$, $d_L=552.15344\,{\rm Mpc}$,
$\theta_{JN}=2.91410$, $\phi_c=3.44024$, $\psi=1.47285$,
$(\alpha,\delta)=(0,0)$, and geocentric time $1126259462.44179$; the analysis
uses $f_{\min}^{\rm template}=10\,{\rm Hz}$, $f_{\min}^{\rm analysis}=20\,{\rm Hz}$,
$f_{\max}=895.875\,{\rm Hz}$, $4096\,{\rm Hz}$ sampling, and the same H1/L1
PSDs as Figure~\ref{fig:likelihood_matched_taper}.  The 60 template
points are sampled in the nearby chirp-mass--symmetric-mass-ratio region.  The rms
paired difference is smaller than the typical quoted Monte Carlo uncertainty; the mass dependence of the correction is discussed in the text.\llmfig
}
\label{fig:likelihood_ile_marginal_time_interp}
\end{figure}

As RIFT's previous success demonstrates, in almost all circumstances extremely high-precision ilkelihoods are not
required: usually, any likelihood interpolation error fluctuates periodically across the broad sky posterior and
therefore averages out when performing the marginal likelihood.
Figure~\ref{fig:likelihood_ile_marginal_time_interp} demonstrates this tendency quantitatively, by
comparing the RIFT marginal likelihoods computed using two interpolation schemes: nearest-neighbor and cubic.
The difference in marginal likelihoods is  smaller than the Monte Carlo uncertainty.

\subsection{Sampling rate as a comparison variable}
\label{ap:likelihood_sampling_rate}

Comparisons with other codes add the data sampling rate $f_s$ as a variable.
Bilby applies detector arrival times as a continuous Fourier phase, so its
fixed-time likelihood does not depend on $f_s$ once the analysis band is fixed.
RIFT instead interpolates each detector correlation on a grid of spacing
$\delta t_Q=1/(N_Q f_s)$, where $N_Q$ is the spectral refinement factor of the $Q$ table.
For a Fourier component $e^{i\omega n}$ with $\omega=2\pi f\,\delta t_Q$, the four-point
cubic stencil evaluated half a sample from the grid has response
\begin{equation}
 H_{1/2}(\omega)
 =\frac98\cos\frac\omega2-\frac18\cos\frac{3\omega}{2}
 =1-\frac{3}{128}\omega^4+O(\omega^6)
\label{eq:cubic-transfer}
\end{equation}
relative to exact translation, the $p=4$ case of Eq.~(\ref{eq:stencil-scaling}).
This is an interpolation error of an adequately sampled series, not aliasing, and
band-limited quadrature of the network correlation does not remove it. At fixed
signal and analysis band the error depends on $f_s$ and $N_Q$ only through
$N_Q f_s$.

\begin{figure}
\includegraphics[width=\columnwidth]{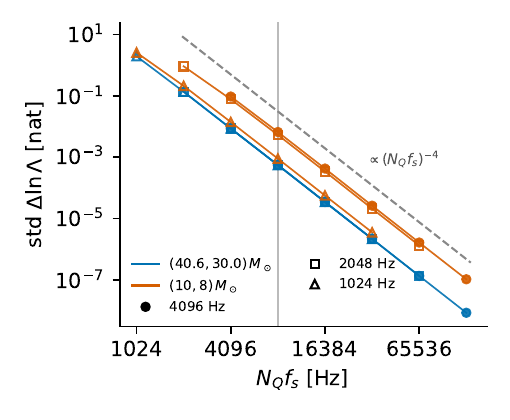}
\caption{\label{fig:likelihood_q_rate_scaling}Scatter of the fixed-time
$\ln\Lambda$ error from cubic interpolation of the $Q$ table versus the refined
rate $N_Q f_s$. Colored curves: zero-noise single-detector \texttt{IMRPhenomD}
correlation at $\rho=\irQScanRho$ with the aLIGO design PSD, for two mass pairs
and three data rates (markers), with $f_{\max}=7f_s/16$. RIFT's own refinement
and GPU kernel reproduce these values. Dashed line: the quartic scaling of
Eq.~(\ref{eq:cubic-transfer}). Vertical line: the default
$N_Q f_s=8192\,\mathrm{Hz}$.\llmfig}
\end{figure}

Figure~\ref{fig:likelihood_q_rate_scaling} confirms the quartic scaling
from $1024$ to $4096\,\mathrm{Hz}$. The data rates agree at fixed $N_Q f_s$ when the signal has little
power above the lowest band edge; otherwise a lower rate gives a smaller error.

To match a bilby analysis performed at a reduced rate, choose $N_Q$ from the
product $N_Q f_s$, not from $f_s$. If a configuration with $N_Q f_s=R_0$ has
scatter $\sigma_0$, a target $\epsilon$ requires
\begin{equation}
 N_Q f_s \gtrsim R_0\left(\sigma_0/\epsilon\right)^{1/4} .
\end{equation}
Because the error grows as $\rho^2$, the required product grows as $\rho^{1/2}$.
Our default is $N_Q f_s=8192\,\mathrm{Hz}$: $N_Q=2$ for $4096\,\mathrm{Hz}$ data and $N_Q=8$
for bilby analyses at $1024\,\mathrm{Hz}$.

Three conditions accompany the rule. Both codes must analyze the same samples
and band: refinement cannot restore content above the data's Nyquist frequency.
The waveform-generation rate must be recorded, since lowering $f_s$ or $f_{\max}$
can change it. The rule controls only RIFT's fixed-time error; bilby's time
marginalization at low $f_s$ has a separate discretization error.

\subsection{Matching SEOB waveform conditioning}

Figure~\ref{fig:likelihood_matched_taper} separates waveform-conditioning
differences from errors in the time representation. Matching the SEOB
conditioning removes the waveform-dependent offset in the cross-code comparison;
resolving the detector correlations then controls the remaining interpolation
error. More accurate interpolation alone cannot correct a difference in the
waveforms being compared.

The optional native prescription extends the waveform below the requested
start frequency and applies the SEOB start taper, high-pass filter and end
conditioning to a common mode grid before projection. This matches bilby's
native SEOB polarizations while preserving the physical waveform epoch.
It is selected with \texttt{--gwsignal-mode-conditioning native};
\texttt{legacy} retains the normal mode taper.
Waveform-generation rates must also be matched when the analysis cutoff
selects lower-rate frequency-domain generation.

The two references distinguish these effects. Comparison with a converged
Fourier reconstruction of bilby's waveform tests cross-code agreement, whereas
comparison with RIFT's own waveform isolates the time treatment. Matching
conditioning improves the former without eliminating the latter's interpolation
error. A cross-code likelihood comparison therefore requires both a common
conditioned waveform and a converged time representation.

\subsection{Windowing, frequency masks and PSDs}

Beyond time representation, four conventions differ between the codes and must
be matched before a cross-code residual is interpreted.

\emph{Waveform support.} The largest waveform effect in these checks is the
time-domain start. \texttt{GenerateTDWaveform} with \texttt{condition=1} begins
earlier than the native mode output, so a direct-polarization and a mode-sum
likelihood integrate different signals. With matched support
(\texttt{condition=0}) and no taper, the two agree to roundoff in
$\langle h|h\rangle$, $\langle h|d\rangle$ and $\ln\Lambda$; with production
tapering and unmatched support they differ by order one nat. This is the
within-RIFT counterpart of the cross-code conditioning offset discussed above.

\emph{Data window.} Equal parameter names do not imply equal tapers. A RIFT
Tukey window with \texttt{window\_shape}~$\beta$ on a segment of duration $T$
matches a bilby \texttt{roll\_off} of $\beta T/2$. Any mean-square window
correction applied to the PSD in one code must be applied, or removed, in the
other.

\emph{Frequency mask.} When $f_{\max}$ lies on a Fourier bin, bilby includes
that bin and RIFT's inner product excludes it. With noise the bin adds
$4\,\Delta f\,\mathrm{Re}[\tilde d^*\tilde h]/S_n$ to $\ln\Lambda$ (with
$\tilde h$ the detector-projected strain and $S_n$ the one-sided PSD); for
zero-noise data this is
\begin{equation}
 \delta\ln\Lambda = 2\,\Delta f\,\frac{|\tilde h(f_{\max})|^2}{S_n(f_{\max})} .
\end{equation}
For the binary black hole tested here, which has little power at $f_{\max}$,
the term is a few parts in $10^{15}$ of $\ln\Lambda$.

\emph{PSD abscissae.} Bilby interpolates a tabulated PSD at the frequencies
given in the file. ASCII rounding above $1\,\mathrm{kHz}$ can produce duplicate
abscissae; remove them on import.

\section{Detector-network sky coordinates: an inverse-mapping correction}
\label{ap:review:skyframe}
Our  GW250114 investigation identified a small implementation
error in the optional detector-network sky coordinates used by batch adaptive-volume
integration. If $R$ maps an equatorial unit sky vector to the network frame,
$\boldsymbol n_{\rm net}=R\boldsymbol n_{\rm eq}$, the likelihood and serialized
sky samples must use $\boldsymbol n_{\rm eq}=R^{-1}\boldsymbol n_{\rm net}$.
The driver previously reused $R$ in that second conversion. The correction uses
the inverse frame for both likelihood evaluation and CPU/XML export, with explicit
names for the two directions.

\section{Calibration marginalization error budget}
\label{ap:calmarg}

A separate post-hoc GW250114 diagnostic illustrates why reweighting needs
its own coverage and weight checks. The completed nearest-interpolation
calculation produced 300,000 conditional draws, but the calibration
workflow reweighted only its first 60,000 rows. Of those weights, 107 are
nonfinite; the finite weights have effective sample size $\pqCalEss$, and
rejection sampling leaves four output rows. This partial, severely
degenerate calibrated output is excluded from the posterior figures.
Successful workflow completion does not establish a usable calibrated
posterior. This diagnostic concerns post-hoc reweighting, not the in-loop
Monte-Carlo error budget below.

\mysub{Monte-Carlo error budget for in-loop marginalization}
When calibration is marginalized \emph{inside} the extrinsic integral, the marginalized evidence is a
Monte-Carlo average $\hat Z = N_{\rm cal}^{-1}\sum_c w_c Z_c$ over a single, fixed set of $N_{\rm cal}$
calibration draws, with $Z_c$ the extrinsic-marginalized likelihood under draw $c$ (and $w_c$ optional
importance weights).  Because the draw set is held fixed for the whole evaluation, the extrinsic
integrator's internal variance estimate is structurally blind to the spread over $c$: the calibration
contribution to the error of $\ln\hat Z$ must be budgeted separately.  We estimate it from the
normalized per-draw contributions $a_c = w_c Z_c / (N_{\rm cal}\hat Z)$ (computed on a small
extrinsic-prior batch, since the $a_c$ are nearly extrinsic-independent), via the delta-method relation
${\rm Var}(\ln\hat Z) \simeq N_{\rm cal}\,{\rm Var}_c(a_c)$, and report it in quadrature with the
extrinsic sampling error, together with the effective draw count $n_{\rm eff,cal} = 1/\sum_c a_c^2$.
Points with $n_{\rm eff,cal} \lesssim 10$ are flagged: there the marginalization is dominated by a
handful of draws and the quoted error is only a lower bound.  Omitting this term is not benign: in
early testing with deliberately broad envelopes and $N_{\rm cal}=20$, the marginalized $\ln L$ surface
for a two-parameter zero-spin source carried point-to-point scatter of order unity while the
(extrinsic-only) error column claimed $\sigma \simeq 0.2$.

\begin{figure*}
  \includegraphics[width=\textwidth]{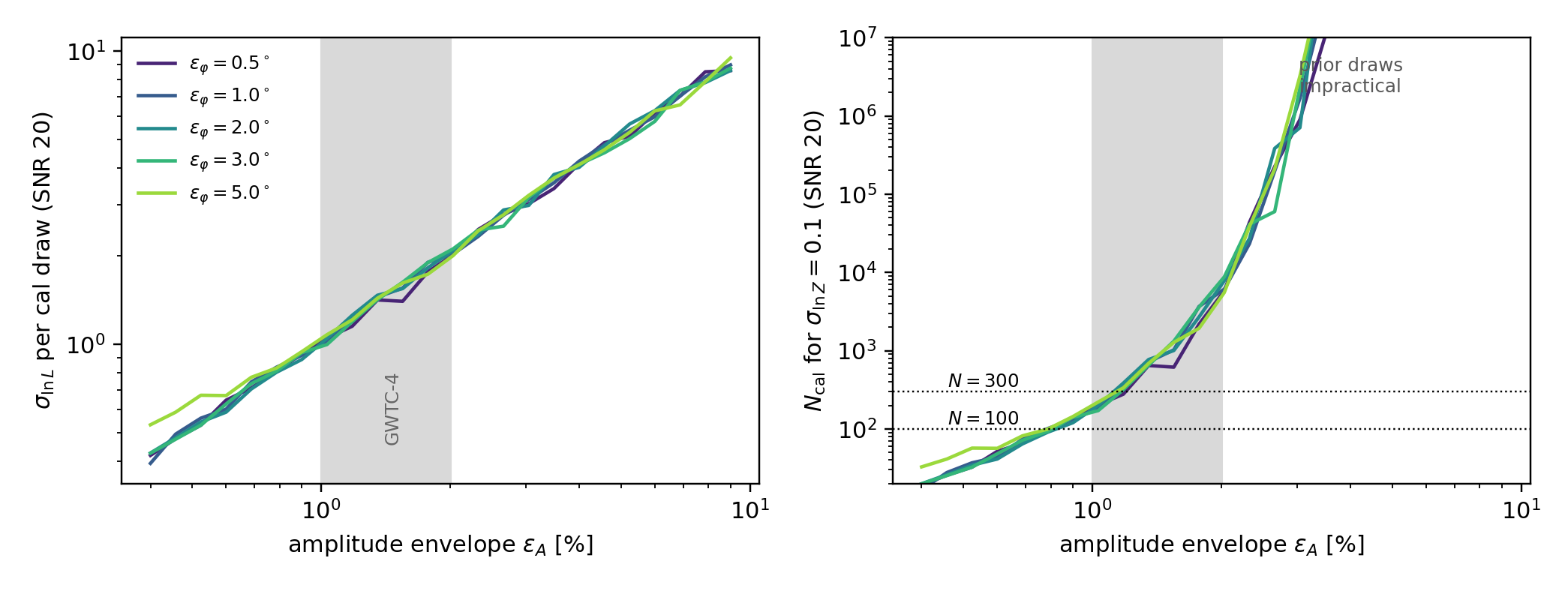}
  \caption{\label{fig:calmarg:scaling}
    Cost scaling of brute-force in-loop calibration marginalization, from a spline-envelope toy model
    of a network of $\cmScalingNdet$ detectors at network SNR $\cmScalingSnr$ (even SNR split, independent per-detector draws,
    per-detector absorption of the amplitude/phase/time degeneracies; see
    \texttt{demos/calmarg/cal\_envelope\_scaling.py}).
    \emph{Left:} per-draw log-likelihood spread $\sigma_{\ln L}$ versus the amplitude envelope width
    $\epsilon_A$, for several phase envelope widths $\epsilon_\varphi$: the spread is linear in
    $\epsilon_A$ ($\sigma_{\ln L} \simeq \cmSigmaPerPercent$ per $\cmNcalRefAmp\%$ at SNR
    $\cmScalingSnr$, scaling as $\rho^2$) and nearly
    independent of $\epsilon_\varphi \le \cmPhaseIndepMax^\circ$.
    \emph{Right:} draws required for $\sigma_{\ln Z} = \cmSigmaTarget$, Eq.~(\ref{eq:calmarg:ncal}).  The shaded
    band marks the GWTC-4-scale calibration error (${\lesssim}\cmCalBoundAmp\%$
    amplitude~\cite{2025arXiv250818079O4aOpenData}), where
    $N_{\rm cal} \sim \cmNcalBandLow$--$\cmNcalBandHigh$ suffices;
    beyond ${\sim}\cmCeilingHi\%$ the exponential takes over and prior draws become impractical.\llmfig}
\end{figure*}

\mysub{Envelope widths the method can handle}
The cost of brute-force (prior-draw) in-loop marginalization is controlled by the per-draw spread
$\sigma_{\ln L} = {\rm std}_c[\ln L(c)]$.  For an approximately log-normal draw set,
${\rm Var}(\ln\hat Z) \simeq (e^{\sigma_{\ln L}^2}-1)/N_{\rm cal}$, so reaching a target accuracy
$\sigma_{\rm t}$ on $\ln Z$ requires
\begin{equation}
N_{\rm cal} \;\simeq\; \frac{e^{\sigma_{\ln L}^2}-1}{\sigma_{\rm t}^2}\,,
\label{eq:calmarg:ncal}
\end{equation}
i.e.\ \emph{exponentially} sensitive to the envelope width.  The spread $\sigma_{\ln L}$ is dominated
by the band-averaged \emph{amplitude} envelope: a constant phase offset is absorbed by the reference
phase, a linear-in-frequency phase by the coalescence time, and per-detector amplitude and timing
offsets largely by the distance, orientation, and sky-position freedoms, while the residual
band-averaged amplitude shift moves $\ln L$ at first order, so $\sigma_{\ln L} \propto \rho^2 \epsilon_A$.
We quantify this with a spline-envelope toy model of a network of $\cmScalingNdet$ detectors that
mirrors our realization generator and applies the degeneracy absorption per detector
(Figure~\ref{fig:calmarg:scaling}): at network SNR $\rho = 20$ we find
$\sigma_{\ln L} \simeq \cmSigmaPerPercent\,(\epsilon_A/\cmNcalRefAmp\%)$, nearly independent of the
phase envelope up to $\cmPhaseIndepMax^\circ$.  Through Eq.~(\ref{eq:calmarg:ncal}), envelopes at the
GWTC-4 scale---LIGO O4a calibration error is bounded by ${\lesssim}\cmCalBoundAmp\%$ in amplitude and
${\lesssim}\cmCalBoundPhase^\circ$ in phase below
$\cmCalBoundFmax$\,kHz~\cite{2025arXiv250818079O4aOpenData}---require
$N_{\rm cal} \sim \cmNcalBandLow$--$\cmNcalBandHigh$ for $\sigma_{\rm t} = \cmSigmaTarget$ at SNR
$\cmScalingSnr$: $N_{\rm cal} \simeq \cmNcalRef$ at $\cmNcalRefAmp\%/\cmNcalRefPhase^\circ$, rising
steeply to $\simeq\cmNcalGwtcFour$ at the $\cmCalBoundAmp\%$ edge of that same bound.  The ILE default
is $N_{\rm cal} = \cmNcalDefault$, which we consider too low; we recommend
$N_{\rm cal} = \cmNcalRecommended$, which in production has proved comfortable well beyond the
$\cmNcalRefAmp\%$ end.
That is better than Eq.~(\ref{eq:calmarg:ncal}) would suggest, and the reason is that the equation is
a \emph{lognormal worst case}: it assumes the calibration weights are as heavy-tailed as
$\sigma_{\ln L}$ alone implies, whereas the realized weights are considerably more stable.  The
equation should therefore be read as a conservative ceiling on the requirement rather than a
prediction of it.  Taken at face value it calls for $\simeq\cmNcalGwtcFour$ draws at the
$\cmCalBoundAmp\%$ edge --- more than the adaptive escalation could reach, since that caps at
$\cmNcalMaxFactor\times$ its starting value --- but we have not seen that bound bind in practice.  Per-evaluation cost is linear and GPU
memory mildly linear in $N_{\rm cal}$ (the realization blocks add ${\sim}\cmMemPerDraw$\,MB per draw,
$\cmMemAtRecommended$\,MB at $N_{\rm cal} = \cmNcalRecommended$, in our reference configuration).  In practice the draw count is chosen \emph{adaptively}: the same responsibility
probe that feeds the error budget measures the effective draw count at each intrinsic point, and the
draw set is doubled (with an incremental precompute of only the new blocks) until a target
$n_{\rm eff,cal}$ is met or a cap is reached.  Beyond $\epsilon_A \sim \cmCeilingLo$--$\cmCeilingHi\%$ at this SNR---or
equivalently at proportionally higher SNR, since $\sigma_{\ln L} \propto \rho^2$---the required
$N_{\rm cal}$ grows astronomically and prior draws are impractical; an adaptive calibration proposal
(importance sampling from a proposal learned over the spline nodes), which would remove the bulk of
$\sigma_{\ln L}$ before the exponential in Eq.~(\ref{eq:calmarg:ncal}) is taken, is under development
for that regime but requires validation against the brute-force path before production use.

\mysub{The template self-term and the global-norm approximation}
\label{ap:calmarg:selfterm}
The likelihood in Section~\ref{sec:calmarg} uses the template factor $\alpha_k=\eta_k^{-1}$ at fixed released-strain PSD.
The complete construction weights the template--template blocks by $|\alpha_k(\bm c_c)|^2$ [Eq.~(\ref{eq:cal:selfterm})].
The global-norm approximation instead retains $\qmstateproduct{h}{h}_k$, with $U,V$ computed once for the draw set.
The omitted term is $\qmstateproduct{h}{(|\alpha_k|^2-1)h}_k$.
A uniform rescaling $\alpha_k\to\lambda\alpha_k$ is degenerate with luminosity distance in the signal model.
With unrestricted source amplitude, maximizing over that amplitude leaves the complete profile likelihood invariant.
The cross-term scales as $\lambda$ and the template norm as $\lambda^2$.
With the global norm, the maximized log likelihood ratio instead grows as $\lambda^2$.
Distance marginalization also depends on the distance prior, its bounds, and the change-of-variable Jacobian; it need not be invariant.

The bias is an amplitude effect.  Summarizing realization $c$ by its band-averaged amplitude
$a_c=1+\mu_c$ (the phase cancels in $|\alpha_c|^2$), the amplitude-profiled per-realization log-likelihood is
$\simeq L_0 a_c^2$ under the global norm but $\simeq L_0$ (the invariant) under the complete norm, with
$L_0\simeq\tfrac12\qmstateproduct{h}{h}\simeq\tfrac12\rho^2$.  In this profiled model, the global-norm calibration integral
up-weights realizations as $\exp(2L_0\mu_c)$; removing that reward gives
\begin{equation}
\ln Z_{\rm complete}-\ln Z_{\rm global}\;\simeq\;-2L_0^2\,\sigma_\mu^2\;=\;-\tfrac12\rho^4\,\sigma_\mu^2\,,
\label{eq:calmarg:selfterm:bias}
\end{equation}
where $\sigma_\mu$ is the signal-power-weighted, band-averaged amplitude spread --- the same quantity
that sets $\sigma_{\ln L}\propto\rho^2\epsilon_A$ above (Fig.~\ref{fig:calmarg:scaling}).  The bias is
quartic in SNR and quadratic in amplitude width, signed so that the global norm inflates the evidence, and
independent of the phase envelope.

We verify Eq.~(\ref{eq:calmarg:selfterm:bias}) two ways.  A semi-analytic testbed built on the exact
realization draws and the factored likelihood reproduces both evidences over a two-dimensional sweep in
amplitude width ($\cmSelfTermWidthLo$--$\cmSelfTermWidthHi\%$) and network SNR
($\cmSelfTermSnrLo$--$\cmSelfTermSnrHi$).  Because the bias scales as $\rho^4\sigma_\mu^2$ it varies by
more than three orders of magnitude across that box, so a single interval does not characterize it: the
global norm is biased by $+\cmSelfTermBiasLo$~nats at the quiet, narrow corner
($\cmSelfTermWidthLo\%$, SNR~$\cmSelfTermSnrLo$), by
$+\cmSelfTermBiasMidLo$--$\cmSelfTermBiasMidHi$~nats across all widths at SNR~$\cmSelfTermBiasMidSnr$,
and by $+\cmSelfTermBiasHi$~nats at the loud, wide corner ($\cmSelfTermWidthHi\%$,
SNR~$\cmSelfTermSnrHi$).  The complete norm tracks a template-side (\texttt{bilby}-convention) reference
throughout, to $\le\cmSelfTermCompleteTol$~nat for SNR~$\le\cmSelfTermBiasMidSnr$ at every width; its
residual grows only in the same loud/wide corner, reaching $\cmSelfTermCompleteWorst$~nat there, which
is where the fixed draw set itself becomes the limiting approximation.  End-to-end in the production
pipeline, a single intrinsic point of GW240426\_031451 (network SNR $\approx\cmSelfTermGWSnr$, ${\sim}2\%$
amplitude envelope) yields $\ln Z_{\rm complete}-\ln Z_{\rm global} = \cmSelfTermGWShift\pm
\cmSelfTermGWShiftErr$~nats (six-seed average), consistent with the $\simeq\cmSelfTermGWPred$~nats
predicted by Eq.~(\ref{eq:calmarg:selfterm:bias}) at this SNR.  Because the bias grows as $\rho^4$, it
stays well below a nat at contemporary detector SNRs and envelopes --- so the calibration-posterior
shapes and the evidence alike are essentially unaffected here --- and reaches the tens-of-nats regime
only for loud sources or wide amplitude envelopes.

The two constructions have identical per-\emph{sample} kernel cost --- both index a precomputed norm ---
so the difference lives entirely in the precompute, and there it need not be small.  The global norm
computes $U,V$ once.  The complete norm additionally forms the amplitude basis --- a singular-value
decomposition of the in-band $\{|\alpha_c(f)|^2\}$, of cost $\mathcal{O}(N_{\rm band}N_{\rm cal}^2)$ and
cached across the intrinsic grid --- and, for a spline model of $|\alpha_k|$, rank $\le M(M+1)/2$ basis-weighted template blocks, each an
$\mathcal{O}(N_{\rm band})$ frequency sum (for the default $M=\cmSplineNodes$ nodes,
${\sim}30$ blocks rather than the $N_{\rm cal}=\cmNcalDefault$ a naive per-draw evaluation
would require).  Both terms scale with the number of in-band frequency samples $N_{\rm band}$, which is
large for long, low-mass templates; combined with a large draw set this added precompute is \emph{not}
negligible and can dominate the per-intrinsic-point cost, even though the sample-loop integration that
usually sets the wall-clock is untouched.  The choice is therefore a quantitative tradeoff set by Eq.~(\ref{eq:calmarg:selfterm:bias}): the
global-norm approximation is accurate to a target $\sigma_{\rm t}$ on $\ln Z$ --- and the cheaper of the
two --- as long as $\tfrac12\rho^4\sigma_\mu^2 \lesssim \sigma_{\rm t}$, which holds at low to moderate
SNR unless the amplitude envelope is unusually wide, so it suffices for the bulk of a contemporary
catalog.  The complete self-term becomes necessary as the SNR rises (the bias grows as $\rho^4$) or for
wide amplitude envelopes --- the same loud/wide regime that drives up the draw-count requirement of
Eq.~(\ref{eq:calmarg:ncal}), and precisely where its more expensive precompute is most likely to
dominate.  Because the term depends only on $|\alpha|^2$, phase-envelope width never triggers it.  The
pipeline exposes both routes at runtime (the complete self-term by default, the global-norm
approximation as an opt-in that skips the amplitude-basis precompute), so this tradeoff can be made
per analysis.

\section{Exploiting redundancy for validation}
\label{ap:checks}
\noindent \emph{Calibration marginalization or not}: The current implementation marginalizes calibration inside
the extrinsic likelihood integral (Section~\ref{sec:calmarg}), not by reweighting a completed posterior.
A diagnostic comparison therefore runs matched analyses with and without in-loop calibration
marginalization and compares their fairly drawn source-parameter samples. For a narrow envelope in a
well-calibrated event the posteriors can agree closely, as in the GW240426\_031451 demonstration of
Subsection~\ref{sec:results:calmarg}; loud signals or broad envelopes need not do so. This paired
comparison checks the impact of calibration uncertainty and the stability of the sampling procedure,
not an assumed universal agreement between the two posteriors.

\noindent \emph{Before and after extrinsic resampling}: RIFT generates its final extrinsic samples from a final set of
intrinsic-only posterior samples: each such sample is passed to a fast run of ILE to produce one (or sometimes several) fair extrinsic
samples. Therefore, the intrinsic-only and extrinsic samples should agree.  This consistency test can diagnose whether
ILE sampling has been ineffective for some intrinsic configurations, which could for example  involve a particularly exceptional extrinsic configuration (e.g., an edge-on strongly
precessing binary, whose parameters are all tightly constrained).  Prior to adopting the AV integrator, some
exceptional combinations of intrinsic and extrinsic configurations were difficult to sample.

\noindent \emph{Before and after convergence testing}: As normally operated, RIFT usually iterates to convergence with
relatively low-resolution sampling, merely sufficient for exploration and convergence testing.  Once converged, the full
posterior generation code is run with many CIP worker instances, to generate a full extrinsic posterior.  The
low-resolution and high-resolution sampling should agree.  This consistency test allows us to diagnose whether RIFT's settings are
well-adapted to interpret a specific observation.

\noindent \emph{Overall and per-iteration marginal likelihoods: agreement and error}: The most common RIFT consistency test is checking whether posterior
credible intervals encose all regions of largest marginal likelihood, or otherwise providing a compelling reason why that should
not occur (e.g., strong priors).   In the best-converged analyses, the marginal likelihood colormap smoothly transitions in
scale over a region surrounding the posterior.

\section{Automated waveform symmetry checks}
\label{ap:waveform_checks}

RIFT consumes waveform models maintained upstream, and a model revision can
change the templates without any change to our code. Several released
precessing models violate the parity symmetry of general relativity;
Ref.~\cite{gwastro-pe-EstellesParitySymmetryDiagnostic2026} constructs
waveform-level parity tests, traces the violations to their origins and
corrects several models. We add two tests built on RIFT's likelihood: exact
identities on the cross-term matrices, and the marginal-likelihood difference
between mirror-image sources. They test upstream models rather than our code, so
we run them when we adopt a model release or a new waveform interface, not in
continuous integration.

Reflection through the orbital plane maps both spins as
$(s_x,s_y,s_z)\rightarrow(-s_x,-s_y,s_z)$ and leaves the masses and orbital
phase unchanged. Writing this map as $\lambda\rightarrow R\lambda$, general
relativity requires
\begin{equation}
h_{\ell m}(R\lambda;t) = (-1)^\ell \, h^*_{\ell,-m}(\lambda;t)\,.
\label{eq:ap:parity}
\end{equation}
The factored likelihood exposes the modes and the cross-term matrices $U,V$ of
Eq.~(\ref{eq:QUV}), on which Eq.~(\ref{eq:ap:parity}) implies exact identities.

For one detector, reflecting the
source is compensated by reflecting the viewing geometry,
$(\iota,\phi_{\rm ref},\psi)\rightarrow(\pi-\iota,\phi_{\rm ref},-\psi-\gamma)$,
with $\gamma=\arg(F_+ - iF_\times)|_{\psi=0}$. This map preserves the uniform
prior on $(\cos\iota,\phi_{\rm ref},\psi)$, so a parity-respecting model gives
equal marginal likelihoods at a mirror pair on any data:
\begin{equation}
\Delta \equiv \ln{\cal L}_{\rm marg}(\lambda\,|\,d) - \ln{\cal L}_{\rm marg}(R\lambda\,|\,d) = 0\,.
\label{eq:ap:parity_lnL}
\end{equation}
A network has no common $\psi$ shift, so we use one detector. Mapping the
extrinsic samples through the reflection makes $\Delta$ vanish exactly for a
parity-respecting model at any sample size. A nonzero $\Delta$ biases the
posterior odds between mirror-image configurations by $e^{\Delta}$.

\begin{figure}
  \includegraphics[width=\columnwidth]{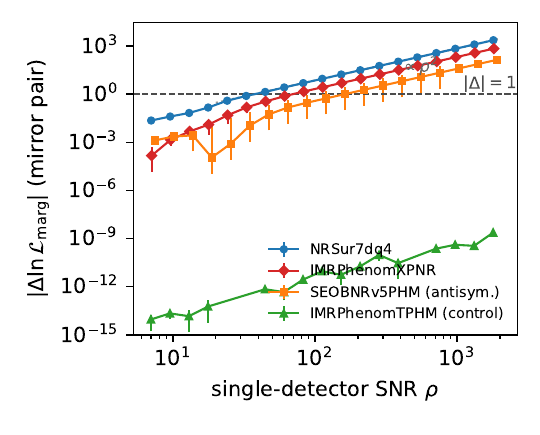}
  \caption{\label{fig:parity_lnL}Marginal-likelihood parity defect $|\Delta|$
    [Eq.~(\ref{eq:ap:parity_lnL})] versus single-detector SNR, for an edge-on,
    near-superkick binary, from one zero-noise dataset rescaled in SNR, for
    NRSur7dq4, IMRPhenomXPNR, SEOBNRv5PHM with equatorial-asymmetric modes, and
    the IMRPhenomTPHM control (IMRPhenomXPHM, not shown, tracks IMRPhenomXPNR).
    A parity-respecting model gives $\Delta=0$; the control does so to round-off.
    Error bars are bootstrap estimates over extrinsic samples; at the highest
    SNRs a single sample dominates each estimate.%
    \llmfig}
\end{figure}

Figure~\ref{fig:parity_lnL} shows $|\Delta|$ against SNR for an unfavorable,
edge-on near-superkick source. The parity-clean control stays at round-off, and
the resolved defects grow as $\rho^2$, as a fixed fractional model error should.
For NRSur7dq4, $|\Delta|$ reaches unity at about half the SNR of
IMRPhenomXPNR, below the single-detector SNR of the loudest current events for
this configuration. The SEOBNRv5PHM defect is not resolved at our sample size.
The IMRPhenomXPNR curve also contains an artifact of RIFT's frequency-domain
mode conditioning, which affects every model generated through that path. The
checks are packaged in \texttt{demos/waveform\_symmetry/}.

\section{Additional transverse-spin coordinates}
\label{ap:rf:additional:coordinates}
These alternatives distinguish physical precession coordinates from empirical
features for RF likelihood fitting. In geometrized units $G=c=1$, let
$M=m_1+m_2$, $q=m_2/m_1\leq1$, $\eta=m_1m_2/M^2$ and
$v=(\pi M f_{\rm ref})^{1/3}$. We use the standard angular momenta
\begin{align}
 \boldsymbol S_i&=m_i^2\boldsymbol\chi_i,&
 L_N&=\frac{\eta M^2}{v},\\
 \boldsymbol S&=\boldsymbol S_1+\boldsymbol S_2,&
 \boldsymbol J&=\boldsymbol L_N+\boldsymbol S.
\end{align}
\noindent\emph{Single-spin coordinates.}
Single-spin precession motivates cone and phase coordinates.
All projections below refer to the instantaneous orbital direction
$\hat{\boldsymbol L}_N$ at the configured reference frequency. In particular,
$J_\parallel=L_N+S_\parallel$ and
$\boldsymbol J_\perp=\boldsymbol S_\perp$. The geometric cone angle obeys
\begin{equation}
 \cos\beta=\frac{J_\parallel}{J},\qquad
 \sin\beta=\frac{|\boldsymbol S_\perp|}{J}.
 \label{eq:rf:precession:cone}
\end{equation}
It is undefined at $J=0$, where the simple-precession description also
breaks down. In the single-significant-spin simple-precession limit,
$\boldsymbol L_N$ precesses about an approximately fixed $\boldsymbol J$;
for a significant primary spin its leading-order phase evolution is
\begin{equation}
 \dot\alpha=\left(2+\frac{3q}{2}\right)\frac{J}{r^3},\qquad
 r=\frac{M}{v^2},\qquad
 \dot\gamma=-\cos\beta\,\dot\alpha.
 \label{eq:rf:precession:phase}
\end{equation}
Here $\alpha$ is the accumulated precession phase about $\boldsymbol J$
and $\gamma$ is the minimal-rotation phase. These are the conventional
angles entering the precessing waveform and its Fisher derivatives
\cite{gwastro-SpinTaylorF2-2013,gwastro-mergers-Prakash-PrecessingFisherMatrix-2014}.
Equation~\eqref{eq:rf:precession:phase} is not a general two-spin evolution
law: nutation and motion of $\boldsymbol J$ require the fuller dynamics.

For two spins the leading spin-orbit torque involves the effective spin
\begin{align}
 \boldsymbol S_{\rm eff}
 &=\left(1+\frac{3q}{4}\right)\boldsymbol S_1
  +\left(1+\frac{3}{4q}\right)\boldsymbol S_2,\\
 \dot{\boldsymbol L}_N\big|_{\rm SO}
 &=\frac{2}{r^3}\boldsymbol S_{\rm eff}\mathbin{\times}\boldsymbol L_N.
 \label{eq:rf:precession:torque}
\end{align}
This convention agrees with the leading spin-orbit part of the usual
precession equations \cite{2008PhRvD..78d4021R}. Thus the total transverse
spin determines the instantaneous cone geometry, while a different,
mass-weighted combination determines the torque. Their interference can
contain information absent from either individual transverse-spin magnitude.

Three scalar fitting features motivated by this geometry are:
\begin{align}
 C&=\frac{|\boldsymbol S_\perp|^2}
                  {J_\parallel^2+(\epsilon L_N)^2},\\
 Q&=\frac{J-J_\parallel}{L_N},\\
 A&=\frac{|2\boldsymbol S_{{\rm eff},\perp}/M^2|^2}
                         {(\eta v^2)^2}.
 \label{eq:rf:precession:features}
\end{align}
The cone feature $C$ is related to the standard angle by
\begin{equation}
 C=\frac{1-\cos^2\beta}
          {\cos^2\beta+\epsilon^2(L_N/J)^2},\qquad J\ne0.
\end{equation}
Away from the regularized region, $C\simeq\tan^2\beta$, or
$|\cos\beta|\simeq(1+C)^{-1/2}$. Thus it expresses a familiar cone property
rather than a new physical coordinate. It does not retain the sign of
$\cos\beta$; the native spin components retained by the augmented fit still
do. Direct use of the bounded quantity $\cos\beta$ is a natural alternative.
Neither parameterization is selected by the geometry alone.

The prototype choice $\epsilon=0.1$ is \emph{ad hoc}: it avoids a divergent
feature near $J_\parallel=0$, but is not derived from precession dynamics,
a Fisher metric or the detector noise spectrum. Its value has not been
established as optimal. The second is an instantaneous geometric function,
not an accumulated phase. In the single-primary-spin limit, with leading
quadrupole radiation reaction,
$d(\alpha+\gamma)/d\ln v=[5(2+3q/2)/(32v^3)]Q$.
At the same leading order, the third is proportional to the squared rate
of change of the orbital direction,
$|d\hat{\boldsymbol L}_N/d\ln v|^2=(25/1024)A$.
These relations motivate the feature shapes; they do not derive an optimal
RF representation or an exact two-spin Fisher metric. The approximation
and the empirical choice must be assessed on evaluated likelihood data.

Used alone, these scalars compress the spin information: they do not
specify both transverse spin vectors. Such dimensional reduction could itself
help interpolation, but would discard information relevant to two-spin
inference. A complete representation must supplement the scalars with
angular and, where necessary, signed residual coordinates. The relative
azimuth $\phi_{12}=\phi_2-\phi_1$ is a useful candidate, but is not by itself
sufficient: it also leaves the common transverse azimuth unspecified, and
$C$ and $Q$ are redundant functions of the total transverse magnitude at
fixed masses and aligned spins. Counting four entries therefore does not
establish four independent coordinates.

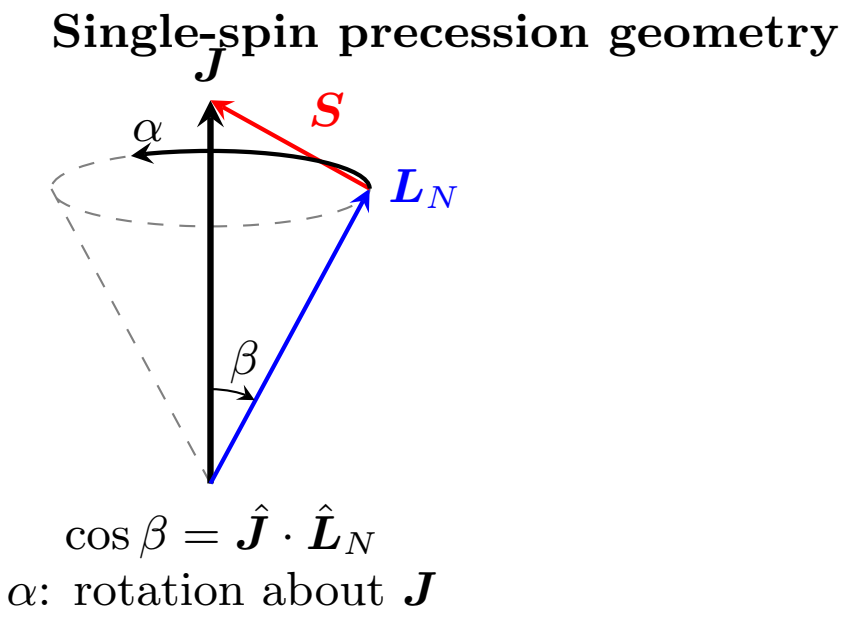
\begin{figure}[t]
\centering
\begin{tikzpicture}[>=stealth,scale=0.78]
\begin{scope}[xshift=0cm]
\node[anchor=west] at (-1.5,3.8) {\textbf{Single-spin precession geometry}};
\draw[dashed,gray] (0,0)--(-1.35,2.5);
\draw[dashed,gray] (0,2.5) ellipse (1.35 and 0.32);
\draw[->,very thick] (0,0)--(0,3.25) node[above] {$\boldsymbol J$};
\draw[->,thick,blue] (0,0)--(1.35,2.5) node[right] {$\boldsymbol L_N$};
\draw[->,thick,red] (1.35,2.5)--(0,3.25) node[midway,above right] {$\boldsymbol S$};
\draw[->] (0,0.8) arc[start angle=90,end angle=62,radius=0.8];
\node at (0.29,1.03) {$\beta$};
\draw[->,thick] (1.35,2.5) arc[start angle=0,end angle=120,x radius=1.35,y radius=0.32];
\node at (-0.53,2.99) {$\alpha$};
\node[align=center] at (0.1,-0.6) {$\cos\beta=\hat{\boldsymbol J}\cdot\hat{\boldsymbol L}_N$\\$\alpha$: rotation about $\boldsymbol J$};
\end{scope}
\end{tikzpicture}
\caption{Simple-precession geometry. The orbital angular momentum sweeps
an approximately fixed cone about $\boldsymbol J$; $\beta$ is its opening
angle and $\alpha$ the precession phase. The double-spin coordinates in
Sec.~\ref{sec:rf:transverse:methods} instead use the instantaneous $L$-frame
transverse plane. Arrow lengths are schematic.\llmfig}
\end{figure}

\noindent\emph{A phase-excess radial coordinate.}
The double-spin chart of Eq.~\eqref{eq:rf:geometric4} can replace $R$ by
\begin{equation}
 H=\frac{J-|J_\parallel|}{L_N}.
\end{equation}
The azimuth and signed residuals are unchanged. At fixed masses, aligned
spins and reference frequency, the inverse radial map is
$R=(L_N/M^2)\sqrt{H[H+2|J_\parallel|/L_N]}$.
This is an aligned-spin-dependent radial rescaling, distinct from the raw
spin-sum radius and from $Q$ when $J_\parallel<0$. It preserves the four
transverse degrees of freedom with the same axis convention.

\paragraph{Separating fitting and sampling coordinates.}
Fitting coordinates and sampling coordinates have distinct roles. The
physical spin prior can be sampled using spin magnitudes, tilt cosines and
azimuths while the likelihood surrogate uses Cartesian transverse components
or derived precession features. Changing the surrogate's inputs can change
its approximation even when the physical prior and evaluated likelihoods
are identical. Conversely, a change of sampling coordinates requires the
corresponding prior correction and the same physical spin bound. A coordinate
comparison must separate these two interventions.

\section{Radial-phase coordinates for strong eccentric encounters}
\label{ap:eccentric:high_e}

The phase transport in Sec.~\ref{sec:eccentric:coordinates} admits a
semianalytic extension to relativistic bound orbits. We formulate it with
a planar effective-one-body (EOB) Hamiltonian and consistent radiation
reaction \cite{2017PhRvD..96j4048H,Bini:2012eccRR}. The published
SEOBNRv6EHM model \cite{Gamboa:2026SEOBNRv6EHM} motivates the intended
physical scope; our prototype independently implements the public
3PN nonspinning EOB Hamiltonian \cite{gw-mergers-approx-pn-nonspinning-NRCompare-AB}
with the public leading spin--orbit and black-hole spin--spin terms of
Ref.~\cite{2021arXiv210411705K}, including their momentum dependence.
It uses no non-public model code. Let
$E,J$ denote the real energy and orbital angular momentum, $T_r$ the
conservative radial period, and $D_E,D_J>0$ the losses integrated over one
radial cycle. All quantities use the same physical coordinate time.
At leading adiabatic order, the uniform radial angle $q_r$ satisfies
\begin{align}
 \dot q_r&=\frac{2\pi}{T_r}, &
 \dot E&=-\frac{D_E}{T_r}, &
 \dot J&=-\frac{D_J}{T_r}.
 \label{eq:ecc:cycle:balance}
\end{align}
Eliminating time gives a phase coordinate at a later slow-state landmark,
\begin{align}
 \theta_a&=\operatorname{wrap}_{2\pi}\!\left[
 q_{r0}+2\pi\int_{E_a}^{E_0}\frac{dE}{D_E(E,J(E))}\right],
 \label{eq:ecc:cycle:phase}\\
 \frac{dJ}{dE}&=\frac{D_J}{D_E}.
 \label{eq:ecc:cycle:path}
\end{align}
Turning-point quadratures determine the cycle losses and convert the
native anomaly to $q_{r0}$ through the elapsed radial flight time.
The Newtonian quadrupole limit recovers Eq.~\ref{eq:ecc:advance} at
identical mean-frequency endpoints. An osculating pericenter-frequency
surface can instead select an in-band strong encounter, but changes the
endpoint convention. An actual pericenter event fixes $q_r=0$ and cannot
serve as an informative radial-phase landmark.

We pair the aligned-spin Hamiltonian with the relative second-PN-order
radiation-reaction force of Ref.~\cite{2021arXiv210411705K}, including its
local leading-tail approximation and leading spin couplings. The radial
and azimuthal forces, spin Schott energy and angular momentum use one gauge.
Canonical-to-tortoise conversion is applied to the radial force, and the
same mechanical power determines cycle losses and direct evolution.
Radiative balance is tested through the retained spin orders. This
uncalibrated, PN-truncated construction does not reproduce SEOBNRv6EHM.

The Supplementary Material repeats the larger-eccentricity Danilo
comparison with each archived draw's two aligned spins. Full-e Peters
transport remains the control; a zero-spin calculation with the same
radiation reaction separates spin effects from the change in force gauge.
Aligned spin removes most of the residual ridge tilt left by the zero-spin
control, with smaller posterior dependence than Peters at every smoothing
examined. This geometric diagnostic does not establish phase accuracy:
direct evolution at the same mechanical-energy section differs from the
averaged advance by $\eccSpinDirectOffsetA$ and $\eccSpinDirectOffsetB$ radians
at the tested starting phases.
For routine use, full-e Peters is the default when the coordinate feature
is enabled; the public EOB coordinate remains optional. Its local chart
is generated at workflow build time from the coincidence and configured
priors, with a smooth Peters exterior. Frozen coefficients travel in worker
arguments, requiring no separate map-file transfer. Native sampling and
priors are preserved. Geometric eccentricity
and Kepler-mean initialization remain proxies for the backend state, and
this example does not probe near-separatrix matching.

Near an unstable circular orbit, write the real Hamiltonian as
\begin{equation}
 H=H_u(J)+\frac{a p_r^2}{2}-\frac{b(r-r_u)^2}{2}+\cdots,
 \qquad a,b>0.
 \label{eq:ecc:saddle}
\end{equation}
Here $p_r$ is canonical radial momentum and $r_u$ the unstable radius.
For detuning $\Delta=E-H_u(J)$ and instability rate
$\lambda_u=\sqrt{ab}$, the local flight time contains
$-\lambda_u^{-1}\log(|\Delta|/\Delta_{\rm match})$.
The energy scale $\Delta_{\rm match}$ and regular part follow by matching
to the outer radial integral. The prolonged whirl contributes to the cycle losses;
its azimuthal phase does not add radial cycles to
Eq.~\ref{eq:ecc:cycle:phase}.
Nearly identical orbit-averaged parameters can leave the final detuning
unresolved within one dissipative step and hence admit different final-whirl
durations \cite{OShaughnessy:2003eccTransition}. We retain the actual
detuning, radial phase, and branch, and require small barrier drift and
small potential changes across the encounter. A flat barrier or loss of
radial return requires a separate transition treatment.

For an isolated encounter, absolute arrival time and azimuth are already
extrinsic parameters. Duration, frequency evolution, relative burst timing,
and plunge attachment can retain intrinsic information. We therefore use
branch-resolved encounter-shape features when a repeating radial phase
ceases to exist. Native sampling and priors are retained: conversion to
$q_r$ generally has a nontrivial Jacobian, unlike the angle-independent
shear in Eq.~\ref{eq:ecc:angular:shear}. This construction specifies a
high-eccentricity extension; its accuracy and effect on likelihood fitting
require validation against the chosen dynamics and initialized states.

\section{Early-time multirate data for long signals}
\label{app:early-multirate}

A binary neutron star observed from a few hertz by a third-generation network spends almost all of its
in-band time at low frequency.
At full rate, the precompute of $Q_{k,\ell m}$, $U$ and $V$ on these long series then sets the
per-point cost and memory of ILE.
RIFT pull request 436 adds an opt-in prototype multirate representation of the data for this case.
It applies the time slicing of LLOID \cite{2012ApJ...748..136C} to the likelihood precompute, and
evaluates $U,V$ with frequency multibanding \cite{2017CQGra..34k5006V,2021PhRvD.104d4062M}.
It uses no reference waveform and no reduced basis.

\emph{Method.}
The data keep the full sample rate for the final $\tau_{\rm tr}$ before the event time.
Earlier data are low-pass filtered (finite impulse response, $\etcAttenDb\,\mathrm{dB}$ stop-band
attenuation) and decimated to an early rate $f_e$.
The early rate exceeds twice the highest frequency that any retained mode reaches before the
transition, over the declared chirp-mass range.
The noise weight is applied to the data once per job, $\bar d = C^{-1}d$.
Each early inner product then needs the raw template only at $f_e$, and no full-length weighted
template is formed per point.
The early and late segments together give $Q_{k,\ell m}(t)$ on the full-rate arrival-time grid, so
time interpolation and quadrature are unchanged.
$U$ and $V$ are evaluated on a multibanded frequency grid with a $\etcUVMarginS\,\mathrm{s}$ margin per
band, reading each bin from the early or late template series already built for $Q$.

\emph{Accuracy.}
The acceptance test uses a binary neutron star in Cosmic Explorer noise from $\etcAccFmin\,\mathrm{Hz}$ in
$\etcAccDurationS\,\mathrm{s}$ of data, with one detector, $f_e=\etcEarlyRate\,\mathrm{Hz}$ and
$\tau_{\rm tr}=\etcAccTailS\,\mathrm{s}$.
It evaluates the injection, a nearby template and both ends of the chirp-mass range.
This test measures the error from segmentation alone.
The full-rate reference uses the same template conventions as the multirate path: each $|m|$ group
starts at $(|m|/2)f_{\rm min}$ and \textsc{IMRPhenomXHM} multibanding is off.
At $\rho=\etcAccRho$, over all extrinsic draws and lags near the likelihood maximum, $\ln{\cal L}$
differs from the full-rate value by at most $\etcDlnLMax$ nats (registered gate $\etcDlnLGate$ nats).
With whitened noise of unit variance, the largest error in any $Q_{k,\ell m}$ sample is $\etcQErr$,
and in any $U$ or $V$ element $\etcUVErr$ (gate $\etcDThreeGate$).
The peak lag sample does not move, and the peak time shifts by at most $\etcPeakShift\,\sigma_t$.
The GPU implementation reproduces the CPU implementation to a relative $\etcDeviceErr$.

\emph{Template conventions: not ready for high-SNR or 3G use.}
The multirate template conventions differ from RIFT's production default.
Multirate starts each $|m|$ group at $(|m|/2)f_{\rm min}$, turns multibanding off, and uses a different
high-pass standoff.
We compared the two on three aligned-spin binary black holes ($\etcConvMassesA$, $\etcConvMassesB$ and
$\etcConvMassesC\,M_\odot$) in zero-noise data with O4 sensitivity, evaluating $\ln{\cal L}$ at fixed extrinsic samples.
Without segmentation, the largest $|\Delta\ln{\cal L}|$ between them is $\etcConvFracMin$ to $\etcConvFracMax$ of $\rho^2/2$.
This offset scales as $\rho^2$: about $\etcConvNatsLoMin$ to $\etcConvNatsLoMax$ nats at $\rho=\etcConvRhoLo$,
and $\etcConvNatsHiMin$ to $\etcConvNatsHiMax$ nats at $\rho=\etcConvRhoHi$.
Most of it comes from the per-$|m|$ start frequency (up to $\etcConvPerMFrac$ of $\rho^2/2$), which drops
the in-band content of modes with $|m|\ge 3$ below each mode's start frequency.
Multibanding contributes up to $\etcConvMbandFrac$ and the standoff up to $\etcConvStandoffFrac$.
With the multirate conventions, production and multirate agree to $\etcConvSameFrac$.
In its present form the method is therefore a prototype and a first step toward 3G inference.
It is not ready for high-SNR or 3G production use.
High-SNR use needs a template convention that keeps the in-band content of every mode, so high- and
low-SNR analyses will need different template conventions.

\emph{Cost.}
Table~\ref{tab:early-multirate} compares production and multirate ILE on a binary neutron star in
Cosmic Explorer noise ($\etcMassOne+\etcMassTwo\,M_\odot$, aligned spins, \textsc{IMRPhenomXHM} with
$\ell\le\etcLmax$) from $\etcFmin\,\mathrm{Hz}$, in $\etcDurationS\,\mathrm{s}$ of $\etcSrate\,\mathrm{Hz}$ data.
The second and third detectors use Einstein Telescope and Cosmic Explorer noise at other sites,
with a static response.
The multirate schedule uses $f_e=\etcEarlyRate\,\mathrm{Hz}$ and $\tau_{\rm tr}=\etcTailS\,\mathrm{s}$.
Each pair ran both paths on one \etcGpuModel{} GPU ($\etcGpuMemGB\,\mathrm{GB}$), with the same seed,
NoLoop likelihood, AV integration and distance marginalization.
The multirate per-point preparation runs on the GPU; waveform generation runs on the CPU for both
paths.
At $\etcMultiSMin$ to $\etcMultiSMax\,\mathrm{s}$ per point, multirate is
$\etcRatioMin$ to $\etcRatioMax$ times faster than production.
The largest remaining stage is CPU waveform generation ($\etcWaveformMinS$ to $\etcWaveformMaxS\,\mathrm{s}$),
followed by the GPU integration.
Generating the early and late template pieces with the GPU port of Appendix~\ref{app:gpu-xphm} removes
the CPU waveform stage.
Steady-state cost then falls to $\etcXpMultiSMin$ to $\etcXpMultiSMax\,\mathrm{s}$ per point,
$\etcXpRatioMin$ to $\etcXpRatioMax$ times below production (lower rows of Table~\ref{tab:early-multirate}).
Peak host memory falls by a factor of $\etcRssRatioMin$ to $\etcRssRatioMax$.
With the precompute alone on one CPU core, one detector and $\etcGFivePairs$ paired jobs, the
per-point speedup is $\etcGFiveRatio$.

\begin{table}
\caption{\label{tab:early-multirate}
Steady-state ILE cost per intrinsic point for production and multirate data, by number of detectors.
Times and peak host memory (RSS) are medians of $\etcRepeats$ paired repeats; the steady state excludes
once-per-job data and PSD loading.
Lower rows: a separate set of $\etcXpRepeats$ paired repeats with the multirate template pieces generated on the GPU.}
\begin{ruledtabular}
\begin{tabular}{ccccc}
Detectors & Production (s) & Multirate (s) & Ratio & Host RSS (GB) \\
\hline
1 & $\etcProdSOne$ & $\etcMultiSOne$ & $\etcRatioOne$ & $\etcProdRssOne\to\etcMultiRssOne$ \\
2 & $\etcProdSTwo$ & $\etcMultiSTwo$ & $\etcRatioTwo$ & $\etcProdRssTwo\to\etcMultiRssTwo$ \\
3 & $\etcProdSThree$ & $\etcMultiSThree$ & $\etcRatioThree$ & $\etcProdRssThree\to\etcMultiRssThree$ \\
\hline
\multicolumn{5}{l}{GPU waveform pieces} \\
1 & $\etcXpProdSOne$ & $\etcXpMultiSOne$ & $\etcXpRatioOne$ & \\
2 & $\etcXpProdSTwo$ & $\etcXpMultiSTwo$ & $\etcXpRatioTwo$ & \\
3 & $\etcXpProdSThree$ & $\etcXpMultiSThree$ & $\etcXpRatioThree$ & \\
\end{tabular}
\end{ruledtabular}
\end{table}

A job that evaluates a single point gains only $\etcJobRatioMin$ to $\etcJobRatioMax$ times, and
$\etcXpJobRatioMin$ to $\etcXpJobRatioMax$ times with GPU waveform pieces.
Reading frames and PSD files costs both paths the same and dominates such a job.
The full gain needs workers that evaluate many points per job.

\emph{Limits.}
The template conventions limit the method to low SNR (above).
The method has been tested for quasi-circular, aligned-spin signals only.
RIFT builds the late-segment modes from its own waveform calls, which a precessing model with one start
frequency per inertial mode does not support.
The detector response is static: Earth rotation and finite-arm response are not modelled.
For current-detector signals the integration dominates the per-point cost and the saving is small.
The code path is opt-in, and RIFT defaults are unchanged.

\section{Batched GPU evaluation of IMRPhenomXHM modes}
\label{app:gpu-xphm}

RIFT generates waveform modes on the CPU through \texttt{lalsimulation} \cite{lalsuite}, one intrinsic point
per call.
RIFT pull requests 453 and 465 add a batched implementation of parts of \textsc{IMRPhenomXPHM}
\cite{gwastro-mergers-IMRPhenomXP} that evaluates the modes of many points in one call, on a GPU with CuPy or
on the CPU with NumPy.
It is translated from LALSimulation \gxpLalVersion{} and has no JAX or Ripple dependency.

\emph{Scope.}
The port covers the \textsc{IMRPhenomXAS} $(2,2)$ mode \cite{gwastro-mergers-IMRPhenomXAS}, the
\textsc{IMRPhenomXP} NNLO twisting-up of the $\ell=2$ modes (PrecVersion \gxpPrecVersion), and the aligned-spin
\textsc{IMRPhenomXHM} modes $(2,1)$, $(3,3)$, $(3,2)$ with mode mixing, and $(4,4)$
\cite{gwastro-mergers-IMRPhenomXHM}.
Only LAL's default model versions are ported.
The per-point setup (fits, collocation solves) runs on the host, and the frequency-grid evaluation on the
device.

\emph{Validation.}
The reference is \texttt{SimInspiralChooseFDModes} with multibanding off.
The test uses $\gxpValPoints$ aligned-spin binaries ($\gxpValBbh$ binary black holes, $\gxpValNsbh$ neutron
star--black hole, $\gxpValBns$ binary neutron stars), with $f_{\rm min}$ from $\gxpValFminLo$ to
$\gxpValFminHi\,\mathrm{Hz}$ and durations from $\gxpValDurLo$ to $\gxpValDurHi\,\mathrm{s}$ at
$\gxpValSrate\,\mathrm{Hz}$.
Each mode is compared on its own norm for both signs of $m$.
The mismatch is maximized over time and phase with the aLIGO design noise curve.
The $(2,2)$, $(2,1)$, $(3,3)$ and $(4,4)$ modes agree at round-off (Table~\ref{tab:gpu-xphm}): mismatch at most $\gxpMisExactMax$ and
relative error at most $\gxpErrExactMax$.
Over $\gxpSOnePoints$ points the precessing $\ell=2$ modes reach a mismatch of $\gxpSOneMis$ and a relative
error of $\gxpSOneErr$.
CuPy and NumPy agree to $\gxpCupyNumpy$ of the largest mode amplitude ($\gxpSOneCupyNumpy$ for the
precessing modes).

The $(3,2)$ mode agrees less closely: relative error at most $\gxpErrThreeTwo$ (median $\gxpErrMedThreeTwo$).
LAL computes collocation values of its intermediate phase from a second finite difference with step
$\gxpFdStep$ in long double, which amplifies rounding differences in the setup.
Reproducing LAL's arithmetic on the host (C \texttt{libm}, part-wise complex division, single-precision
mixing coefficients, the GSL LU solver) brings the other modes to round-off, but not the $(3,2)$.
Its mismatch is at most $\gxpMisThreeTwo$.

\emph{Cost.}
A call costs at least $\gxpFloorMs\,\mathrm{ms}$ (host setup, kernel launches), so a single-point call
is slower than LAL.
With $\gxpBatch$ points per call the GPU is $\gxpSpeedShort$ times faster than LAL on the short grid and
$\gxpSpeedLong$ times faster on the long grid.
Memory traffic limits long grids: at $\gxpLongDur\,\mathrm{s}$, ten and $\gxpBatch$ points per call give
$\gxpCupyLongTen$ and $\gxpCupyLongBatch$ points per second.
On a $\gxpXLongDur\,\mathrm{s}$ grid at $\gxpXLongSrate\,\mathrm{Hz}$, ten points use
$\gxpPoolXLongTen\,\mathrm{GB}$ of GPU memory and run at $\gxpCupyXLongTen$ points per second, against
$\gxpLalXLongMin$ to $\gxpLalXLongMax$ for LAL.
$\gxpBatch$ points do not fit, since the output alone is $\gxpOutXLongGB\,\mathrm{GB}$.
A repeat run on the same card was up to $\gxpRepeatMax$ times slower.

\emph{Gaps.}
Not yet ported: twisting-up of the $\ell=3,4$ modes, the MSA and SpinTaylor precession angles, RIFT's
mode conditioning (tapers, inverse FFT), the frame rotation for $\ell>2$, and fused device kernels.
Only the opt-in multirate path of ILE calls the port.
In Appendix~\ref{app:early-multirate}, CPU generation of these aligned modes is the largest remaining
per-point stage.
With this port the stage moves to the GPU (Table~\ref{tab:early-multirate}).

\begin{table}
\caption{\label{tab:gpu-xphm}
Batched \textsc{IMRPhenomXHM} modes against LALSimulation.
Top: largest mismatch and relative error per mode over $\gxpValPoints$ points (NumPy backend).
Bottom: points per second for five modes from $\gxpBenchFmin\,\mathrm{Hz}$, with one and $\gxpBatch$ points per call, on one \gxpGpuModel{} GPU;
NumPy and LAL use one thread of a \gxpCpuModel{} CPU.
GPU memory is the CuPy pool after the $\gxpBatch$-point call.}
\begin{ruledtabular}
\begin{tabular}{lccccc}
Mode & \multicolumn{2}{c}{Max mismatch} & \multicolumn{3}{c}{Max relative error} \\
\hline
$(2,2)$ & \multicolumn{2}{c}{$\gxpMisTwoTwo$} & \multicolumn{3}{c}{$\gxpErrTwoTwo$} \\
$(2,1)$ & \multicolumn{2}{c}{$\gxpMisTwoOne$} & \multicolumn{3}{c}{$\gxpErrTwoOne$} \\
$(3,3)$ & \multicolumn{2}{c}{$\gxpMisThreeThree$} & \multicolumn{3}{c}{$\gxpErrThreeThree$} \\
$(4,4)$ & \multicolumn{2}{c}{$\gxpMisFourFour$} & \multicolumn{3}{c}{$\gxpErrFourFour$} \\
$(3,2)$ & \multicolumn{2}{c}{$\gxpMisThreeTwo$} & \multicolumn{3}{c}{$\gxpErrThreeTwo$} \\
\hline
Grid & Points/call & CuPy & NumPy & LAL & GPU memory (GB) \\
\hline
$\gxpShortDur\,\mathrm{s}$, $\gxpShortSrate\,\mathrm{Hz}$ & 1 & $\gxpCupyShortOne$ & $\gxpNumpyShortOne$ & $\gxpLalShortOne$ & \\
 & $\gxpBatch$ & $\gxpCupyShortBatch$ & $\gxpNumpyShortBatch$ & $\gxpLalShortBatch$ & $\gxpPoolShortBatch$ \\
$\gxpLongDur\,\mathrm{s}$, $\gxpLongSrate\,\mathrm{Hz}$ & 1 & $\gxpCupyLongOne$ & $\gxpNumpyLongOne$ & $\gxpLalLongOne$ & \\
 & $\gxpBatch$ & $\gxpCupyLongBatch$ & $\gxpNumpyLongBatch$ & $\gxpLalLongBatch$ & $\gxpPoolLongBatch$ \\
\end{tabular}
\end{ruledtabular}
\end{table}

\section{Likelihood-guided transverse-spin seed proposals}
\label{sec:transverse_seed_climb}

An initial grid concentrated near aligned spins can miss likelihood support
at large transverse spin. Changing the transverse components alone can move
a point away from a narrow ridge in mass and aligned spin.
We therefore provide an optional search that follows these correlations
while placing additional intrinsic grid points. The resulting points receive
ordinary ILE evaluations before entering the likelihood interpolation.
This separates the search objective from the likelihood used for inference.

Let $\lambda$ denote intrinsic parameters and $\xi$ extrinsic parameters.
The target for interpolation remains
\begin{equation}
 {\cal L}_{\rm marg}(\lambda)
 =\int d\xi\,p(\xi){\cal L}(d\mid\lambda,\xi).
 \label{eq:seed_target}
\end{equation}
For the search, a faster approximation integrates time, distance, orbital
phase and polarization with the JAX likelihood kernel. An importance-weighted
batch covers sky position and inclination. Holding that batch fixed across
climbs reduces random changes in their relative scores. The default proposal
adapts sky position while drawing the cosine of inclination from its prior.
The approximate objective selects locations; its values never replace
Eq.~\ref{eq:seed_target} in the fit or evidence calculation.

The search uses phase coordinates $(\mu_1,\mu_2)$ that reduce correlations
between chirp mass, mass ratio and aligned spin. Its intrinsic chart is
\begin{equation}
 x=(\mu_1,\mu_2,q,\chi_{2z},\chi_{1\perp},\phi_1),
 \qquad \chi_{i\perp}=\sqrt{\chi_{ix}^2+\chi_{iy}^2},
 \label{eq:seed_chart}
\end{equation}
where $q=m_2/m_1\leq1$ and $\phi_i$ is the in-plane spin azimuth.
The phase-coordinate inverse determines chirp mass and primary aligned spin
at fixed $q$ and $\chi_{2z}$. An optional extension also varies
$(\chi_{2\perp},\phi_2)$; otherwise the secondary transverse spin vanishes.
Points violating the Kerr spin bound are excluded.

A covariance matrix adaptation evolution strategy first performs free climbs
from promising input points. Further climbs pin $\chi_{1\perp}$ at selected
values and optimize the remaining coordinates. Each pinned climb starts
from the nearest completed solution, projected into the physical region.
This supplies transverse-spin coverage even when the free optimum lies
near aligned spin. Seed placement draws from broadened final climb
covariances, with a spread in the pinned coordinate between adjacent rungs.
Only climbs within a chosen objective difference from the best climb
contribute. These draws supplement the training grid and do not constitute
posterior samples or modify the physical prior.

The implementation, \texttt{util\_TransverseSeedClimb.py}, supports setup-time
placement, placement before the first ILE batch, a parallel seed branch,
and placement from an already evaluated composite. In composite mode,
the initial ILE consolidation supplies the starting points. The climb runs
alongside the first fit, and separate ILE jobs evaluate its proposals.
The next consolidation waits for both ordinary and seed evaluations.
Container execution allows the climb and production ILE to use separate
images, since their numerical dependencies differ.

The search should retain the analysis sampling rate and frequency band.
Any approximation used for placement must preserve the likelihood support
that the production analysis evaluates. Production ILE re-scoring remains
necessary even when the search uses a different waveform model.

\section{Warm-starting adaptive extrinsic integration on a loud event}
\label{ap:review:warmstart}
Warm starting reuses samples from an earlier extrinsic integration to
seed the next one \cite{gwastro-RIFT-roboto-scaling}. We provide it as a
proof-of-concept tool for future development, not as a validated
acceleration: it \emph{can} bias the evidence. For an importance-sampling
proposal, correct weighting preserves the integral provided support is
retained. Adaptive-volume integration instead contracts the live volume
from the seed, so likelihood support the seed omits is never recovered.

\begin{figure}
  \includegraphics[width=\columnwidth]{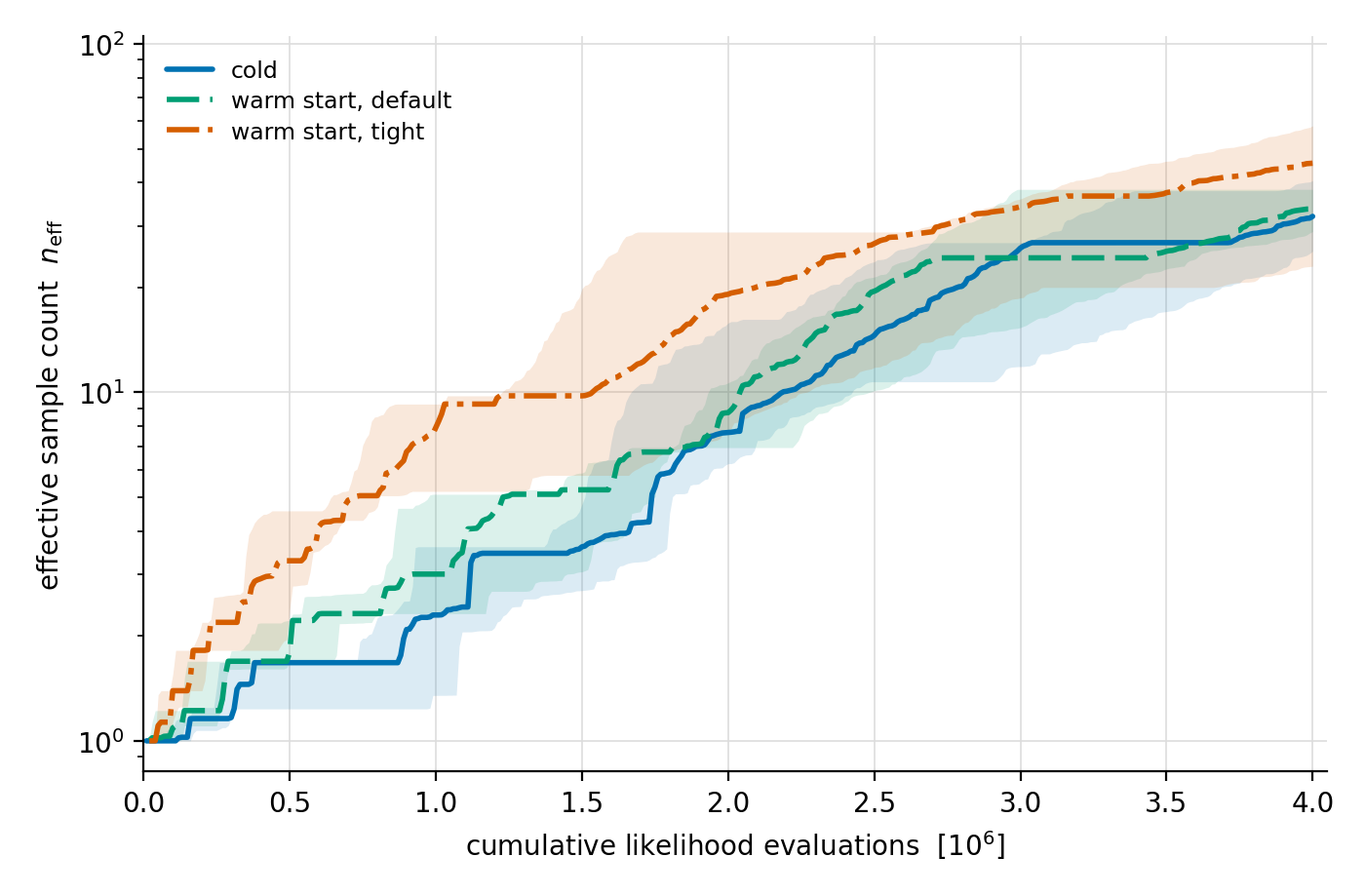}
  \caption{\label{fig:int:warmstart}Adaptive-volume integration for GW250114:
  effective sample count versus likelihood evaluations. Lines show the median
  of three seeds and bands their range.
  Cold (solid) and default warm-start (dashed) runs have comparable growth;
  tighter initialization (dash-dotted) accelerates growth but lowers the
  evidence.\llmfig}
\end{figure}

Figure~\ref{fig:int:warmstart} shows this tradeoff for GW250114. At the
default coverage setting, warm and cold runs have comparable
effective-sample growth. Tighter initialization accelerates growth but
lowers the evidence, consistent with under-coverage. Faster growth alone
therefore does not show an accurate acceleration, and these runs do not
establish the cold result as an unbiased reference. These runs sample the
sky in equatorial (inertial-frame) coordinates; we expect better
performance with detector-frame sky coordinates
(App.~\ref{ap:review:skyframe}), which we have not tested here.

\section{Use of large language models}
\label{app:llm}

We used large language model (LLM) agents throughout this work, following the
APS policy on the appropriate use of AI tools. We directed these agents in
a role qualitatively equivalent to that of research staff. They carried out
assigned calculations, returned preliminary reports and findings, and
drafted initial proposals for material to incorporate into the text,
including first drafts of many sections. LLM
agents operated the analysis and plotting tools that produced every figure
in this paper, and generated those figures with those tools. The authors
revised all drafted text and decided what entered the paper. All
calculations were carried out under exhaustive human direction. Both the
calculations and the code developments behind them were subjected to
adversarial review and to repeated cross-validation tests. The workflow,
its documentation conventions, and its review procedures are described in
Ref.~\cite{whitepaper-transparent-agentic-science}. The authors take full
responsibility for the content of this paper.

\end{document}